\documentclass[preprint,authoryear]{elsarticle}
\usepackage{color}
\usepackage{a4wide}
\usepackage{amsmath}
\usepackage{amsfonts}
\usepackage{nicefrac}
\usepackage{graphicx}
\usepackage{color}
\usepackage{setspace}
\usepackage{newtxtext,newtxmath} 
\usepackage{bm}
\usepackage{tcolorbox}
\usepackage{float}
\usepackage{caption}
\usepackage{subfigure}
\usepackage{tikz}
\usepackage{subfigure}
\usepackage{makecell}
\usepackage{setspace}
\usepackage[colorlinks=true, citecolor=blue, linkcolor=black, urlcolor=blue]{hyperref}
\usetikzlibrary{calc}
\parskip \baselineskip

\parskip \baselineskip

\begin{document}
\begin{frontmatter}

\title{On the statistical physics and thermodynamics of polymer networks: a Eulerian theory for entropic elasticity}

\author[first]{Siyu Wang}
\author[second]{Heng Xiao}
\author[first]{Lin Zhan \corref{cor}}
\cortext[cor]{Corresponding author.}
\ead{lzhan@jnu.edu.cn}

\address[first]{Department of Mechanics, School of Mechanics and Construction Engineering, Jinan University, Guangzhou, 510632, China}

\address[second]{School of Intelligent Manufacturing and Future Technologies \& MOE Key Lab for Intelligent Manufacturing and Future Technologies of High-End Equipment, Fuyao University of Science and Technology, Fuzhou, 350100, China}	

\begin{abstract}
This study presents a Eulerian theory to elucidate the molecular kinematics in polymer networks and their connection to continuum deformation, grounded in fundamental statistical physics and thermodynamics, and free from phenomenological assumptions and additional parameters. Three key innovations are incorporated:
\begin{enumerate}
	\item The network behavior is described through a global thermodynamic equilibrium condition that maximizes the number of accessible microstates for all segments, instead of directly dealing with the well-established single-chain models commonly adopted in traditional approaches. A variational problem is then posed in the Eulerian framework to identify this equilibrium state under geometric fluctuation constraints.  Its solution recaptures the classical single-chain model and reveals the dependence of chain kinematics upon continuum deformation.
	
	\item The chain stretch and orientation probability are found to be explicitly specified through the Eulerian logarithmic strain and  spatial direction. 
	The resulting hyperelastic model, with only two physical parameters, outperforms the extant models with same number of parameters. It further provides a physical justification for prior models exhibiting superior predictive capabilities: the model becomes equivalent to the Biot-chain model \citep{zhan2023new} at moderate deformations, while converging to the classical Hencky strain energy in the small strain limit.
	
	\item A novel biaxial instability emerges as a phase transition in chain orientation. At sufficiently large deformation, chains increasingly align with the primary stretched direction, depleting their density in other directions. Consequently, the stresses in non-primary stretched directions would decrease as the loss in chain density outweighs the gain in chain force. For equal biaxial tension, instability is therefore triggered because perfect equality of the two principal stretches without any perturbation is practically unachievable.
\end{enumerate}\vspace{-1em}
The theory establishes a physical picture of the network response under thermodynamic equilibrium:  chain constrains but segment fluctuates. It is the statistical behavior of the latter, under the structural influence of the former, that governs the continuum response. Specifically, the macroscopic response of the network may not emerge from simplistic extensions of the chain-level models, but arises as a natural consequence of the underlying \textit{segment-level} statistics. The theory also necessitates a Eulerian statistical perspective, as random thermal fluctuations prevent continuous tracking of any specified chains (material description) during deformation. Consequently, the Lagrangian framework may not be well suited in this context and chain stretch and orientation probability need be treated as Eulerian/spatial field variables. These perspectives not only advance the theoretical foundations of constitutive modeling for soft polymer networks, but also offer new insights into the microstructural origin of their macroscopic behavior.
\end{abstract}

\begin{keyword}
	hyperelastic model, statistical mechanics, micro-macro transition, chain orientation, logarithmic strain, biaxial instability
\end{keyword}
\end{frontmatter}

\section{Introduction}
Soft polymer networks exhibit highly nonlinear reversible responses under large deformation. From a phenomenological perspective, models based on well-established constitutive theories in continuum mechanics have offered convenient and practical approaches to describe a wide range of polymeric solids and are readily integrated into engineering applications. Representative developments include the Mooney–Rivlin model \citep{mooney1940theory,rivlin1948large}, the Ogden model \citep{ogden1972large}, and non-polynomial models \citep{gent1996new,hartmann2003polyconvexity,xiao2012explicit,anssari2022assessment}, among others. Comprehensive investigation of their performance may be found in review articles such as \cite{boyce2000constitutive}, \cite{marckmann2006comparison}, \cite{steinmann2012hyperelastic}, and \cite{dal2021performance}.

While phenomenological models need not refer to the underlying complex mechanisms of macroscopic behaviors, this simplicity prevents them from distinguishing key qualitative differences between soft and hard materials.  For example, soft elastomers often stiffen upon heating, whereas hard solids (e.g., metals, rocks) typically soften. This phenomenon, known as the Gough-Joule effect, was not addressed until the advent of entropic elasticity theory \citep{kuhn1934gestalt,guth1934innermolekularen,james1943theory,flory1953principles,treloar1975physics,weiner2012statistical}, which laid the foundation for modern statistical theories of polymer elasticity. Within the entropic elasticity framework, the conformational entropy associated with bond orientations in long flexible macromolecular chains constitutes the primary contribution to the free energy of polymers. In contrast, the elasticity in hard solids is predominantly driven by internal energy changes.  Such insights in conjunction with statistical approaches underlie the classical single-chain models, which admit closed-form expressions for the free energy as functions of the chain end-to-end configuration.

Toward extending single-chain models to the continuum level, it is imperative to establish a linkage between chain end-to-end configuration and the continuum deformation. This issue has long constituted a central theme in the microstructurally motivated modeling of soft materials \citep{Flory1943statistical,wang1952statistical,arruda1993three,miehe2004micro,xiang2018general,amores2021network,zhan2023new}.  A conventional strategy to depict the chain configuration entails constructing discrete chain network structures with assumed geometries, exemplified by the three-chain model \citep{wang1952statistical}, the four-chain model \citep{Flory1943statistical}, and the eight-chain model \citep{arruda1993three}, etc. Although these classical models provide practical approaches, they often prove inadequate in capturing the full stress responses under complex multiaxial loading conditions \citep{boyce2000constitutive,marckmann2006comparison,dal2021performance,zhan2023new}. Moreover, their assumption of discrete and specified chain orientations precludes the deformation-induced anisotropic reorientation of chains  —  a key mechanism underlying critical phenomena including birefringence \citep{treloar1975physics} and electromechanical coupling \citep{cohen2016electroelasticity}.

In contrast to discrete chain network models, the full-network approach \citep{treloar1975physics}, also known as the microsphere model \citep{miehe2004micro}, constitutes another major class of micromechanical formulations capable of capturing direction-dependent chain kinematics. This framework assumes molecular chains to be initially distributed uniformly over all spatial directions and to respond independently to the imposed macroscopic deformation. The most prevalent formulation within this class is the affine full-network model, which postulates that the deformation (stretch and orientation) of each chain replicates that of material line element governed by continuum kinematics \citep{treloar1979non,wu1993improved,goktepe2005micro,alastrue2009anisotropic,vernerey2017statistically,vernerey2018statistical,ogouari2024multiscale}. 
However, substantial literature has demonstrated that affine full-network model fail to match the multiaxial experimental data \citep{boyce2000constitutive,marckmann2006comparison,steinmann2012hyperelastic,dal2021performance,zhan2023general,zhan2023new}. Furthermore, their predictions of chain orientations are also conceptually problematic: under affine assumption, chains in all types of soft elastomers subjected to the same macroscopic deformation are expected to undergo identical orientational evolution, a conclusion contradicting experimental observations \citep{treloar1975physics,sun2021nonlinear}.
In a most extreme scenario, polymer solutions with negligible shear stiffness subjected to external deformation would retain their isotropic chain distribution, rather than exhibit a pronounced anisotropic alignment in analogy to those observed in highly cross-linked elastomers. These inconsistencies indicate that the affine assumption may deviate from the physical reality of orientation dynamics in deforming soft materials.

Instead of specifying a direct relation between the chain kinematics and macroscopic deformation, \cite{tkachuk2012maximal} developed a new non-affine approach that constrains the average microscopic chain stretch to be consistent with the macroscopic deformation. Such an assumption indeed yields a stress mapping that relates the chain force to the macro stress \citep{rastak2018non,diani2019fully,govindjee2019fully,xiao2021micromechanical,arunachala2021energy}. However, this non-affine mapping, as shown in \cite{diani2019fully} and \cite{zhan2023new}, does not outperform the affine model in describing the multi-axial elastic behaviors. Alternative non-affine strategies have also been proposed to characterize chain kinematics within the full-network framework \citep{miehe2004micro,khiem2016analytical,zhan2022microstructural,yang2025hyperelastic}. These models demonstrate enhanced flexibility in capturing macroscopic responses. However, their incorporation of supplementary assumptions and parameters may introduce challenges regarding physical plausibility and predictive capability.

Traditional network models do not account for the interactions between individual polymer chains and their surrounding environment. This issue is addressed in the entangled-network models, which augment classical entropic chain models with topological constraint energy \citep{heinrich1997theoretical,mcleish2002tube,khiem2016analytical,xiang2018general,darabi2021generalized,wang2024correlation,yang2025hyperelastic}. In these frameworks, chain kinematics is confined within a deformable tube where topological constraints intensify as tube cross-section diminishes.  While significantly improving multi-axial performance, these models typically require additional parameters and multi-experiment calibration \citep{steinmann2012hyperelastic, dal2021performance}. Moreover, establishing quantitative relationships between tube dimensions and macroscopic observable remains challenging.

In this work, we aim to establish a connection between the chain kinematics and continuum deformation through fundamental principles of thermodynamics and statistical mechanics, without invoking non-physical assumptions or additional parameters. In our theory, the polymer network is conceptualized as a hierarchical ensemble of weakly interacting rod-like segment particles. These segments are treated as the basic statistical units under thermal fluctuation, while the chains act as geometrical constraints that shape the segmental energy landscape. A variational formulation toward maximizing the global network entropy simultaneously recovers classical single-chain models and predicts the macroscopic network response. The Eulerian logarithmic strain naturally emerges as the unique deformation measure that explicitly governs the chain kinematics and orientation probability. The resulting model, with minimal physical parameters, exhibits strong predictive capability across a wide range of multiaxial loading conditions. Furthermore, the theory predicts a previously unreported biaxial instability as a  phase transition in the collective chain orientation.

The remainder of this paper is structured as follows. Section 2 develops the statistical and thermodynamic theory of polymer network reconceptualized as a hierarchical ensemble of segments, along with the associated variational formulation and solutions. Section 3 presents the resulting continuum models and discusses their performance and analytical properties, including comparisons to classical models in limiting deformation regimes. Section 4 investigates the emergence of biaxial instability from the collective alignment of chains. Section 5 presents further discussions on the core perspectives of the current theory. Section 6 summarizes the main findings and proposes potential applications of the theory.

\section{Statistical physics and thermodynamics of polymer network}
Our theory starts with the conception of Kuhn segment, which serves as the fundamental statistical unit in the freely jointed chain model. The preliminary background is briefly revisited as follows.

\begin{figure}[htbp]
	\centering       
	\includegraphics[width=155.00mm,height=44.00mm]{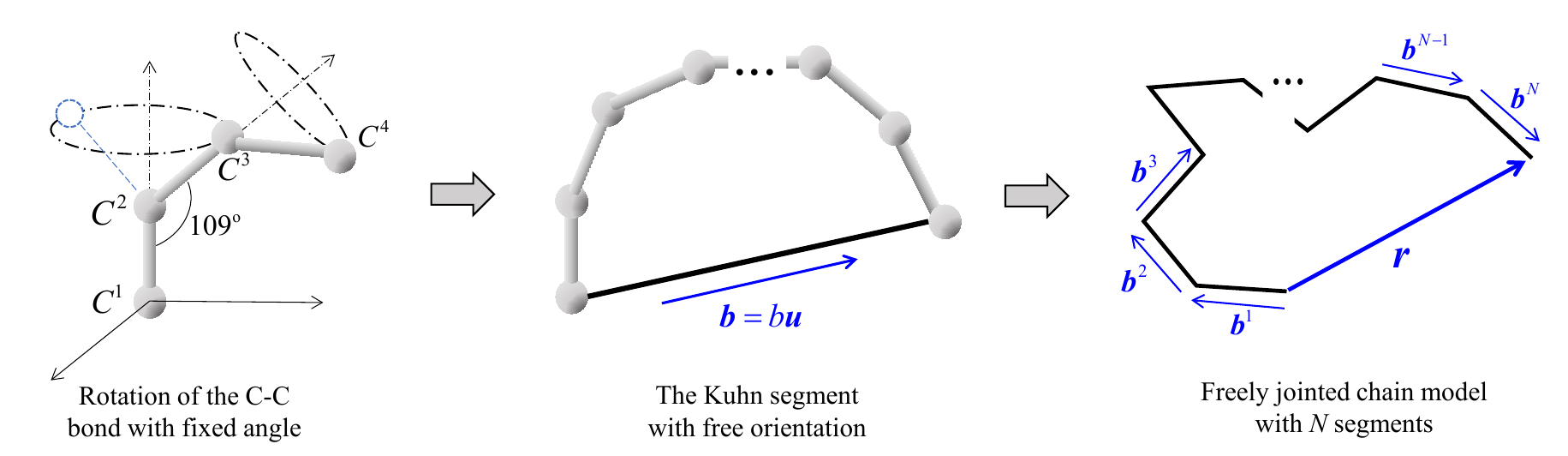}
	\caption{A sketch for the freely jointed chain model.} 
	\label{fig:chain} 
\end{figure}

Consider a macromolecular chain  with a large number of repeating units, say the C-C bonds. In general, the angle between two neighboring C-C bonds remains around 109° so as to minimize electron pair repulsion; see, for example, \cite{weiner2012statistical,clayden2012organic}. As shown in Fig. \ref{fig:chain}, this equilibrium angle cannot be easily changed but allows the C$^2$-C$^3$ bond to adopt any orientation on a conical surface at 109° relative to the C$^1$-C$^2$ bond, with all directions being equally probable. Consequently, as the number of consecutively bonded atoms increases, the end-to-end orientation from the first to the last atom becomes increasingly stochastic and less constrained by the fixed bond angle.

A Kuhn segment \citep{kuhn1934gestalt,flory1953principles} is defined as the effective statistical unit with completely random orientation, uncorrelated with neighboring segments. It can be treated as a freely rotating rigid rod of average length $b$  —  a constant molecular parameter determined by fixed bond lengths and bond angles under statistical equilibrium. A macromolecular chain can then be conceptualized as a sequence of $N$ such segments, each oriented independently and randomly, as illustrated in Fig.~\ref{fig:chain}. This idealized representation constitutes the classical freely jointed chain model.

Let $\bm b^i (i=1, 2, ..., N)$ denote the configuration vector of the $i$-th segment. The chain end-to-end vector $\bm r$ is then expressed as 
\begin{equation}
	\label{r constraint dis}
	\bm r=\bm b^1+\bm b^2+...+\bm b^N=\sum_{i=1}^{N}\bm b^i.
\end{equation}
Due to the randomness of the segment orientation, a specified $\bm r$ can accommodate plenty of different microscopic segment conformations, each representing a distinct microstate. Under a specified force, the most probable state (thermodynamic equilibrium) of the chain corresponds to the configuration $\bm r$ containing the largest number of microstates, i.e., maximal entropy. This observation underlies the classical entropic elasticity theory, wherein the free energy of a chain is expressed as a function of the end-to-end distance \citep{kuhn1934gestalt,james1943theory,flory1953principles,treloar1975physics}.

As outlined in Section 1, extending the single-chain model to continuum level is known as the micro-macro transition problem. It is rendered, under most circumstances, independent of a single chain's internal statistical features and relies on several phenomenological assumptions. In this study, however, we depart from these decoupled treatments and develop a statistical theory that relates the global thermodynamic equilibrium of the whole network directly to the elementary segment conformations. It will be shown that the macroscopic response of the network does not emerge from simplistic extensions of specified chain-level models, but arises as a natural consequence of the underlying segment-level statistics.

\subsection{Polymer network as a hierarchical ensemble of segments}

Consider an incompressible polymer network of unit volume under thermodynamic equilibrium at finite temperature $T$. Let $M$ denote the number of chains per unit volume and $N$ the number of segments per chain \footnote{ In realistic polymer networks, chain  lengths may be polydisperse, i.e., chains have different numbers of segments. In such cases, $N$ can be naturally interpreted as the statistical average of the segment number per chain. This approximation is physically justified, as the macroscopic response of the network primarily depends on collective statistical behavior rather than the precise configuration of each individual chain.}. Due to the incompressibility and mass conservation constraints, both $M$ and $N$ remain constant in the elastic region.
Rather than describing the network behavior based on predefined chain models,  we here treat the network as a global system composed of $MN$ segments. Each segment is idealized as a rod-like particle, whose microscopic configuration is fully characterized by its orientation. Let $\bm{u}^k$ be the unit vector denoting the orientation of the $k$-th segment. Then a complete specification of the network conformation is given by the set $\{ \bm{u}^1, \bm{u}^2, \dots, \bm{u}^{MN} \}$, which corresponds to a definite \textit{microstate} of the network. In general, a given macrostate \footnote{In the current problem, this corresponds to a given temperature and stress/deformation.} contains numerous microstates. Per Ludwig Boltzmann’s profound insight \citep{boltzmann1877beziehung,boltzmann2015relationship}, thermodynamic equilibrium of the network corresponds to the very macrostate that maximizes the number of accessible microstates, namely, the one with maximal conformational entropy.

\begin{figure}[htbp]
	\centering       
	\includegraphics[width=120.00mm,height=60.00mm]{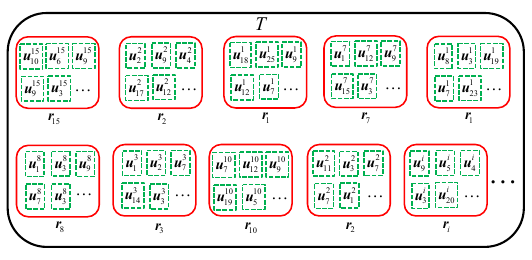}
	\caption{A sketch for the hierarchical ensemble of segments in the network.} 
	\label{fig:ensemble} 
\end{figure}

To properly define the microstate of the network without redundancy or omission, we consider a hierarchical ensemble of segments, as illustrated in Fig. \ref{fig:ensemble}. The segments are treated as identical but distinguishable, thus following the classical Boltzmann statistics. They are first classified according to the configuration of the chains to which they belong. In Fig. \ref{fig:ensemble}, each red box represents a chain, while each green dashed box represents a segment within that chain. The notation $\bm{u}_j^i$ denotes a segment oriented along $\bm{u}_j$, belonging to the chain with configuration $\bm{r}_i$. Segment and chain ordering is arbitrarily assigned  to emphasize their stochastic nature.

Within this classification, two segments $\{\bm{u}_m^i, \bm{u}_n^j\}$ are rendered non-interchangeable if $i \neq j$, regardless of their individual orientations $\bm{u}_m$ and $\bm{u}_n$ are identical ($m = n$) or not ($m \neq n$). Namely, exchanging them yields new microstate. For example, two network conformations $\{\bm{u}_9^1, \bm{u}_4^2, Y\}$ and $\{\bm{u}_9^2, \bm{u}_4^1, Y\}$ are regarded as distinct microstates, where $Y$ denotes the specific conformation of the remaining $MN-2$ segments.

\begin{figure}[htbp]
	\centering       
	\includegraphics[width=120.00mm,height=100.00mm]{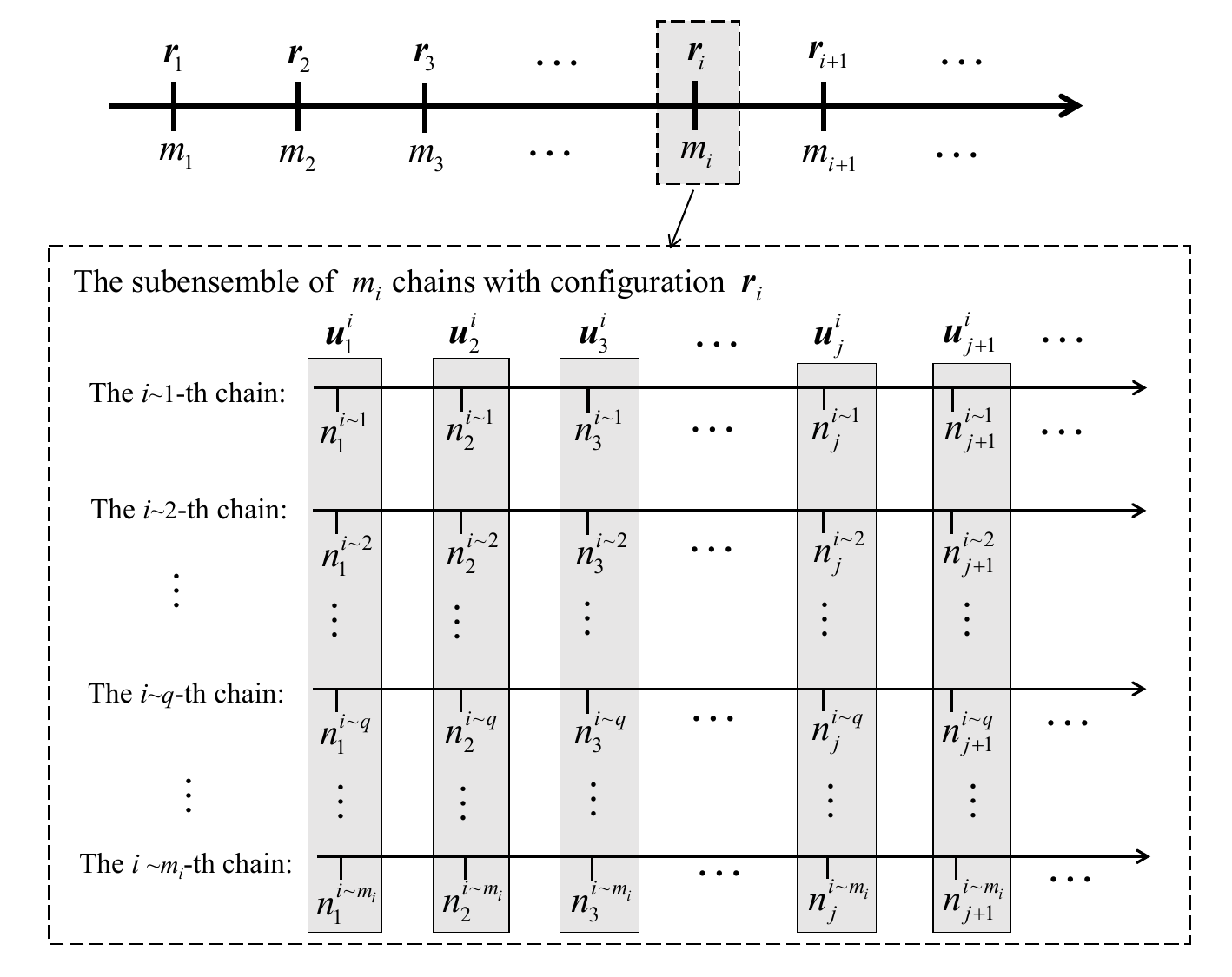}
	\caption{A sketch for the subensemble of segments in chains with specified configuration.} 
	\label{fig:subensemble} 
\end{figure}

On the other hand, let $m_i$ denote the number of chains occupying a specified configuration $\bm{r}_i$. As illustrated in Fig. \ref{fig:subensemble}, these chains collectively form a subensemble,  within  which segments can be further classified according to their individual orientations. Let $n_j^{i\sim q}$ represents the number of segments oriented along $\bm{u}_j$ in the $q$-th chain of the $\bm{r}_i$-subensemble. Since chains in a specified subensemble possess the same configuration, exchanging any two segments of identical orientation within this group does not give rise to a new microstate. Namely, any permutation of segments in a gray shade box in Fig. \ref{fig:subensemble} corresponds to the same microstate.

Let $s_j^i$ denote the number of segments adopting configuration $\bm u_j^i$, such that
\begin{equation}
	\label{sji}
	s_j^i = n_j^{i\sim 1} + n_j^{i\sim 2} + n_j^{i\sim 3} + \cdots + n_j^{i\sim m_i}.
\end{equation}
The total number of segments in the $\bm r_i$-subensemble is then given by
\begin{equation}
	\label{sji}
	\sum_{j=1}^{\infty} s_j^i = N m_i.
\end{equation}
Moreover, summing over all subensembles yields
\begin{equation}
	\label{MN}
	\sum_{i=1}^{\infty} \sum_{j=1}^{\infty} s_j^i = N \sum_{i=1}^{\infty} m_i = MN.
\end{equation}

At specified distribution $\{s_j^i\}$, the number of possible microstates for the $MN$ segments is then given by
\begin{equation}
	\label{Omega}
	\Omega = \dfrac{(MN)!}{\prod_{j=1}^{\infty} \left(s_j^1!\right) \cdot \prod_{j=1}^{\infty} \left(s_j^2!\right) \cdot \prod_{j=1}^{\infty} \left(s_j^3!\right) \cdots} = \dfrac{(MN)!}{\prod_{i=1}^{\infty} \prod_{j=1}^{\infty} \left(s_j^i!\right)}.
\end{equation}
As discussed, the factorials in the denominator account for the interchangeability of segments that share the same configuration $\bm u_j^i$ within each subensemble.

The total entropy of the network is then expressed as
\begin{equation}
	\label{networkentropydiscrete1}
	\begin{split}
		S = k_{\text B} \ln \Omega &= k_{\text B} \left( \ln (MN)! - \sum_i^{\infty} \sum_j^{\infty} \ln s_j^i! \right) \\
		&= k_{\text B} \left( MN \ln MN - MN \right) - k_{\text B}\sum_i^{\infty} \sum_j^{\infty} \left( s_j^i \ln s_j^i - s_j^i \right) \\
		&= -k_{\text B} MN \sum_i^{\infty} \sum_j^{\infty} \left( \dfrac{s_j^i}{MN} \ln \dfrac{s_j^i}{MN} \right)
	\end{split}
\end{equation}
with $ k_{\text B}$ the Boltzmann's constant. Here, the Stirling’s approximation $\ln x!= x\ln x-x$ has been applied.

Define the proportion of chains adopting configuration $\bm r_i$ as 
\begin{equation}
	P_i = \dfrac{m_i}{M},
\end{equation}
and  the proportion of segments orienting direction $\bm u_j$ within the $\bm r_i$-subensemble as
\begin{equation}
	p_j^i = \dfrac{s_j^i}{N m_i}.
\end{equation}
Then Eq.  (\ref{networkentropydiscrete1}) can be reformulated as
\begin{equation}
	\label{networkentropydiscrete2}
	S = -k_{\text B} M N \sum_i^{\infty} \sum_j^{\infty} \left( P_i p_j^i \ln P_i p_j^i \right).
\end{equation}

In the absence of quantum effects, we may take the spacing between neighboring chain configurations and segment orientations to be infinitesimal. Then Eq.  (\ref{networkentropydiscrete2}) can be expressed as an integral over all possible configurations:
\begin{equation}
	\label{networkentropycontinuous}
	S = -k_{\text B} M N \iint P_{\bm r} p_{\bm u}^{\bm r} \ln \left( P_{\bm r} p_{\bm u}^{\bm r} \right) \, \text{d} \bm u \, \text{d} \bm r.
\end{equation}
Here and in what follows, $\iint f(\bm r, \bm u) \text{d} \bm u \text{d} \bm r$ denotes integration over all admissible chain configurations $\bm r$ and segment orientations $\bm u$. In the above, 
\begin{equation}
	\label{chain probability}
	P_{\bm r} \equiv P_i= \dfrac{m_i}{M}
\end{equation}
can be also interpreted, according to standard statistical arguments, as the probability of finding a chain in configuration $\bm r$ if the chain number $M$ is large. Attention must be drawn to the term
\begin{equation}
	\label{segment probability}
	p_{\bm u}^{\bm r} \equiv p_j^i = \dfrac{s_j^i}{N m_i},
\end{equation}
which denotes the probability for a segment adopting orientation $\bm u_j$ given that it belongs to the $\bm r_i$-subensemble. Owing to the identity of chains and segments, $p_{\bm u}^{\bm r}$ may equivalently be regarded as the probability of a segment having orientation $\bm u_j^i$ in each chain with configuration  $\bm r_i$. These interpretations become increasingly accurate for large $N$.

Conservation of the total number of chains and segments is then reformulated to the following normalization conditions:
\begin{equation} 
	\label{chainnumbercon}
	1 = \int P_{\bm r} \, \text{d} \bm r,
\end{equation}
and
\begin{equation} 
	\label{segmentnumbercon}
	1 = \iint P_{\bm r} p_{\bm u}^{\bm r} \, \text{d} \bm u \, \text{d} \bm r = \int p_{\bm u}^{\bm r} \, \text{d} \bm u.
\end{equation}
The second equality holds since $P_{\bm r}$ and $p_{\bm u}^{\bm r}$ are statistically independent.

Eq.  (\ref{networkentropycontinuous}) can be further rewritten as
\begin{equation}
	\label{entropy}
	\begin{split}
		S&=-k_{\text B}MN\iint P_{\bm r}p_{\bm u}^{\bm r}\left(\ln  P_{\bm r}+\ln p_{\bm u}^{\bm r}\right)\text d\bm u\text d\bm r\\
		&=-k_{\text B}MN\int p_{\bm u}^{\bm r}\int P_{\bm r}\ln  P_{\bm r}\text d\bm r\text d\bm u-k_{\text B}MN\int P_{\bm r}\int p_{\bm u}^{\bm r}\ln p_{\bm u}^{\bm r}\text d\bm u\text d\bm r\\
		&=-k_{\text B}MN\int P_{\bm r}\ln  P_{\bm r}\text d\bm r-k_{\text B}MN\int P_{\bm r}\int p_{\bm u}^{\bm r}\ln p_{\bm u}^{\bm r}\text d\bm u\text d\bm r.\\
	\end{split}
\end{equation}
Several comments are in order regarding the simplifications above. If $N$ is large, the individual segment orientation $\bm u$ and the overall chain configuration $\bm r$ can be treated as statistically independent. However, ensemble-averaged properties of the segments within a chain may still exhibit dependence on the chain configuration $\bm r$. For example, the average segment orientation is directly related to $\bm r$, as made explicit in Eqs. (\ref{r constraint dis}) and   (\ref{constraint r}). As a consequence, the second term on the right-hand side of Eq.  (\ref{entropy}) cannot be reduced to a single integral, due to the $\bm r$-dependence of $\int p_{\bm u}^{\bm r}\ln p_{\bm u}^{\bm r}\text d\bm u$. In contrast, the first term simplifies because $\int P_{\bm r}\ln  P_{\bm r}\text d\bm r$ is treated as independent of an individual segment orientation $\bm u$.

\subsection{Thermodynamic and mechanical equilibrium}
Since the internal energy $U_0$ of the network remains constant in the entropic elasticity framework, the First and Second Laws of Thermodynamics for reversible process reduce to:
\begin{equation}
	\label{law}
	\mathrm{d}U_0 = T\,\mathrm{d}S + \bm{\sigma} : \mathrm{d}\bm{h} = 0
\end{equation}
which directly yields
\begin{equation}
	\label{law_S}
	\mathrm{d}S = -\frac{1}{T}\, \bm{\sigma} : \mathrm{d}\bm{h} \quad \Rightarrow \quad S = S(\bm{h}).
\end{equation}
Here $\bm{\sigma}$ and $\bm{h}$ are respectively the Cauchy stress and Eulerian logarithmic strain defined by $\bm h=\frac{1}{2}\ln\bm F\bm F^T$ with $\bm F$ the deformation gradient tensor. Eq.  (\ref{law_S}) implies that under thermodynamic equilibrium the network entropy can be fully determined by the deformation.

\textbf{Remark:} Eq. (\ref{law}) employs the Cauchy stress as the continuum stress measure, as it naturally characterizes the stress field in the current configuration and without the need to track any specified chains, in line with the Eulerian framework adopted in this study. For incompressible isotropic materials, the Eulerian logarithmic strain serves as the work-conjugate strain measure to the Cauchy stress \citep{hoger1987stress, ogden1997non, xiao2002hencky}. As will be shown later, an explicit Eulerian micro-macro connection can be established only through the logarithmic strain.

Using Eq. (\ref{entropy}), the Helmholtz free energy of the network can be expressed as
\begin{equation}
	\label{total free energy}
	\mathscr F = U_0-TS=M \int P_{\bm r} w_{\bm r}  \text{d} \bm r +k_{\text B} TMN \int P_{\bm r} \ln P_{\bm r}  \text{d} \bm r,
\end{equation}
with 
\begin{equation}
	\label{chain helmholtz}
	w_{\bm r} =u_0 - T s_{\bm r} = u_0 + k_{\text B}TN \int p_{\bm u}^{\bm r} \ln p_{\bm u}^{\bm r} \text{d} \bm u
\end{equation}
being the chain Helmholtz free energy at configuration $\bm r$.  Here $u_0$ is the chain internal energy satisfying  $U_0=Mu_0$, and $s_{\bm r}=-k_{\text B}N\int p_{\bm u}^{\bm r} \ln p_{\bm u}^{\bm r} \text{d} \bm u$ is the chain entropy.

With Eqs. (\ref{law} - \ref{total free energy}), the Cauchy stress at isothermal condition is expressed as
\begin{equation}
	\label{thermodynamic cauchy}
	\bm{\sigma} =-T\dfrac{\partial S}{\partial \bm{h}}+p \bm I= \dfrac{\partial \mathscr F}{\partial \bm{h}} +p \bm I= M \int  \left(P_{\bm r}  \dfrac{\partial w_{\bm r}}{\partial \bm{h}}+w_{\bm r}\dfrac{\partial P_{\bm r}}{\partial \bm{h}}\right)\text d\bm r +k_{\text B}T MN\int  \left(1+P_{\bm r}\right) \dfrac{\partial P_{\bm r}}{\partial \bm{h}}\text d\bm r+p \bm I.
\end{equation}
Here $p$ is a scalar Lagrangian multiplier accounting for the incompressible condition $\text d J=\text d\bm h:\bm I=0$ with $J=\text{det}\bm F=1$.

On the other hand, the continuum elastic energy is given by the sum of individual chain free energy
\begin{equation}
	\label{elastic energy}
	W = M \int P_{\bm r} w_{\bm r}  \text{d} \bm r.
\end{equation}
The above expression is widely regarded as the Helmholtz free energy in classical entropic elasticity theory, regardless of whether the chain orientation is assumed to be isotropic \citep{miehe2004micro, alastrue2009anisotropic, zhan2023new} or anisotropic \citep{wu1993improved, vernerey2017statistically, vernerey2018statistical}. This interpretation is also supported by \cite{treloar1975physics}, who articulated the foundational assumption for Gaussian networks that “\textit{the entropy of the network is the sum of the entropies of the individual chains} (see Page 61 of \cite{treloar1975physics}).”

However, the Eq. (\ref{elastic energy}) appears mathematically distinct from our theromodynamic Helmholtz free energy in Eq.  (\ref{total free energy}). 
The difference stems from the structure of the network entropy in Eq. (\ref{entropy}). A fundamental fact is that \textit{ the true total entropy of the network is not simply the sum of individual chain entropy}, denoted by $S^* = M \int P_{\bm r} s_{\bm r}  \text{d} \bm r$. Rather, the way in which $S^*$ is distributed across the network gives rise to an additional orientation entropy. This entropy emerges due to weak interactions among chains with different configurations, and it vanishes only in the limiting scenario where chains in the network are treated as non-interacting. In fact, the cross-linking junctions of the network undergo random motions coupled with the attached chain segments, which leads to statistical fluctuations of relative positions between junctions, namely the chain configuration \citep{treloar1975physics, xing2007thermal, zhan2025statistical}. This effect is captured by the additional orientation entropy in our theory.

The Cauchy stress, according to continuum thermodynamics, is given as 
\begin{equation}
	\label{elastic pk}
	\bm{\sigma}=\dfrac{\partial W}{\partial \bm{h}}+p \bm I.
\end{equation}
This relation essentially describes a purely mechanical equilibrium, in which each individual chain is locally balanced, as it neglects interactions between chains. In contrast, Eq. (\ref{thermodynamic cauchy}) characterizes a full thermodynamic equilibrium, where not only each chain satisfies mechanical equilibrium, but the entire network also attains thermal equilibrium.

We will show that the two sets of energy and stress expressions in Eqs. (\ref{total free energy} \& \ref{elastic energy}) and (\ref{thermodynamic cauchy} \& \ref{elastic pk}) can be consistent with each other, based on strict thermodynamic arguments.  Notably, the orientation entropy of Gaussian network is of higher-order infinitesimal compared to the elastic energy, in agreement with Treloar's original assumption. In fact, the orientation entropy does not directly contribute to the continuum stress in either Gaussian or non-Gaussian networks, but rather determines how the macroscopic stress is distributed across spatial directions.

\subsection{Maximum entropy under kinematic constraints }

As is shown in Eq.   (\ref{r constraint dis}), the segment orientations in a specified chain is subjected to the following constraint
\begin{equation}
	\label{constraint r}
	\bm{r} =\sum_{i=1}^{N}\bm b^i= \iint P_{\bm r} p_{\bm u}^{\bm r} \bm b  \text{d} \bm u \text{d} \bm r = N \int p_{\bm u}^{\bm r}  b \bm u  \text{d} \bm u.
\end{equation}
This expression serves as a geometric constraint that confines the segment fluctuation of a chain.

On the network level, the stress formulation can be alternatively treated as another geometrical constraint of derivative form for the relation between microscopic configuration and the strain. However, the thermodynamic expression in Eq. (\ref{thermodynamic cauchy}) cannot be employed as a valid constraint, as it is only satisfied when the system has already reached thermodynamic equilibrium. An appropriate constraint must hold for all possible variations, including those depart from thermodynamic equilibrium.

Hence, we adopt the stress formulation given in Eq. (\ref{elastic pk}) as the variational constraint. This choice implies that the corresponding variation is performed while maintaining local chain equilibrium throughout the network. As discussed, local equilibrium does not necessarily imply thermodynamic equilibrium, as it permits variations in $P_{\bm r}$ — that is, the distribution of the elastic energy $W$ over all chains can be distinct.  Substitution of Eq.  (\ref{elastic energy}) into (\ref{elastic pk}) yields
\begin{equation}
	\label{constraint cauchy}
	\bm{\sigma}=M \int  \left(P_{\bm r}  \dfrac{\partial w_{\bm r}}{\partial \bm{h}}+\bm{P}'_{\bm r}w_{\bm r}\right)\text d\bm r+p \bm I.
\end{equation}
with $\bm{P}'_{\bm r} = \partial P_{\bm r}/\partial \bm{h}$. Physically, this constraint can be interpreted as an additional geometrical condition that restricts the global network fluctuations.

Now a variational problem is set toward determining the probability distributions $P_{\bm r}$, $\bm{P}'_{\bm r}$ and $p_{\bm u}^{\bm r}$ that maximize the total network entropy given in Eq.  (\ref{networkentropycontinuous}), subject to two geometrical constraints in Eqs. (\ref{constraint r} - \ref{constraint cauchy}), and two normalization conditions given in Eqs. (\ref{chainnumbercon} - \ref{segmentnumbercon}). The corresponding Lagrangian functional takes the form:
\begin{equation}
	\label{network lagrangian}
	\begin{aligned}
		L(P_{\bm r},\bm{P}'_{\bm r}, p_{\bm u}^{\bm r}, \bm a, \bm E, \alpha, \eta) = &-k_{\text B} T MN \int P_{\bm r} \ln P_{\bm r} \text{d} \bm r - k_{\text B} T MN \iint P_{\bm r} p_{\bm u}^{\bm r} \ln p_{\bm u}^{\bm r} \, \text{d} \bm u \, \text{d} \bm r \\
		&+ \bm a \cdot \left( N \int p_{\bm u}^{\bm r} b \bm u  \text{d} \bm u - \bm r \right)+ \bm E : \left[ M \int  \left(P_{\bm r}  \dfrac{\partial w_{\bm r}}{\partial \bm{h}}+\bm{P}'_{\bm r} w_{\bm r}\right)\text d\bm r+p \bm I- \bm{\sigma} \right] \\
		&+ \alpha \left( \int p_{\bm u}^{\bm r}  \text{d} \bm u - 1 \right) + \eta \left( \int P_{\bm r}  \text{d} \bm r - 1 \right),
	\end{aligned}
\end{equation}
where $\bm a$ and $\bm E$ are respectively vectorial and tensorial multipliers associated with the geometrical constraints, while $\alpha$ and $\eta$ are two scalar Lagrange multipliers enforcing the normalization conditions. Here, the entropy is multiplied by the absolute temperature $T$ for dimensional consistency, which does not affect the results.

\subsection{The single-chain model}

In this subsection, we demonstrate that the variational formulation in Eq.  (\ref{network lagrangian}) naturally recovers the single-chain model of classical entropic elasticity theory. 

Taking the variation of the Lagrangian with respect to $p_{\bm u}^{\bm r}$, and noting that the chain-level quantities $\bm r$ and $P_{\bm r}$ are independent of $p_{\bm u}^{\bm r}$, the stationarity condition becomes
\begin{equation}
	\label{chainstation}
	\dfrac{\partial L}{\partial p_{\bm u}^{\bm r}} \delta p_{\bm u}^{\bm r} = \int\left[- k_{\text B} T MN \left( 1 + \ln p_{\bm u}^{\bm r} \right)  + N \bm a \cdot b\bm u + \alpha \right] \delta p_{\bm u}^{\bm r} \text{d} \bm u = 0.
\end{equation}
Here the constraint in Eq. (\ref{constraint cauchy}) does not come into play because the term $w_{\bm r}$ is already used as the equilibrium chain free energy, and it only allows variation of $P_{\bm r}$ and $\bm{P}'_{\bm r}$.  Eq. (\ref{chainstation}) leads to the following expression for the segment orientation distribution:
\begin{equation}
	\label{segmentpro1}
	p_{\bm u}^{\bm r} = \dfrac{1}{Z} \exp\left[ \dfrac{\bm a \cdot b\bm u}{Mk_{\text B} T}\right],
\end{equation}
where $Z$ is a normalization constant (also known as the partition function in statistical physics) determined by the condition in Eq.  (\ref{segmentnumbercon}).

With the help of Eqs. (\ref{constraint r}) and (\ref{segmentpro1}), the entropy of a chain is calculated as
\begin{equation}
	\label{chain entropy}
	s_{\bm r} = -Nk_{\text B}\int p_{\bm u}^{\bm r}\ln p_{\bm u}^{\bm r}\text{d} \bm u=-\dfrac{\bm a \cdot \bm r}{MT}+k_{\text B}\ln Z,
\end{equation}
and the chain free energy writes
\begin{equation}
	\label{chainfree2}
	w_{\bm r} = u_0 -T s_{\bm r}=u_0+\dfrac{\bm a}{M} \cdot \bm r - k_{\text B}T\ln Z.
\end{equation}
Under reversible conditions, the chain force is defined as the gradient of the free energy with respect to the chain configuration. Eq. (\ref{chainfree2}) then gives rise to
\begin{equation}
	\label{vector a}
	\bm{f} = \dfrac{\partial w_{\bm r}}{\partial \bm r} = \dfrac{\bm a}{M}.
\end{equation}
Substituting the above relation into Eq.  (\ref{segmentpro1}) yields
\begin{equation}
	\label{chainpro2}
	p_{\bm u}^{\bm r} = \dfrac{1}{Z} \exp\left[\dfrac{\bm{f} \cdot b \bm u}{k_{\text B} T}\right],
\end{equation}
where the partition function $Z$ can be analytically calculated as
\begin{equation}
	\label{chainpartition}
	Z = \int_{|\bm u|=1} \exp\left[\dfrac{\bm{f} \cdot b \bm u}{k_{\text B} T} \right] \text{d} \bm u = 4\pi \dfrac{\sinh(\xi)}{\xi},
\end{equation}
with $\xi \equiv f b / k_{\text B} T$. Here $f \equiv |\bm{f}|$ is the force magnitude.

Substitution of Eqs. (\ref{chainpro2} - \ref{chainpartition}) into Eq.  (\ref{constraint r}) yields the configuration-force relation:
\begin{equation}
	\label{distance force}
	\bm r = N \int p_{\bm u}^{\bm r} b\bm u \text{d} \bm u = N b \dfrac{\bm{f}}{f} \mathcal{L}(\xi),
\end{equation}
where $\mathcal{L}(x) \equiv \coth x - 1/x$ denotes the Langevin function. The inverse force-configuration relation then takes the form:
\begin{equation}
	\label{force distance}
	\bm{f} = \dfrac{k_{\text B} T}{b} \dfrac{\bm r}{r} \mathcal{L}^{-1}\left( \dfrac{r}{N b} \right).
\end{equation}

Inserting Eqs. (\ref{vector a}) and (\ref{force distance}) into Eq. (\ref{chainfree2}), together with the partition function in Eq.  (\ref{chainpartition}), gives the Helmholtz free energy of a chain:
\begin{equation}
	\label{chain helmholtz energy}
	w_{\bm r} = u_0 -N k_{\text B} T \left( \dfrac{r \beta}{N b} + \ln  \dfrac{\beta}{\sinh \beta}  \right)
\end{equation}
with $\beta \equiv \mathcal{L}^{-1}(r / N b)$.

The corresponding Gibbs free energy, obtained via Legendre transformation, reads
\begin{equation}
	\label{chain gibbs energy}
	 g_{\bm{f}} = w_{\bm r} - \bm{f} \cdot \bm r = u_0-N k_{\text B} T\ln Z =u_0+ N k_{\text B} T \ln  \dfrac{\xi}{\sinh \xi} 
\end{equation}
where the constant term is neglected.

Eqs. (\ref{distance force} - \ref{chain gibbs energy}) collectively represent the Langevin chain model in entropic elasticity theory. They specify a single chain's behavior at its local equilibrium state. In the limit of large segment number $N$ or small end-to-end distance $r$, the model reduces to the quadratic Gaussian form with linear force-extension response.

\subsection{Chain orientation probability}

Given that the individual chains are at local equilibrium, i.e., governed by the results derived in the previous subsection, the variation of the Lagrangian in Eq. (\ref{network lagrangian}) with respect to $P_{\bm r}$  then yields
\begin{equation}
	\label{variation Pr}
	\dfrac{\partial L}{\partial P_{\bm r}} \delta P_{\bm r} = MN \int \left( -k_{\text B}(1+ \ln P_{\bm r}) - \dfrac{w_{\bm r}}{ N} + \dfrac{1}{N} \bm E : \dfrac{\partial w_{\bm r}}{\partial \bm{h}} +  \dfrac{\eta}{MN} \right) \delta P_{\bm r} \text{d} \bm r = 0,
\end{equation}
which leads to the following Boltzmann-type distribution:
\begin{equation}
	\label{Pr}
	P_{\bm r} = \dfrac{1}{\mathbb Z} \exp \left[ -\dfrac{g_{\bm r}}{k_{\text B} T N} \right], \quad \mathbb Z = \int \exp \left[ -\dfrac{g_{\bm r}}{k_{\text B} T N} \right]  \text{d} \bm r,
\end{equation}
where
\begin{equation}
	\label{g}
	g_{\bm r} = w_{\bm r} - \bm E : \dfrac{\partial w_{\bm r}}{\partial \bm{h}}.
\end{equation}

On the other hand, variation of the Lagrangian with respect to $\bm{P}'_{\bm r}$ writes
\begin{equation}
	\label{variation P'}
	\dfrac{\partial L}{\partial \bm{P}'_{\bm r}} :\delta \bm{P}'_{\bm r} = \bm E : M \int \delta \bm{P}'_{\bm r} w_{\bm r}  \text{d} \bm r = 0.
\end{equation}
Since $\bm E\neq \bm O$ and $ w_{\bm r}> 0$, Eq.  (\ref{variation P'}) implies $\delta \bm{P}'_{\bm r} = \bm O$, and thus $\bm{P}'_{\bm r}=\partial P_{\bm r}/\partial \bm{h}$ must be independent of $\bm r$.
Then the consistency of two stress formulas in Eqs. (\ref{thermodynamic cauchy}) and (\ref{constraint cauchy}) requires
\begin{equation}
	\label{consistency pk}
\int \left(1 + P_{\bm r}\right)\dfrac{\partial P_{\bm r}}{\partial \bm{h}} \text{d} \bm r = \dfrac{\partial P_{\bm r}}{\partial \bm{h}} \int \left(1 + P_{\bm r}\right) \text{d} \bm r = \bm O.
\end{equation}
Given that $P_{\bm r} > 0$, it follows that $\partial P_{\bm r} / \partial \bm h = \bm O$. Using Eqs. (\ref{Pr} - \ref{g}), one obtains
\begin{equation}
	\label{mini Pr}
	\dfrac{\partial P_{\bm r}}{\partial \bm{h}} = \dfrac{\partial P_{\bm r}}{\partial g_{\bm r}} \dfrac{\partial g_{\bm r}}{\partial \bm{h}} = -\dfrac{1}{k_{\text B} T N}P_{\bm r} \dfrac{\partial g_{\bm r}}{\partial \bm{h}} = \bm O,
\end{equation}
which leads to the identity
\begin{equation}
	\label{mini gr}
	\dfrac{\partial g_{\bm r}}{\partial \bm{h}} = \bm O.
\end{equation}
This condition implies that $g_{\bm r}$ should maintain an extremum with respect to $\bm h$ at thermodynamic equilibrium. Combining with the expression in Eq. (\ref{g}), it is concluded that $g_{\bm r}$ should be given by the following Legendre transformation
\begin{equation}
	\label{legendre gn}
	g_{\bm r} = w_{\bm r} - \bm{h}:\dfrac{\partial w_{\bm r}}{\partial \bm{h}},
\end{equation}
and the tensorial Lagrange multiplier $\bm E$ turns out to be the logarithmic strain $\bm h$.

\subsection{Eulerian micro-macro connection}
For a local observer aligned with a specified spatial direction $\bm n$, the local time rate of chain free energy is given by
\begin{equation}
	\label{dot w}
	\dot w|_{\bm n}=fr\dot{\overline {\ln\lambda}}=fr(\bm n\otimes\bm n:\bm d),
\end{equation}
where $\lambda \equiv r/r_0$ defines the relative chain stretch with $r_0$ representing the initial chain length. The term $\dot{\overline{\ln \lambda}}$ represents the time rate of the logarithmic stretch arising from macroscopic deformation-induced convection, and is given by the Eulerian kinematic relation $\dot{\overline{\ln \lambda}} = \bm n \bm d \bm n$ in continuum mechanics with $\bm d = \frac{1}{2}(\bm{F}\dot{\bm{F}}^{-1}+\dot{\bm{F}}^{-T}\bm{F}^T)$ being the deformation rate tensor.

It should be emphasized that Eq. (\ref{dot w}) does not represent the material time derivative of a specific chain's stretch, as thermal fluctuations may cause the chain to randomly migrate to any spatial direction. Such fluctuations result in additional convected motion of the chain, beyond what is captured by a local observer. In general, the material time derivative may not be deterministic due to the randomness of the thermal motion.

From the perspective of an external observer, the reversible chain behavior is of Green elasticity, i.e., the chain energy can be expressed as a function $w = w(\bm n, \bm h)$, which must satisfy both frame-invariance and transverse isotropy. Thus, the relation in Eq. (\ref{dot w}) should be exactly integrable to yield the energy $w(\bm n, \bm h)$.  As discussed in \cite{xiao1999natural} and \cite{bruhns2003some}, these conditions are simultaneously satisfied if and only if the following objective log-rotation frame is adopted
\begin{equation}
	\label{log dotw}
	\dot w=\dfrac{\partial w}{\partial \bm h}:\dot{\bm h}^{\text{log}}=\dfrac{\partial w}{\partial \bm h}:\bm d,
\end{equation}
where $\dot{\bm h}^{\text{log}} = \dot{\bm h} + \bm h \bm \Omega - \bm \Omega \bm h = \bm d$, and $\bm \Omega$ denotes the logarithmic spin tensor \citep{xiao1997logarithmic,xiao1999natural,bruhns2003some}.

With Eqs. (\ref{dot w} - \ref{log dotw}), one obtains
\begin{equation}
	\label{dwdh}
	\dfrac{\partial w}{\partial \bm h}=\dfrac{\partial w}{\partial \ln\lambda} \dfrac{\partial \ln\lambda}{\partial \bm h}
	=fr\dfrac{\partial \ln\lambda}{\partial \bm h}= f r \bm n \otimes \bm n,
\end{equation}
and the corresponding solution is
\begin{equation}
	\label{micro-macro}
	\ln\lambda = \ln(r / r_0) = \bm h : \bm n \otimes \bm n,
\end{equation}
where the integration constant is eliminated by the initial condition $\ln(r_0/r_0) = 0$, corresponding to the undeformed state $\bm h = \bm O$.

The initial chain length is given by \citep{treloar1975physics,weiner2012statistical,doi2013soft}
\begin{equation}
	r_0 \equiv \sqrt{\left<\bm r_0 \cdot \bm r_0\right>} = \sqrt{ \sum_{i=1}^{N} \sum_{j=1}^{N} \left\langle \bm b^i \cdot \bm b^j \right\rangle } 
	= \sqrt{N} b,
\end{equation}
where the relation $\langle \bm b^i \cdot \bm b^j \rangle = \delta_{ij} b^2$ is used, assuming isotropic segment orientation under undeformed conditions. Then Eq.  (\ref{micro-macro}) yields the following expression for the relative chain stretch in spatial direction $\bm n$:
\begin{equation}
	\label{lambdan}
	\boxed{\lambda_{\bm n} = \dfrac{r}{\sqrt{N} b} = e^{\bm n\bm h \bm n }.}
\end{equation}

Since the chain end-to-end length $r$ now depends explicitly on the orientation $\bm n$, Eqs. (\ref{Pr} - \ref{g}) then becomes
\begin{equation}
	\label{Prn}
	\boxed{P_{\bm n} = \dfrac{1}{\mathbb Z} \exp\left[-\dfrac{g_{\bm n}/N}{k_{\text B} T}\right], \quad \mathbb Z = \int_{|\bm n|=1} \exp\left[-\dfrac{g_{\bm n}/N}{k_{\text B} T}\right]  \text{d} \bm n,}
\end{equation}
with
\begin{equation}
	\label{gn}
	g_{\bm n}= w_{\bm n} - \bm h : \bm{f} \otimes \bm r = w_{\bm n} - fr \ln \lambda_{\bm n}=w_{\bm n} - \dfrac{\partial w_{\bm n}}{\partial\ln \lambda_{\bm n} } \ln \lambda_{\bm n}.
\end{equation}
As the force and configuration vectors of a locally equilibrated chain are both aligned with $\bm n$, it follows that $\bm f \otimes \bm r = fr\bm n \otimes \bm n$.

\subsection{Thermodynamic consistency}
Substituting Eqs. (\ref{mini Pr}) and (\ref{dwdh}) into Eq.  (\ref{constraint cauchy}), the Cauchy stress is reformulated as
\begin{equation}
	\label{virial cauchy}
	\bm{\sigma} = M \int P_{\bm n} \bm{f} \otimes \bm r \text{d} \bm n + p \bm I.
\end{equation}
This formulation indicates that the Cauchy stress arises from the statistical average of traction-configuration contributions of individual chains. If we conceptualize the network as junctions under thermal fluctuations with interaction potential $w_{\bm r}$, this formulation is then consistent with the classical definition of virial stress in statistical mechanics \citep{clausius1870mechanical,tsai1979virial,subramaniyan2008continuum,tadmor2011modeling}.

\begin{figure}[]
	\centering    
	\includegraphics[width=97.00mm,height=65.00mm]{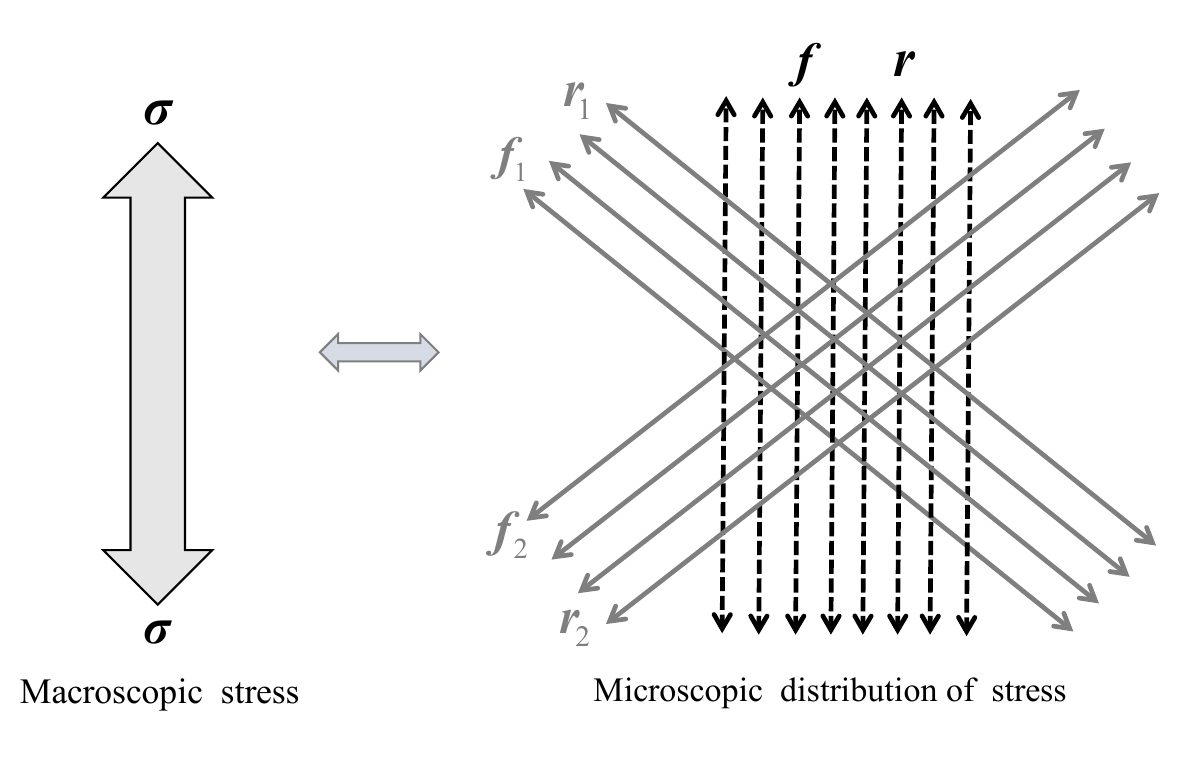}
	\caption{A sketch for the macroscopic stress and admissible microscopic chain orientation.} 
	\label{fig:stress} 
\end{figure}

Eq. (\ref{virial cauchy}) represents a mechanical equilibrium, wherein the total macroscopic stress corresponds to the sum of chain traction-configuration dyads. However, mechanical equilibrium does not necessarily imply thermodynamic equilibrium, as the directional distribution of the stress may remain undetermined. As illustrated in Fig. \ref{fig:stress}, both the chain orientations represented by the black-dashed ($\bm f\otimes\bm r$) and gray-solid ($\bm f_1\otimes\bm r_1+\bm f_2\otimes\bm r_2$) thin line arrows can produce the same macroscopic stress $\bm \sigma$ indicated by the thick gray arrow\footnote{A hydrostatic stress component can be superimposed to eliminate the lateral stress contributions associated with the $\bm f_1\otimes\bm r_1+\bm f_2\otimes\bm r_2$.}.  Given the micro-macro connection in Eq. (\ref{micro-macro}), the black-dashed orientation requires a smaller logarithmic strain $\bm h$, whereas the gray-solid orientation corresponds to a larger one. This ambiguity is resolved by enforcing thermodynamic equilibrium, which entails the complementary energy and the orientation entropy simultaneously retains extremums. This procedure yields a unique distribution of chain orientations, thereby establishing a well-defined strain–stress relationship.

Comparing Eq.  (\ref{gn}) to Eq.  (\ref{chain gibbs energy}), it is found that the Gibbs free energy of an individual chain is reformulated by changing the conjugate variables from $\{f, r\}$ to a new pair of $\{fr, \ln\lambda\}$.  This change forms the conceptual foundation of the alternative approach developed in \cite{zhan2025statistical}. It indeed constitutes a canonical transformation in classical mechanics \citep{arnol2013mathematical}. The new conjugate pair $\{fr, \ln\lambda\}$ matches the continuum Eulerian conjugate pair $\{\bm\sigma, \bm h\}$, where $fr$ serves as the extensive quantity whose directional sum  $\int P_{\bm n}fr\bm n\otimes\bm n\text d\bm n$ equals the Cauchy stress $\bm \sigma$, and $\ln\lambda$ is the intensive quantity directly relating to the Eulerian strain measure $\bm h$.

As seen in Eq. (\ref{law_S}), the network entropy is dictated solely by the deformation and remains independent of temperature. This is also ture for the chain orientation as shown in  Eq. (\ref{Prn}), where the Boltzmann factor $k_{\text B}T$ appears in both the numerator and denominator, and thereby cancels. This contrasts with conventional statistical systems where the contribution of internal energy appears in the numerator, leading to temperature-dependent occupation probabilities. Nevertheless, while temperature does not alter the orientation entropy under fixed strain, it increases the free energy and stress. Conversely, under fixed stress, higher temperature results in lower strain and more isotropic chain orientations.

\subsection{Several remarks}

The expressions in Eqs. (\ref{lambdan} - \ref{gn}) were also derived in \cite{zhan2025statistical} through resolving the inconsistencies in thermodynamic conjugate variables of parallel chain systems. While it offers valuable physical insights by employing canonical transformations between intensive and extensive pairs, the derivation of \cite{zhan2025statistical} may appear somewhat ad hoc and technically opaque to researchers primarily trained in continuum mechanics. Moreover, it may not constitute a systematic framework that can be readily generalized beyond its specific context.

The present study adopts a fundamentally different perspective. By casting the problem as a variational formulation seeking for the maximum entropy of the whole network subject to geometrical constraints, we provide a rigorous and transparent pathway to derive macroscopic network behavior from fundamental segment-level statistics. This paradigm shift enables the construction of microscopically coherent and thermodynamically consistent models without resorting to phenomenological assumptions or externally imposed micro-macro mappings. We are confident that the present study constitutes a conceptual advance in the microstructural modeling of soft materials.

The current problem can be viewed analogously to a trivial exercise in statistical mechanics if we interpret the elastic complementary energy $\tilde W$ as an ``internal energy-like" conserved quantity, the enumeration of ways to assign this energy among the network results in further entropy production. However, the nontrivial (perhaps surprising) aspect is that this distribution is not performed over $M$ chains directly, but effectively over the $MN$ individual segments. It is clearly seen from the factor $MN$ in the term $S_{\text{orien}}=-MNk_{\text B} \int P_{\bm r} \ln P_{\bm r}\text d\bm r$ of Eq. (\ref{total free energy}). This subtle distinction is unraveled within our hierarchical ensemble formulation, wherein thermodynamic equilibrium is imposed at the segment level rather than at the chain level.

\section{The hyperelastic models}

\subsection{Finite strain network model}
The chain Helmholtz free energy in Eq. (\ref{chain helmholtz energy}) can be reformulated in terms of the relative stretch $\lambda$ :
\begin{equation}
	\label{chain w}
	w(\lambda)=Nk_{\text B}T\left(\frac{\lambda}{\sqrt N}\beta+\ln\frac{\beta}{\sinh\beta}\right),
\end{equation}
where the constant internal energy is neglected. The elastic free energy of the network is given by
\begin{equation}
	\label{network free energy}
W=W(\bm h)= M\int P_{\bm n}w\left({\lambda_{\bm n}}\right)\text d\bm n.
\end{equation}
Here, $\lambda_{\bm n}=\bm n\bm h\bm n$ is the relative stretch of a chain aligned along spatial direction $\bm n$, and $P _{\bm n}$ is the probability of a chain appearing in that direction (or equivalently regarded as the normalized density) as given in Eqs. (\ref{Prn} - \ref{gn}).
 
The corresponding Cauchy stress is computed via
\begin{equation}
	\label{cauchy stress Langevin}
	\boxed{\bm\sigma=\dfrac{\partial W}{\partial\bm h}+p\bm I=\dfrac{5}{2}G\int P_{\bm n} \sqrt N \mathcal L^{-1}\left(\dfrac{\lambda_{\bm n}}{\sqrt N}\right)\lambda_{\bm n}\bm n\otimes\bm n\text d\bm n+p\bm I,}
\end{equation}
where $G=2Mk_{\text B}T/5$ denotes the infinitesimal shear modulus, as will be proven later. For numerical efficiency, the sphere integral will be evaluated through an approximation with 37 discrete points \citep{bavzant1986efficient,miehe2004micro,zhan2023new}. 

\begin{figure}[]
 	\centering    
 	\subfigure[] 
 	{
 		\begin{minipage}{7.0cm}
 			\centering       
 			\includegraphics[width=60.00mm,height=60.00mm]{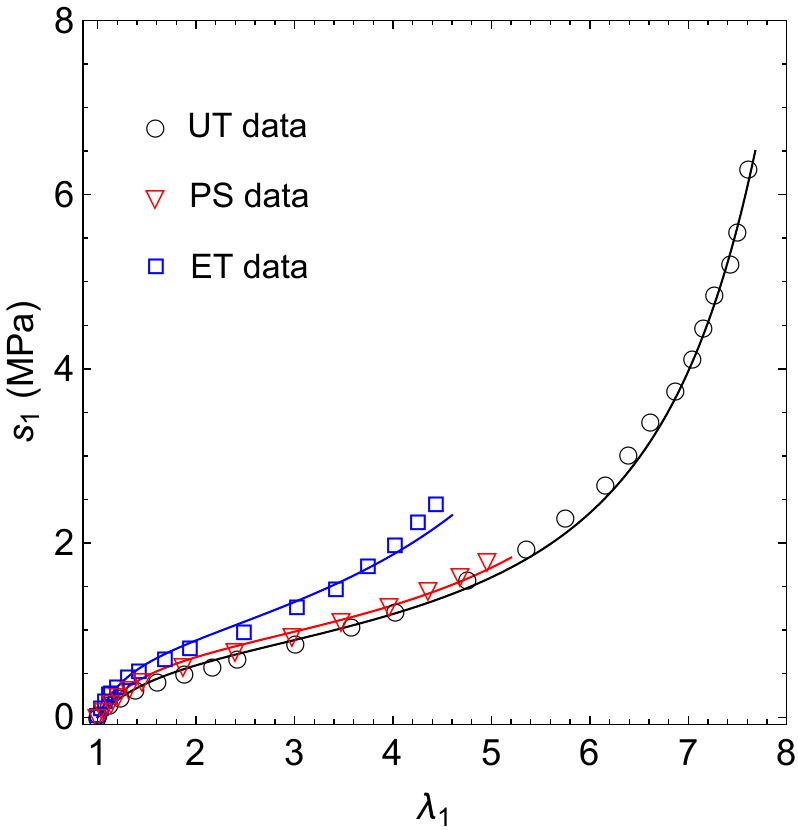}
 		\end{minipage}
 	}
 	\subfigure[] 
 	{
 		\begin{minipage}{7.0cm}
 			\centering       
 			\includegraphics[width=60.00mm,height=60.00mm]{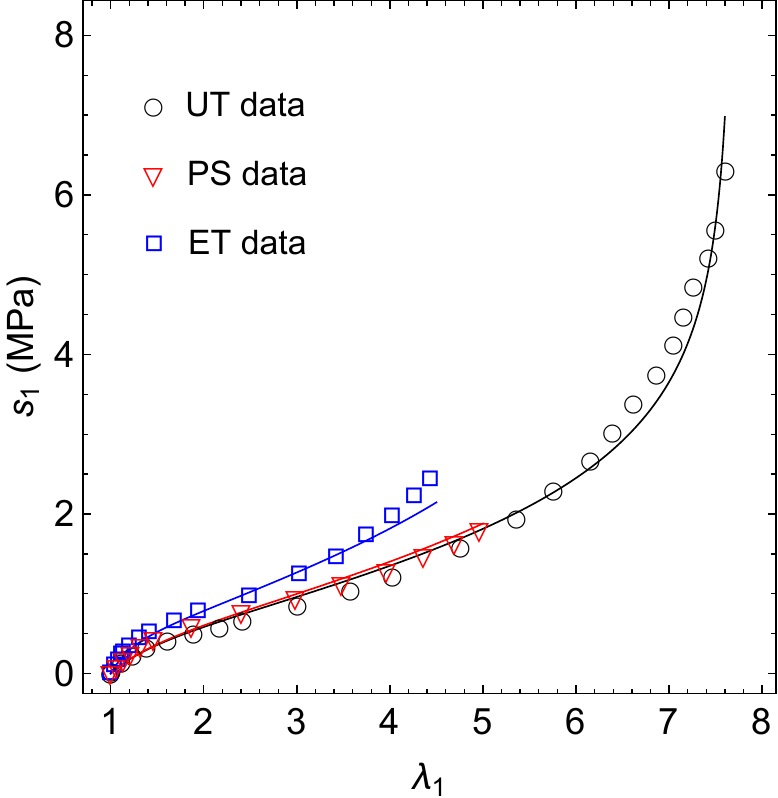}
 		\end{minipage}
 	}   
  	\subfigure[] 
 {
 	\begin{minipage}{7.0cm}
 		\centering       
 		\includegraphics[width=60.00mm,height=60.00mm]{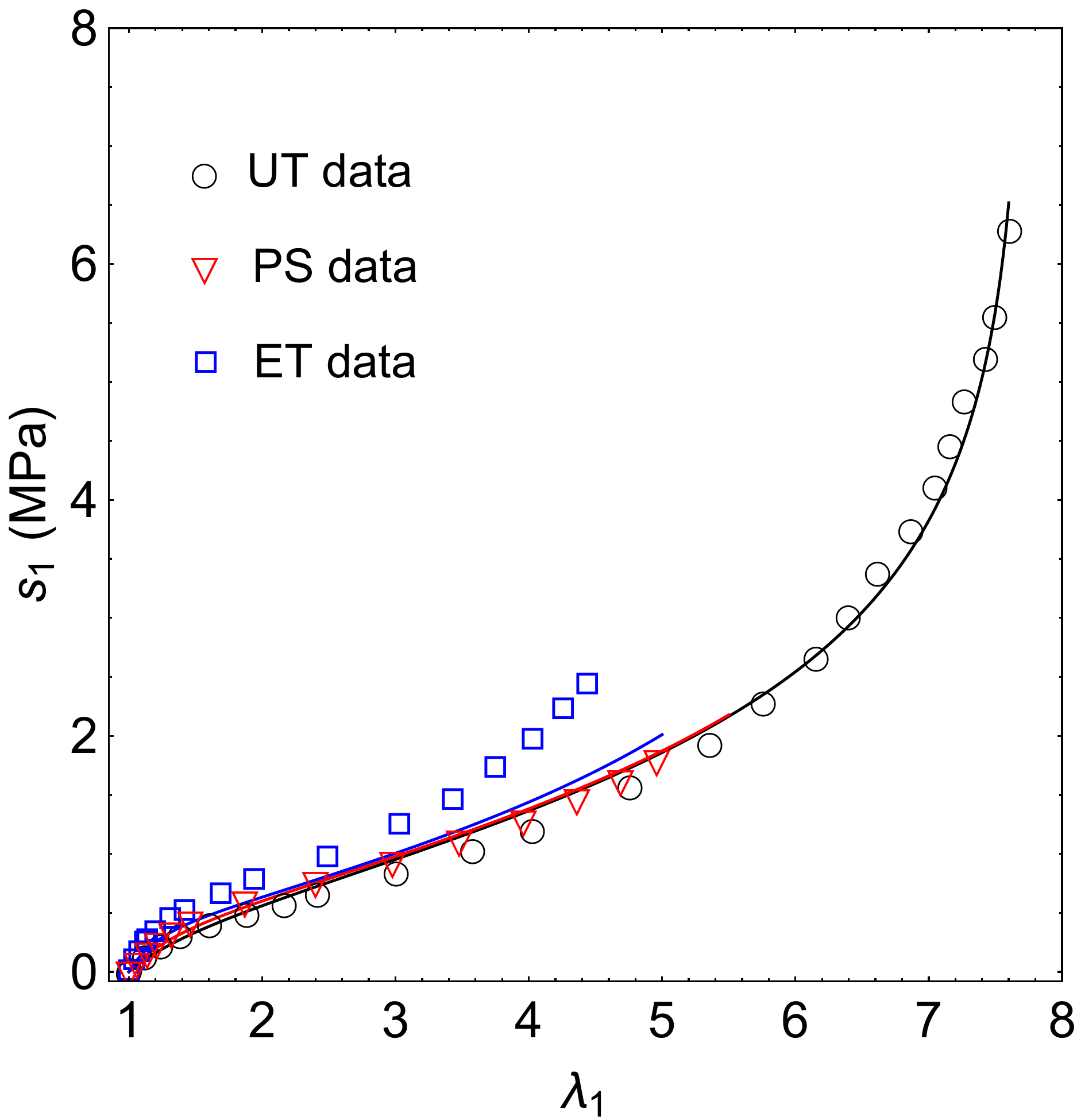}
 	\end{minipage}
 }   
 	\subfigure[] 
{
	\begin{minipage}{7.0cm}
		\centering       
		\includegraphics[width=60.00mm,height=60.00mm]{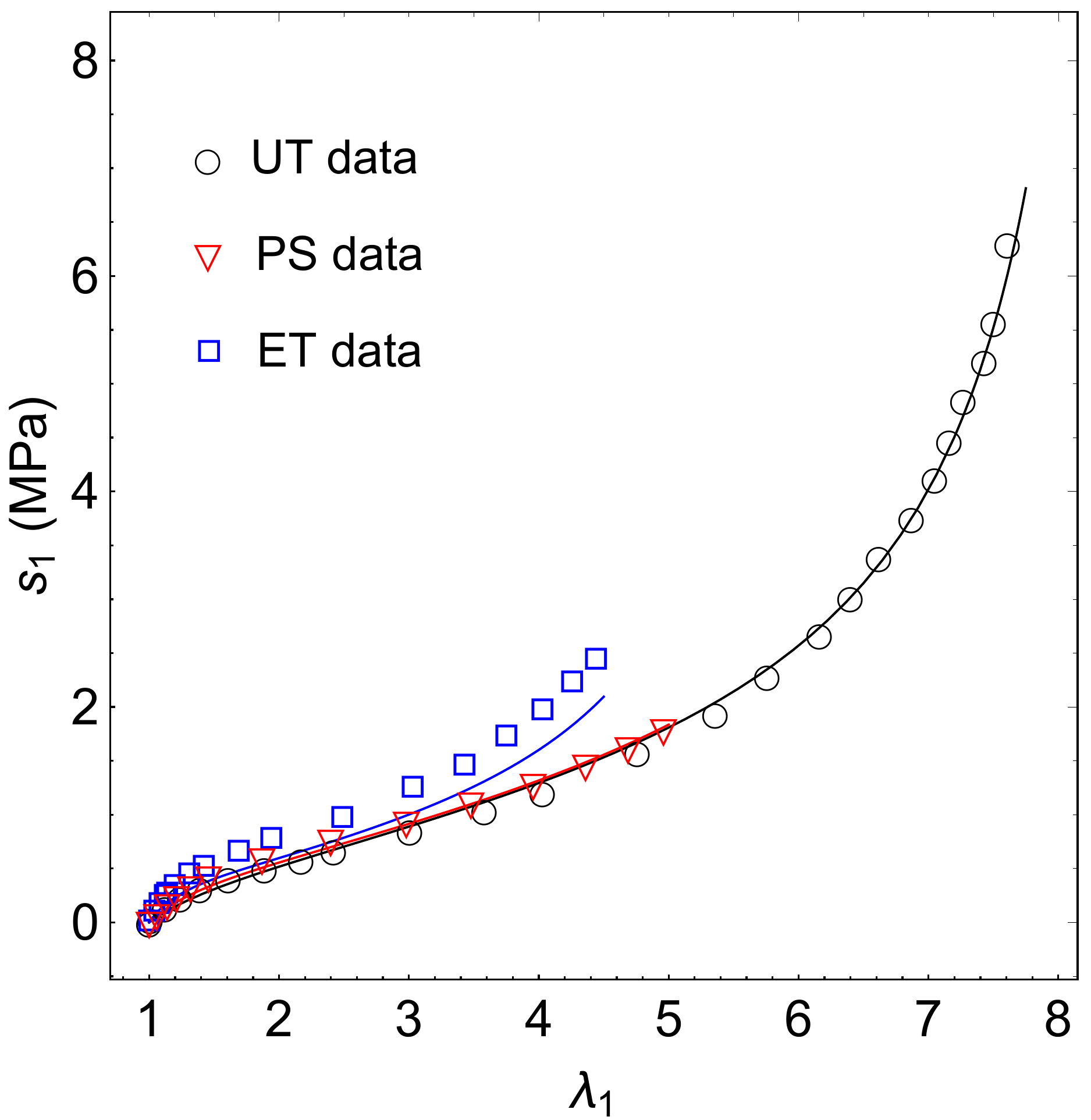}
	\end{minipage}
}   
 	\caption{Fitting Trealor's data using different models: (a) the current model (using Eq.  \ref{cauchy stress Langevin}) with $G=0.407$MPa and $N=168$; (b) the Biot-chain model \citep{zhan2023new} with $G=0.347$MPa and $N=61.9$; (c) the affine model with $G=0.311$MPa and $N=62.3$; and (d) the eight-chain model\citep{arruda1993three} with $G=0.284$MPa and $N=26.2$.} 
 	\label{fig:trealor} 
 \end{figure}

The model of Eq.  (\ref{cauchy stress Langevin}) is used to fit the rubber data of \cite{treloar1944stress} under uniaxial tension (UT), pure shear (PS), and equal biaxial tension (ET). As seen in Fig.~\ref{fig:trealor}, the current model achieves excellent agreement with experiments using only two physical parameters. The accuracy appears to surpass that of existing two-parameter models reported in the literature \citep{boyce2000constitutive,marckmann2006comparison,steinmann2012hyperelastic,dal2021performance,zhan2023new}. A quantitative comparison can be referred to \cite{zhan2025statistical}, wherein numerical approximation with 400 discrete points has been adopted.

\subsection{Gaussian network model} 
For sufficiently large $N$ or small to moderate deformation, the ratio $\lambda / \sqrt{N}$ becomes small. In this limit, the single-chain free energy reduces to the Gaussian form
\begin{equation}
	\label{chain Gaussian}
	w= \dfrac{3}{2} N k_{\text B} T \lambda^2.
\end{equation}
Under this approximation, the contribution of the term $g_{\bm n} / N$ in Eq.  (\ref{Prn}) becomes a higher-order infinitesimal with respect to $\lambda / \sqrt{N}$ and can thus be neglected. Then chain orientation can be treated as isotropic on an unit sphere, i.e., $P_{\bm n} = 1 / 4\pi$. The elastic free energy in Eq.  (\ref{network free energy}) accordingly simplifies to
\begin{equation}
	\label{network energy Gaussian}
	W_{\text{Gussian}} = \dfrac{15}{4} G \cdot \dfrac{1}{4\pi} \int e^{2 \bm n \bm h  \bm n}  \text{d} \bm n,
\end{equation}
and the corresponding Cauchy stress becomes
\begin{equation}
	\label{cauchy stress Gaussian}
	\boxed{\bm \sigma = \dfrac{\partial W}{\partial \bm h} + p \bm I = \dfrac{15}{2} G \cdot \dfrac{1}{4\pi} \int e^{2\bm n \bm h  \bm n} \bm n \otimes \bm n  \text{d} \bm n + p \bm I.}
\end{equation}

\begin{figure}[]
	\centering    
	\subfigure[] 
	{
		\begin{minipage}{7.0cm}
			\centering       
			\includegraphics[width=60.00mm,height=55.00mm]{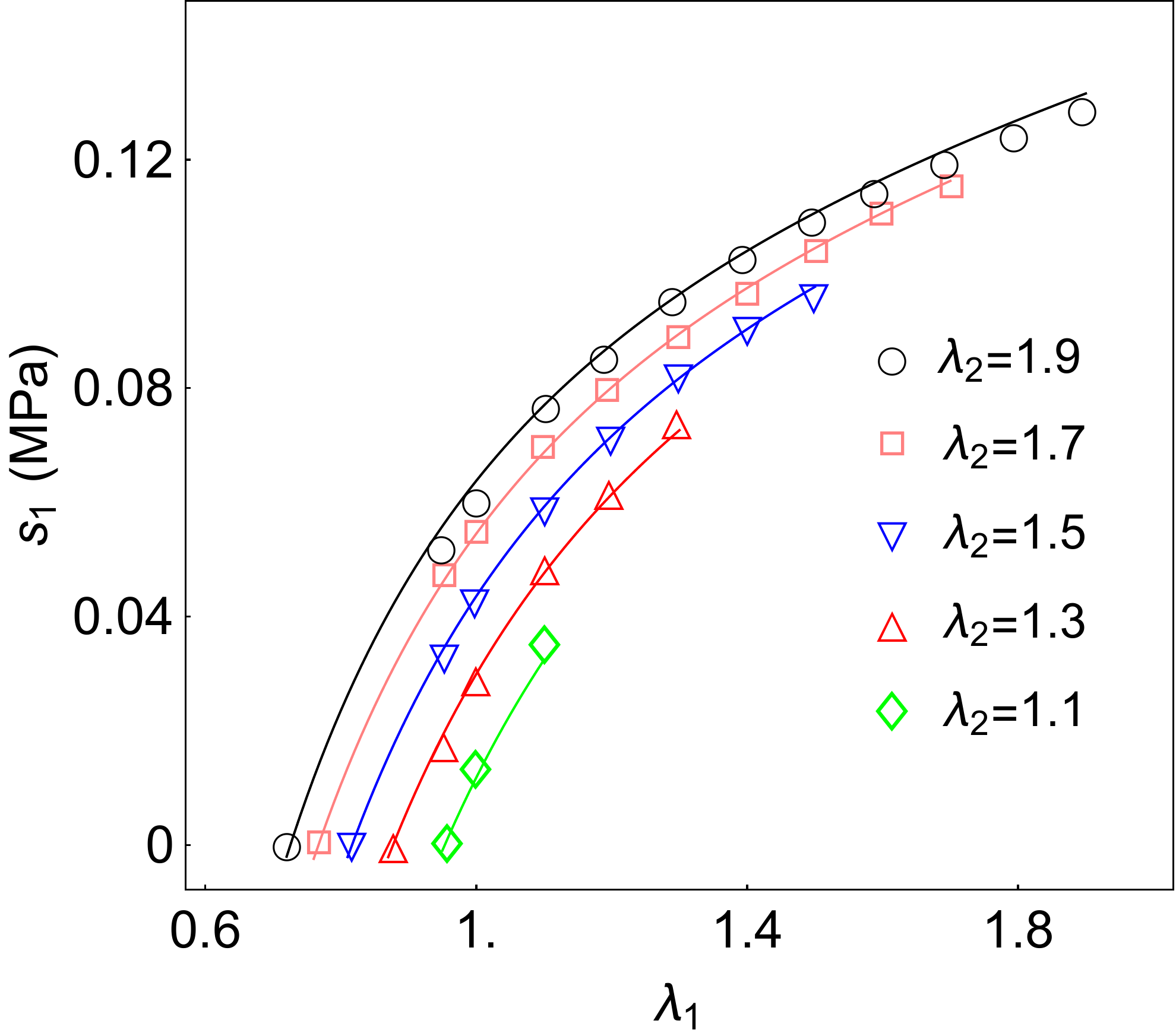}
		\end{minipage}
	}
	\subfigure[] 
	{
		\begin{minipage}{7.0cm}
			\centering       
			\includegraphics[width=60.00mm,height=55.00mm]{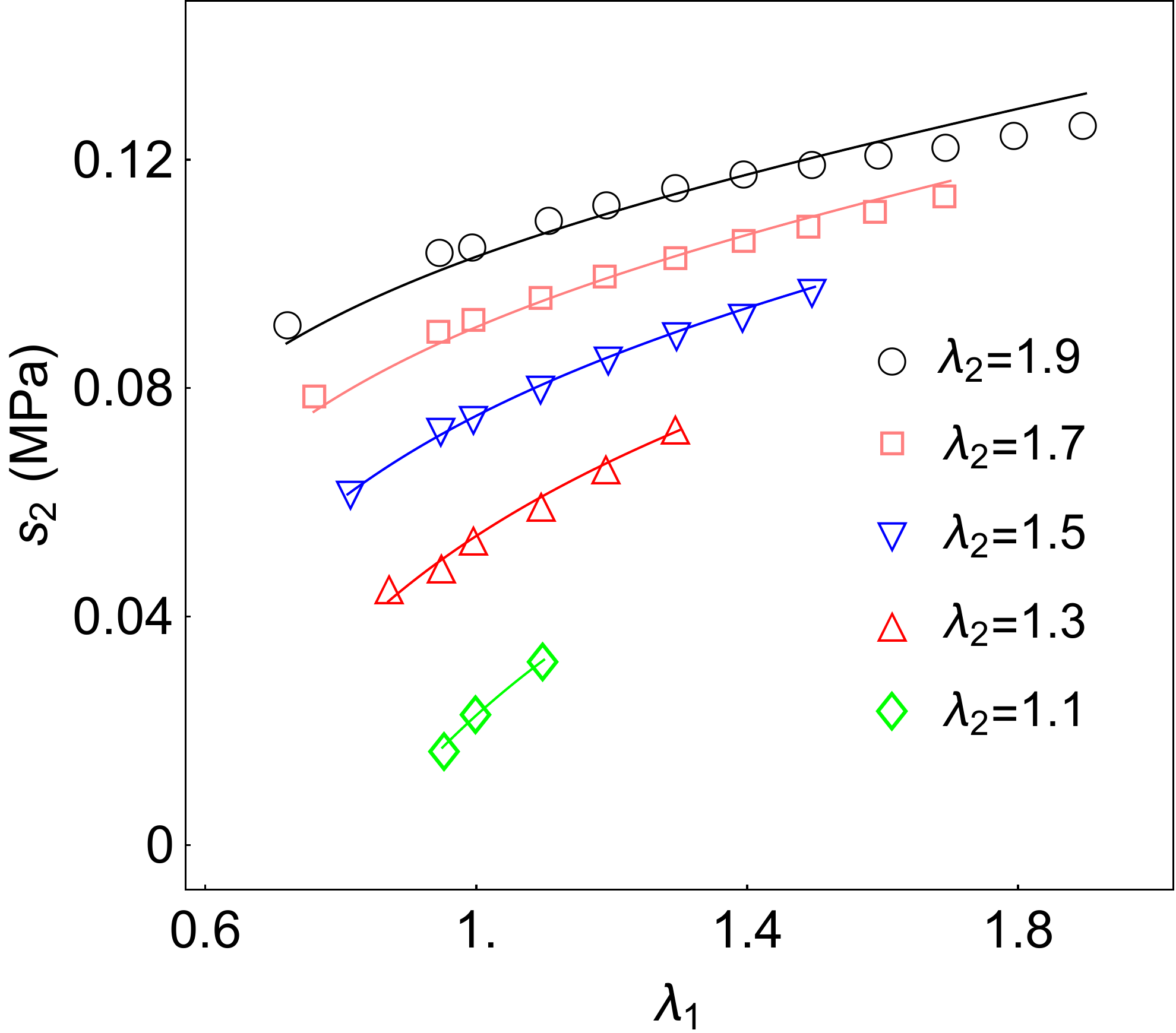}
		\end{minipage}
	}   
	\subfigure[] 
	{
		\begin{minipage}{7.0cm}
			\centering       
			\includegraphics[width=60.00mm,height=55.00mm]{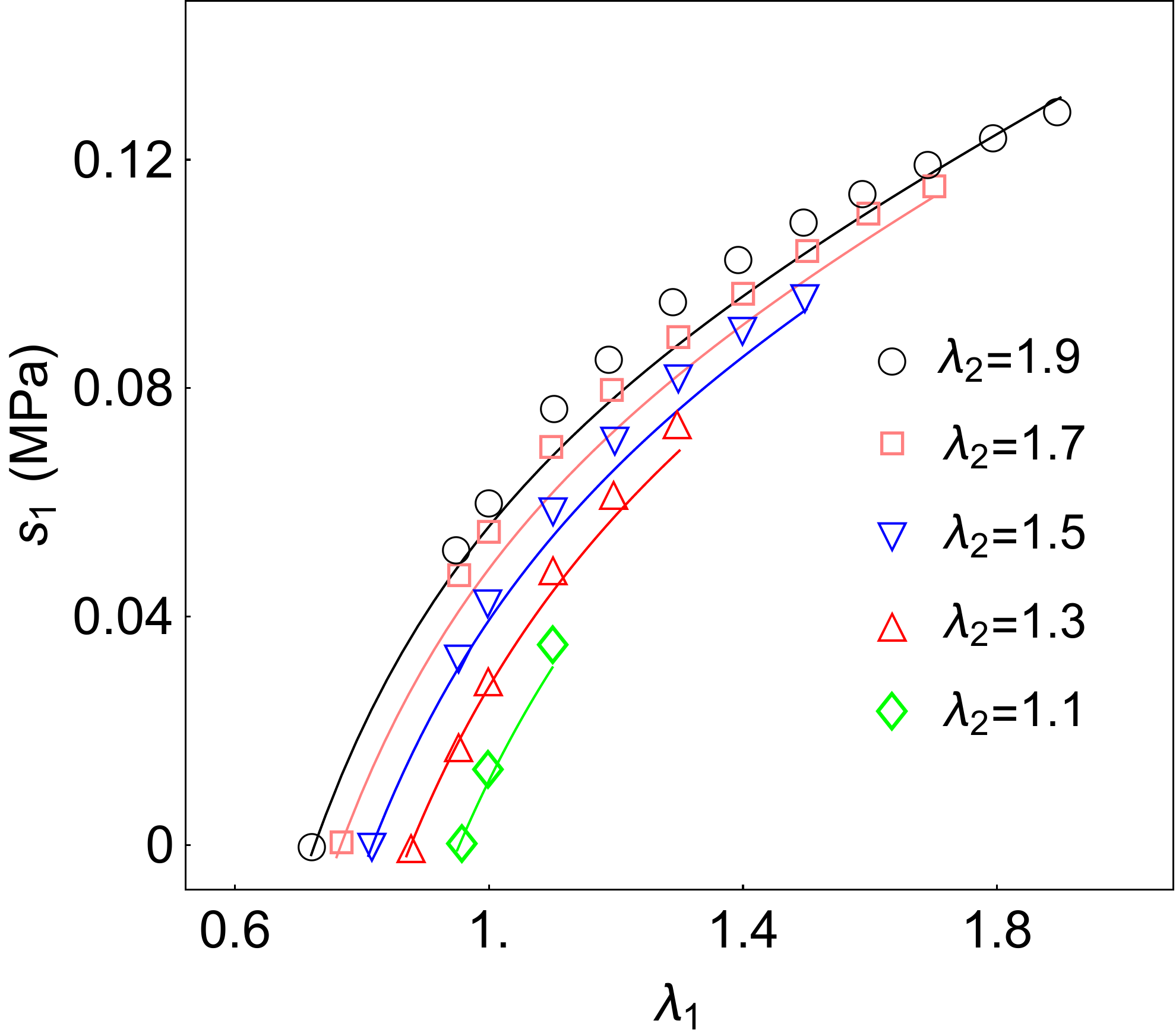}
		\end{minipage}
	}   
	\subfigure[] 
	{
		\begin{minipage}{7.0cm}
			\centering       
			\includegraphics[width=60.00mm,height=55.00mm]{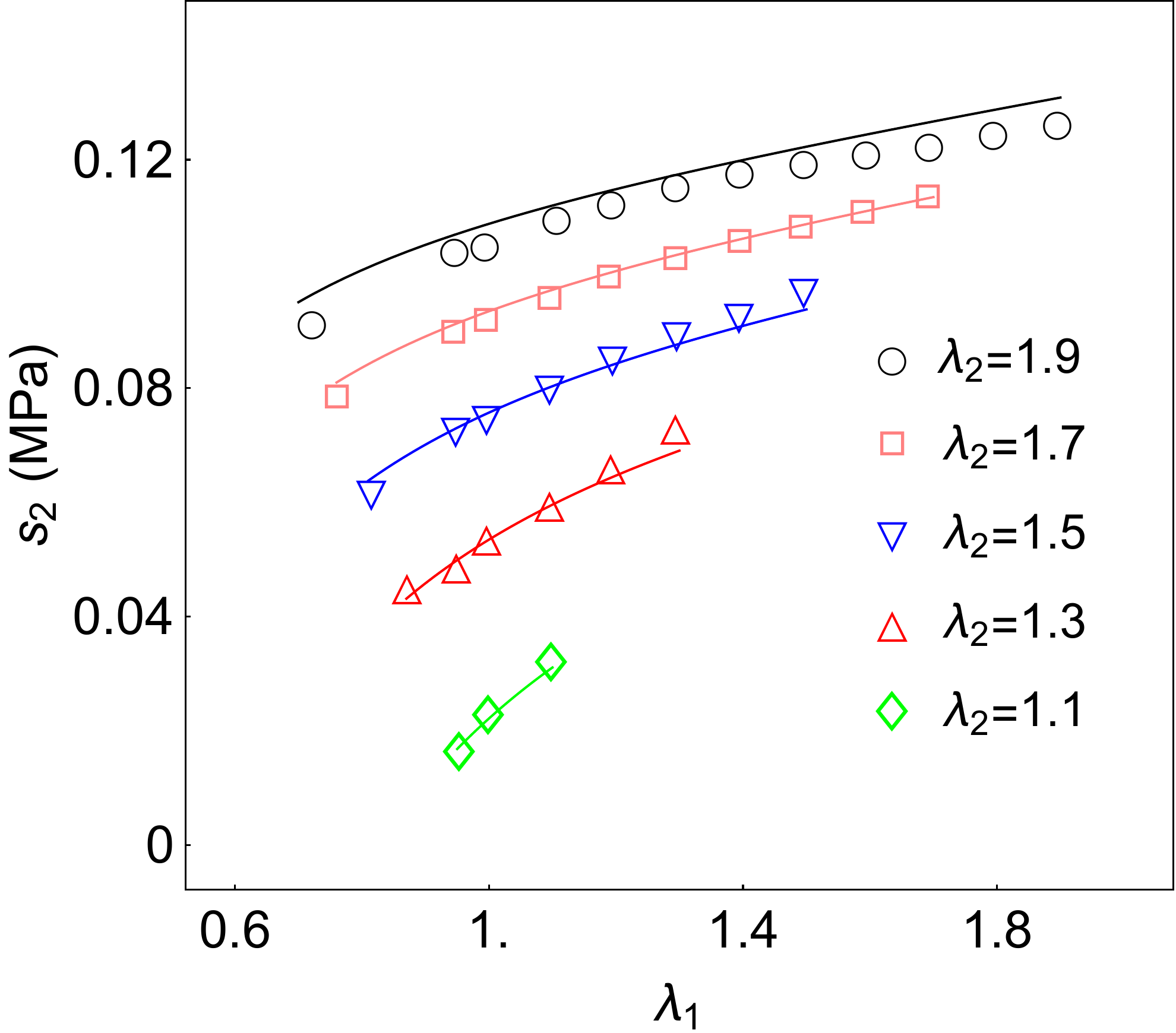}
		\end{minipage}
	}   
	\subfigure[] 
	{
		\begin{minipage}{7.0cm}
			\centering       
			\includegraphics[width=60.00mm,height=55.00mm]{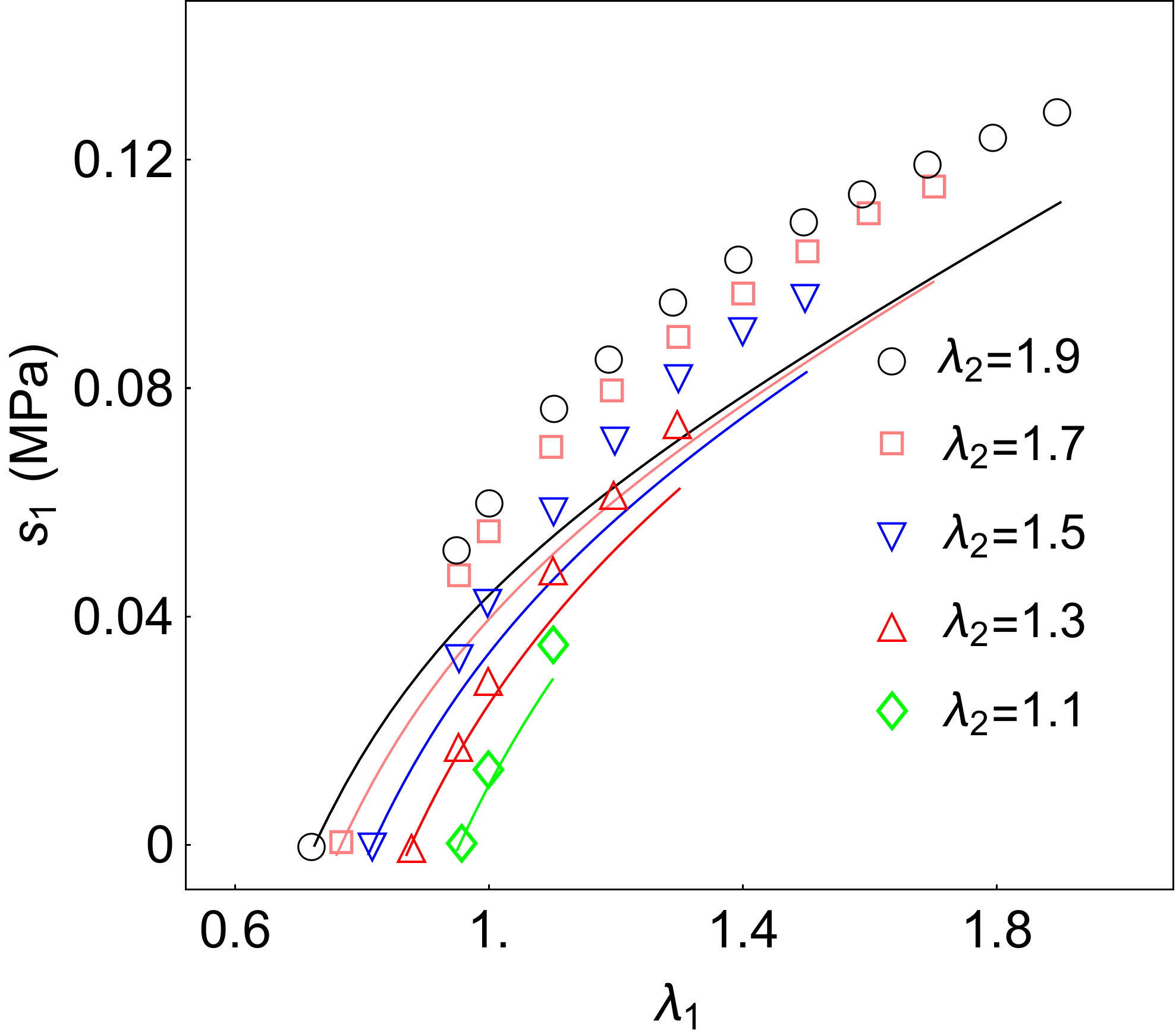}
		\end{minipage}
	}   
	\subfigure[] 
	{
		\begin{minipage}{7.0cm}
			\centering       
			\includegraphics[width=60.00mm,height=55.00mm]{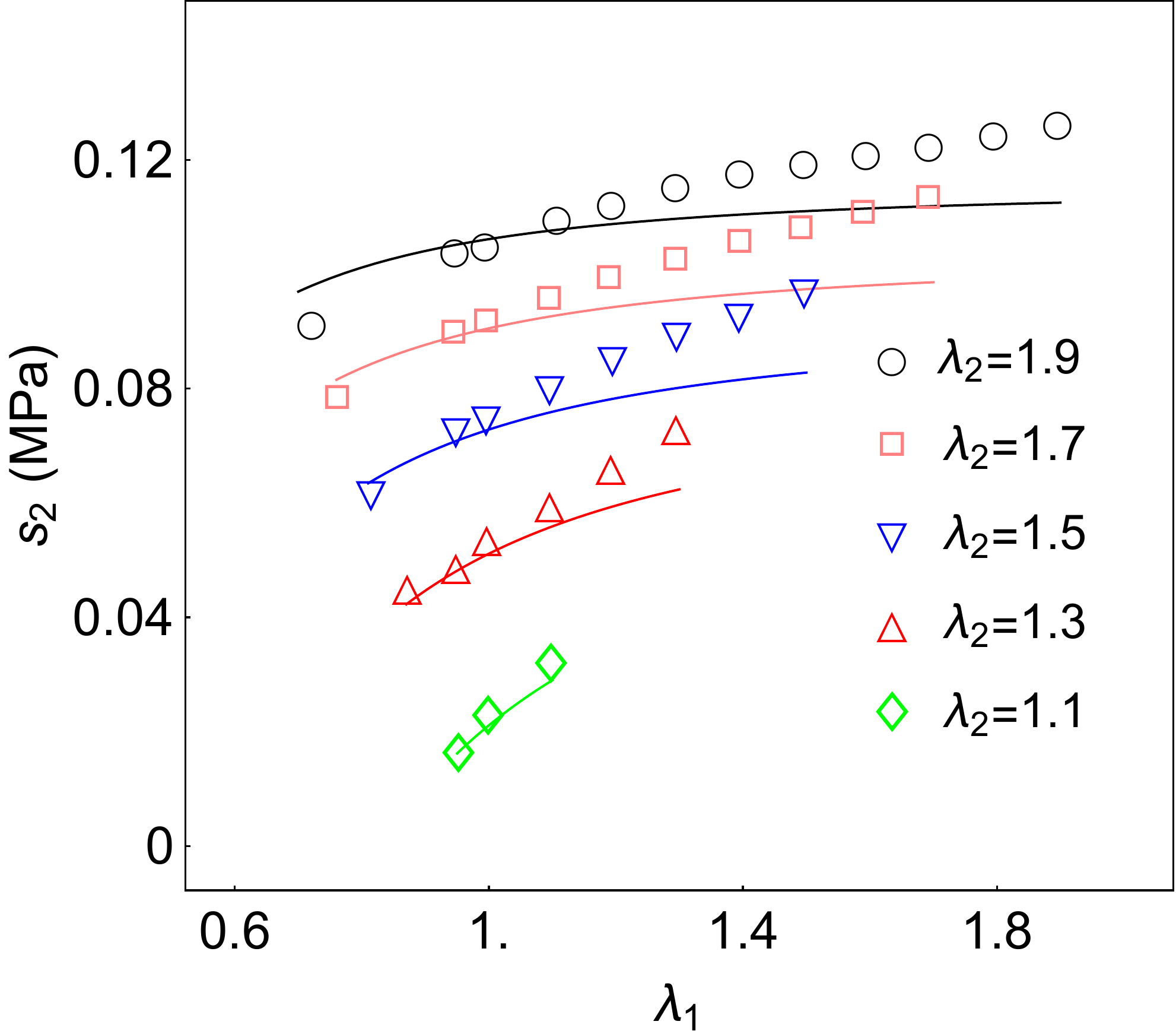}
		\end{minipage}
	}  
	\caption{Comparing the predicted (using Eq.  \ref{cauchy stress Gaussian}) and measured \citep{kawamura2001multiaxial} responses for PDMS in biaxial tests: the nominal stresses $s_1$ and $s_2$ using (a-b) the current model with $G=0.0652$MPa, (c-d) the Biot-chain model\citep{zhan2023new} with $G=0.0636$MPa, and (e-f) the neo-Hookean model with $G=0.0605$MPa. In the experiments, specimens were first subjected to equal biaxial stretching up to a maximum stretch ratio in two orthogonal directions. Then, the stretch ratio $\lambda_1$ was gradually reduced while maintaining $\lambda_2$ constant, until the nominal stress $s_1$ decreased to zero. Throughout the process, both $s_1$ and $s_2$ were evolving with the decrease of $\lambda_1$. } 
	\label{fig:pdms} 
\end{figure}

Fig.~\ref{fig:pdms} compares the simulation of the above Gaussian model with biaxial data of PDMS \citep{kawamura2001multiaxial}. Again, the present model outperforms other existing one-parameter models.

\subsection{Reduced analytical model for small-to-moderate deformation} 

\begin{figure}[]
	\centering    
	\subfigure[] 
	{
		\begin{minipage}{7.0cm}
			\centering       
			\includegraphics[width=60.00mm,height=55.00mm]{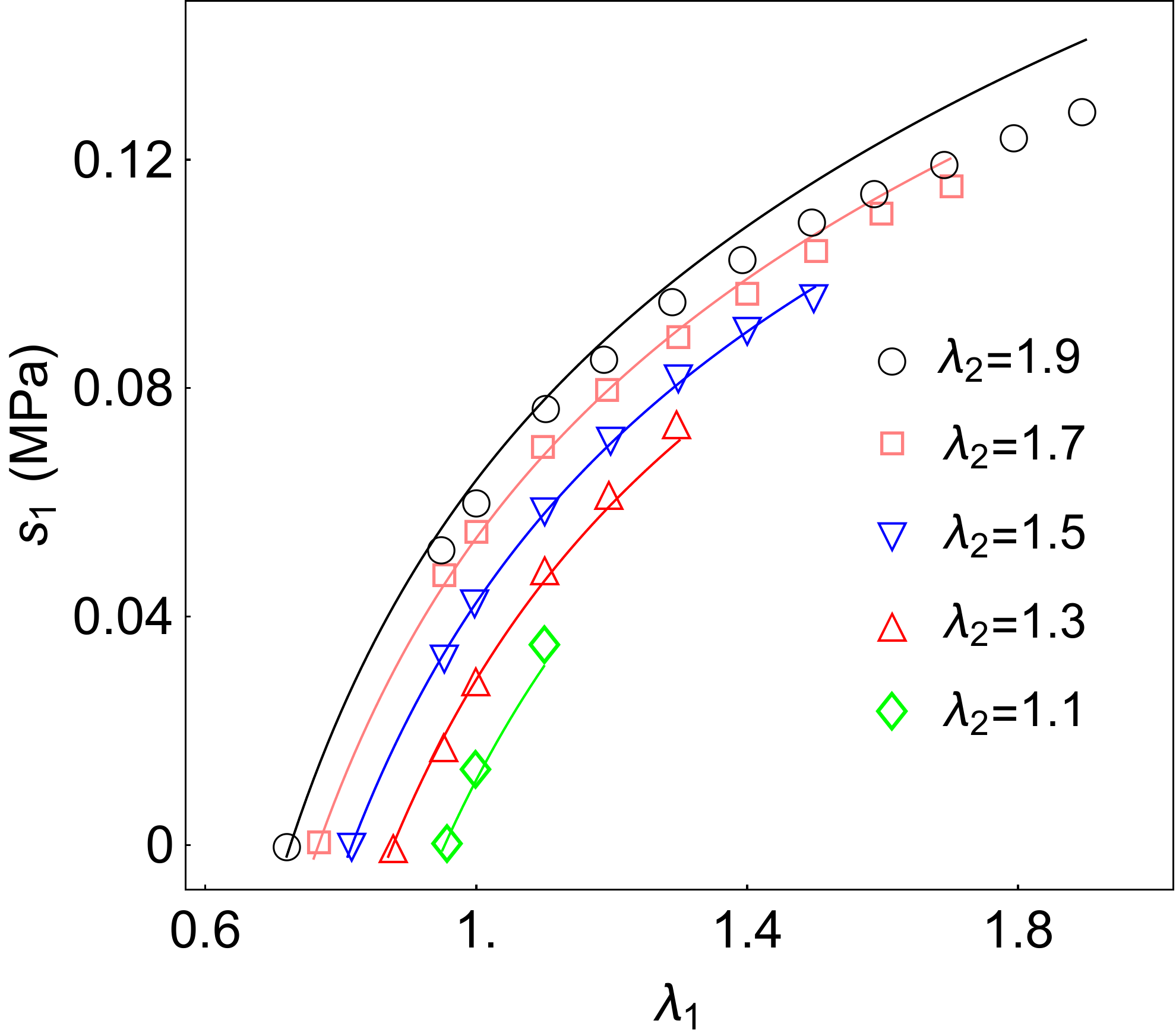}
		\end{minipage}
	}
	\subfigure[] 
	{
		\begin{minipage}{7.0cm}
			\centering       
			\includegraphics[width=60.00mm,height=55.00mm]{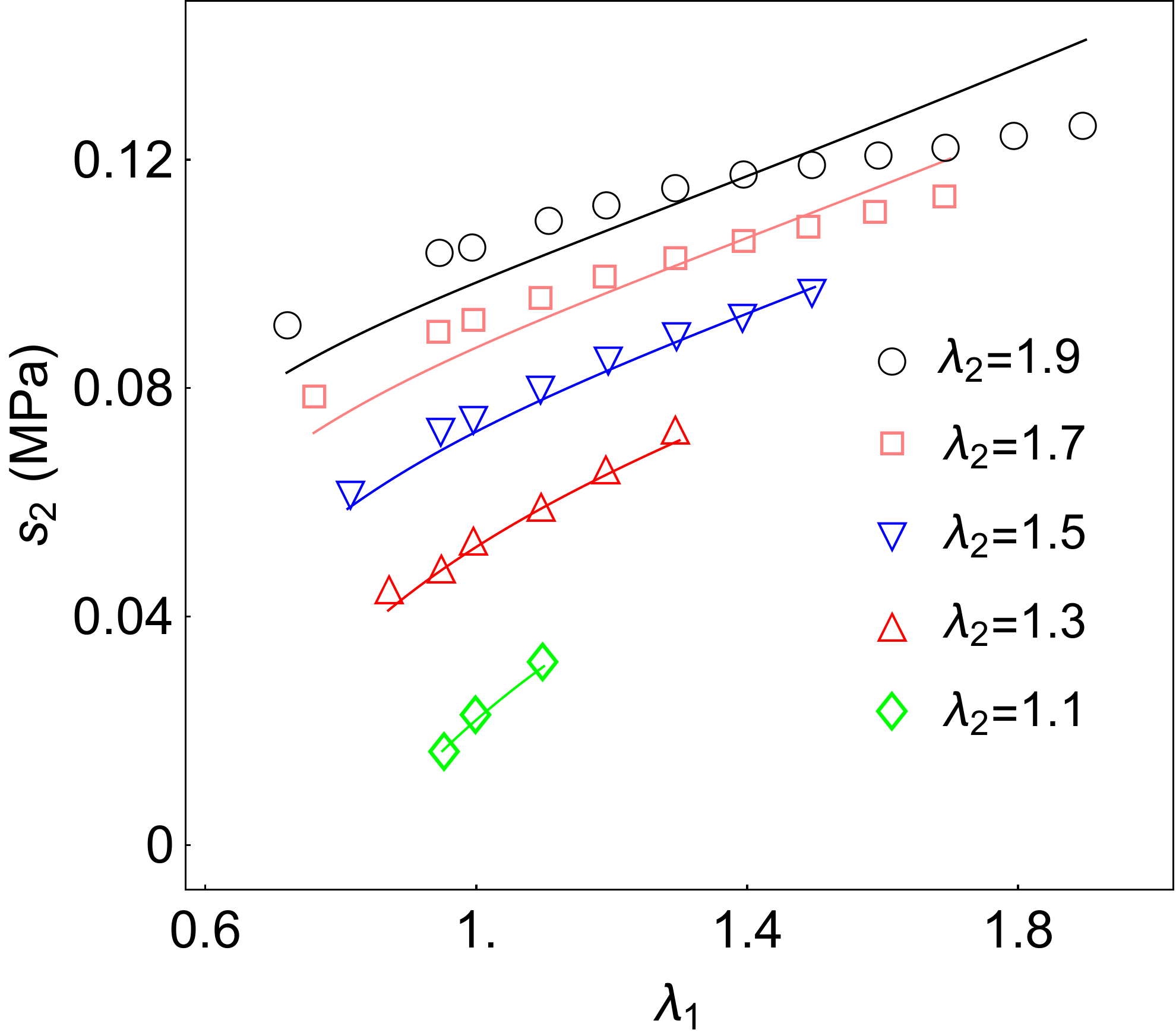}
		\end{minipage}
	}    
	\caption{Comparing the predicted and measured nominal stresses for PDMS using the fourth approximate analytical energy (Eq. \ref{analytical Gaussian}) with $G=0.063$MPa. The experimental data is extracted from \cite{kawamura2001multiaxial}.} 
	\label{fig:pdmsh} 
\end{figure}

Note that the term $e^{2\bm n\bm h\bm n}$ can be expanded as a power series:
\begin{equation}
	\label{eh expanded}
	e^{2\bm n \bm h  \bm n} = 1 + 2\bm n \bm h \bm n + \dfrac{\left(2\bm n \bm h \bm n\right)^2}{2!} + \dfrac{\left(2\bm n \bm h \bm n\right)^3}{3!} + \cdots.
\end{equation}
Using the following integral identities \citep{zhan2023new}:
\begin{equation}
	\dfrac{1}{4\pi}\int\left(\bm n\bm A\bm n\right)^k\text d\bm n=
	\begin{cases}
		\dfrac{1}{3}\text{tr} \bm A,\quad k=1,\\
		\\
		\dfrac{1}{15}\left[\left(\text{tr} \bm A\right)^2+2\text{tr} \bm A^2\right],\quad k=2,\\
		\\
		\dfrac{1}{105}\left[\left(\text{tr} \bm A\right)^3+6(\text{tr}\bm A)\text{tr} \bm A^2+8\text{tr} \bm A^3\right],\quad k=3,\\
		\\
		\dfrac{1}{945} \left[ (\text{tr} \bm A)^4 + 32(\text{tr} \bm A) \text{tr}\bm A^3 +12 (\text{tr} \bm A)^2\text{tr} \bm A^2 + 12 (\text{tr}\bm A^2)^2 + 48 \text{tr}(\bm A^4) \right]\\
		=\dfrac{1}{105} \left[ (\text{tr}\bm A)^4 -4(\text{tr}\bm A)^2\text{tr}(\bm A^2) + 4(\text{tr}\bm A^2)^2 + \dfrac{32}{3}(\text{tr}\bm A)(\text{tr}\bm A^3) \right],\quad k=4,
	\end{cases}
\end{equation}
and the incompressible condition $\text{tr}\bm h = 0$, one obtains the following fourth-order approximation to the Gussian model in Eq.  (\ref{network energy Gaussian}):
\begin{equation}
	\label{analytical Gaussian}
	\boxed{W_{\text{4th-approx}} = G \left[\text{tr} \bm h^2 + \dfrac{8}{21} \text{tr} \bm h^3 + \dfrac{2}{21} \left(\text{tr} \bm h^2\right)^2 \right].}
\end{equation}
The performance of this fourth-order approximation is shown in Fig.~\ref{fig:pdmsh}.

For small deformations, higher-order terms can be neglected, thus yielding the following quadratic form:
\begin{equation}
	\label{Hencky model}
	\boxed{W_\text{2nd-approx} = G \text{tr} \bm h^2.}
\end{equation}
This expression is known as the famous Hencky’s strain energy for incompressible materials (see \cite{neff2014axiomatic} for a English reprint), which has been shown to outperform classical quadratic energy using the infinitesimal strain tensor \citep{anand1979h,anand1986moderate,xiao2002hencky,horgan2009generalization,neff2015exponentiated} . The present derivation provides a clear physical basis for this empirical observation.

In the limit of infinitesimal deformation, the logarithmic strain reduces to infinitesimal strain. In this case, the uniaxial-tension stress is reduced from Eq.  (\ref{Hencky model})
\begin{equation}
	\label{unistressstrain}
	\sigma = 2 G (h_1 - h_2) = 3 G \varepsilon = E \varepsilon,
\end{equation}
where $\varepsilon$ denotes the infinitesimal uniaxial strain and $E$ the Young's modulus. For incompressible materials, the shear modulus is given by $E / 2(1 + \mu) = E / 3$ since the Poisson ratio $\mu$ equals  $0.5$. This confirms that $G$ corresponds to the infinitesimal shear modulus.

\subsection{Comparative discussions with classical models} 

\subsubsection{The Biot-chain model} 
As noted in \cite{dal2021performance} and \cite{zhan2023new}, two-parameter hyperelastic model allows predicting complex loading conditions based on a single test, for example, uniaxial tension. Prior to the present work, the Biot-chain model \citep{zhan2023new} has developed such desirable capability, relying on a phenomenological assumption that the chain stretch in a given direction is the average of affine deformation projected from all orientations, which yields the following chain stretch 
\begin{equation}
	\label{biotchain}
	\lambda_{\bm a} = \bm a\bm U\bm a.
\end{equation}
This relation has been also proposed from different perspective earlier by \cite{amores2021network}. Here, $\bm a$ denotes the unit direction vector of a specified chain in the initial configuration, and $\bm U = \sqrt{\bm{F}^T \bm{F}}$ is the right stretch tensor. This Lagrangian micro-macro mapping has yield constitutive models with outstanding predictive ability compared with classical models \citep{zhan2023general,zhan2023new}.

Under the assumption of Gaussian chain energy, Eq.  (\ref{biotchain}) yields an analytical model:
\begin{equation}
	\label{biotchainW}
	W_\text{Biot} = M \dfrac{1}{4\pi} \int \dfrac{3}{2} N k_{\text B} T \lambda_{\bm a}^2  \text{d} \bm a = \dfrac{1}{7} G \left[ (\text{tr} \bm U)^2 + 2 \text{tr} \bm U^2 \right],
\end{equation}
which is applicable under small to moderate deformations, as shown in Fig. \ref{fig:pdms}c-d. 

Noting that the right stretch tensor $\bm U$ can be expanded as
\begin{equation}
	\label{Uexpand}
	\bm U = e^{\bm H} = \bm I + \bm H + \dfrac{1}{2!} \bm H^2 + \dfrac{1}{3!} \bm H^3 + \cdots,
\end{equation}
where $\bm H = \ln \bm U = \bm R^T \bm h \bm R$ is the Lagrangian logarithmic strain tensor, and $\bm R = \bm{F} \bm U^{-1}$ is the rotation tensor. Substituting Eq.   (\ref{Uexpand}) into Eq.   (\ref{biotchainW}) and retaining the fourth order terms leads to
\begin{equation}
	\label{biotchainh}
	W_\text{Biot-4th-approx} = G \left[ \text{tr} \bm h^2 + \dfrac{8}{21} \text{tr} \bm h^3 + \dfrac{2}{21} \left( \text{tr} \bm h^2 \right)^2 \right],
\end{equation}
where the identity $\text{tr} \bm h^k = \text{tr} \bm H^k$ has been employed.

Comparing Eq.~(\ref{biotchainh}) with Eq.~(\ref{analytical Gaussian}) reveals that the fourth-order approximations of the present model and the Biot-chain model are identical, suggesting nearly equivalent predictive performance in the small to moderate strain regime that can be seen in Fig. \ref{fig:pdms}. This equivalence is not accidental, but rather reflects the inherent statistical randomness in molecular networks. In fact, chains initially aligned along a given direction $\bm{a}$ undergo thermally induced reorientation, redistributing across all spatial directions with non-uniform probabilities. In this context, any effective stretch associated with specified chains in an initial direction $\bm{a}$ must incorporate contributions from chains currently oriented in all directions.
The Biot-chain mapping $\lambda = \bm{a} \bm{U} \bm{a}$ can thus be interpreted as an effective stretch obtained by averaging affine deformations over all orientations under the assumption of equal probability, which coincides with the current model within the Gaussian region.

\subsubsection{The affine model}

The micro-macro connection $\ln\lambda = \bm{n}\bm h \bm n$ in our theory is derived from the relation $\dot{\overline{\ln\lambda}} = \bm n\bm d\bm n$. The latter represents a Eulerian variant of the fundamental kinematic relation $\text d\bm x = \bm F\text d\bm X$, where $\text d\bm x$ and $\text d\bm X$ denote material line elements after and before deformation, respectively. Interpreting a polymer chain as such a line element constitutes the classical affine assumption in nonlinear elasticity, enabling the simultaneous description of chain stretch and rotation. However, this assumption is only approximately valid in crystalline solids, where atomic positions are tightly constrained near fixed lattice sites and co-move with the geometrical background. It may be inappropriate for soft materials, where segment/chain possesses significant freedom with absence of lattice constraint. This perspective is well supported by insights from molecular dynamics simulations; see, for example, \cite{elliott2006stability} and \cite{tadmor2011modeling}.  In the present theory, although the chain stretch along a specified direction is governed by the continuum kinematic relation, the chain itself can randomly orient due to thermal fluctuations. The latter is characterized by an orientation probability function derived from fundamental statistical principles. As a result, the current theory departs significantly from the classical affine model, as evidenced by its excellent predictive performance in the numerical examples.

\subsubsection{The entangled models} 
It is noted that the orientation entropy essentially characterizes the interaction between different chains. It can, to a certain extent, be regarded as analogous to the classical entanglement effects \citep{heinrich1997theoretical,xiang2018general,yang2025hyperelastic}, which accounts for the the topological interaction between chains. The present work, however, lies in its ability to describe this effect through rigorous statistical principles, without introducing additional related parameters. It should be also noted that the present theory treats the network as a  weakly correlated system, that is, different segments are treated as independent but allow weak interactions for energy exchange, so as to reach thermodynamic equilibrium. As is seen in the numerical examples, this classical and simple statistical treatment yields remarkably accurate predictions of multiaxial stress-strain behavior. Our further efforts will be devoted to exploring the potential of using strongly correlated statistical system to characterize highly entangled polymer networks.

\section{The orientation induced instability}

This section is devoted to a novel orientation-induced instability of polymer networks, characterized by a phase-transition-like reconfiguration of the chain orientation. As deformation increases, polymer chains increasingly align with the primary stretched direction, reducing their density in orthogonal directions. Consequently, the stresses in non-primary stretched directions may decrease if the loss in chain density outweighs the gain in chain force. This effect becomes particularly critical in equal biaxial stretching, where large deformation leads to an unstable mechanical state: any small deviation between the two principal stretches, which is inevitable in practical settings, can trigger a sudden reorientation of the chains, thereby disrupting the isotropic response and inducing mechanical instability.

\subsection{Orientation behavior of the network}

Under given stress $\bm \sigma$, thermodynamic equilibrium requires that the strain $\bm h$ minimizes the Gibbs free energy
\begin{equation}
	\label{gibbs mini}
	\mathscr{G} =\tilde W - T S_{\text {orien}}=\tilde W + k_{\text{B}} T M N \int P_{\bm{n}} \ln P_{\bm{n}} \mathrm{d}\bm{n},
\end{equation}
with
\begin{equation}
	\label{mini complementary}
	\tilde W\equiv W-\bm \sigma:\bm h=M\int P_{\bm n}g_{\bm n}  \text{d} \bm n
\end{equation}
being the elastic complementary energy.

Provided the chains are at local equilibrium, the Cauchy stress expressed as the sum of chain-level force–configuration dyadic allows distinct chain orientation distribution. In this setting, minimizing $\tilde W$ alone drives all chains to align along the direction that minimizes $g_{\bm n}$. However, such perfectly aligned configurations correspond to minimal orientation entropy. The true thermodynamic equilibrium, therefore, emerges as a compromise between two competing effects: a tendency to minimize potential $\tilde W$, favoring alignment, and the entropic drive to maximize $S_{\text{orien}}$, favoring isotropy. It is achieved by adjusting the orientations of locally equilibrated chains while ensuring that the sum of force–configuration dyads matches the macroscopic stress. The orientation probability function $P_{\bm n}$ is quantitatively determined in this manner, as detailed in Section 2.  The stationary conditions $\partial \mathscr G/\partial\bm h=\partial \tilde W/\partial \bm h=0$ can be verified by Eqs. (\ref{Prn}) and (\ref{gibbs mini} - \ref{mini complementary}).

We now investigate the chain orientation behavior from another complementary and physically intuitive perspective. Given the  micro-macro connection $\ln\lambda=\bm n\bm h\bm n$, the equilibrium chain orientation corresponds to the orientation that minimizes the total Gibbs free energy for each strain $\bm h$, as shown in Fig.\ref{fig:stress}. Toward quantitative analysis, consider a general incompressible deformation:
\begin{equation}
	\label{logarstrain}
	\begin{aligned}
		\bm h &= h_1 \bm e_1 \otimes \bm e_1 + h_2 \bm e_2 \otimes \bm e_2 + h_3 \bm e_3 \otimes \bm e_3 \\
		&= h \left[ \bm e_1 \otimes \bm e_1 + \gamma \bm e_2 \otimes \bm e_2 - (1 + \gamma) \bm e_3 \otimes \bm e_3 \right].
	\end{aligned}
\end{equation}
Since the above form remains equivalent under permutation of $\{h_1, h_2, h_3\}$, we may, without loss of generality, assume 
$h\geq 0$ and $1\geq\gamma\geq-(1+\gamma)$, which yields the following constraint:
\begin{equation}
	\label{gamma}
	- \dfrac{1}{2} \leq \gamma \leq 1.
\end{equation}
\begin{figure}[]
	\centering    
	\includegraphics[width=50.00mm,height=45.00mm]{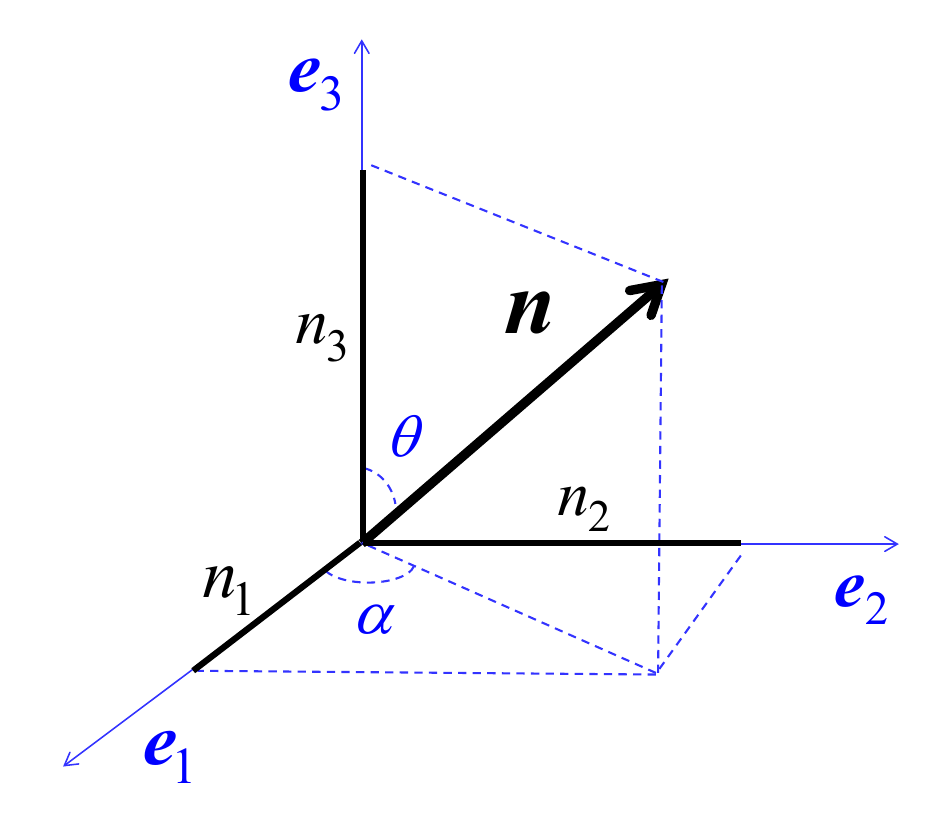}
	\caption{A sketch for the chain direction in the spherical coordinate.} 
	\label{fig:n} 
\end{figure}

Denote the chain orientation as
\begin{equation}
	\label{chainn}
	\begin{aligned}
		\bm n = \sin\theta \cos\alpha \bm e_1 + \sin\theta \sin\alpha \bm e_2 + \cos\theta \bm e_3,
	\end{aligned}
\end{equation}
and the logarithmic chain stretch as
\begin{equation}
	\label{chainthetaphi}
	\begin{aligned}
		h_{\bm n} = \ln \lambda_{\bm n} = h \sin^2\theta \cos^2\alpha + h \gamma \sin^2\theta \sin^2\alpha - h(1+\gamma) \cos^2\theta.
	\end{aligned}
\end{equation}
Here, $\theta \in [0, \pi]$ denotes the polar angle measured from the $\bm e_3$ axis, while $\alpha \in [0, 2\pi)$ is the azimuthal angle in the $\bm e_1 - \bm e_2$ plane, as shown in Fig. \ref{fig:n}.

\begin{figure}[]
	\centering  
	\subfigure[] 
	{
		\begin{minipage}{7.5cm}
			\centering       
			\includegraphics[width=73.00mm,height=58.00mm]{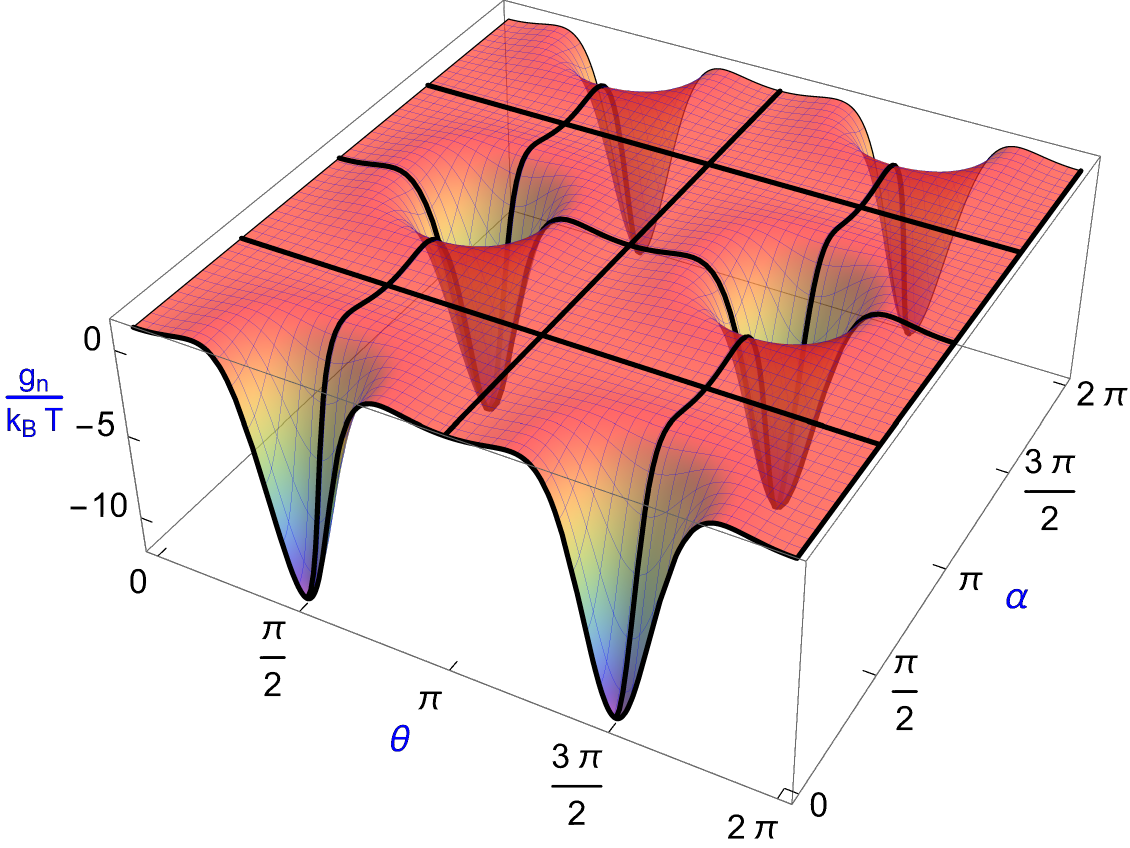}
		\end{minipage}
	}     
	\subfigure[] 
	{
		\begin{minipage}{7.5cm}
			\centering       
			\includegraphics[width=73.00mm,height=58.00mm]{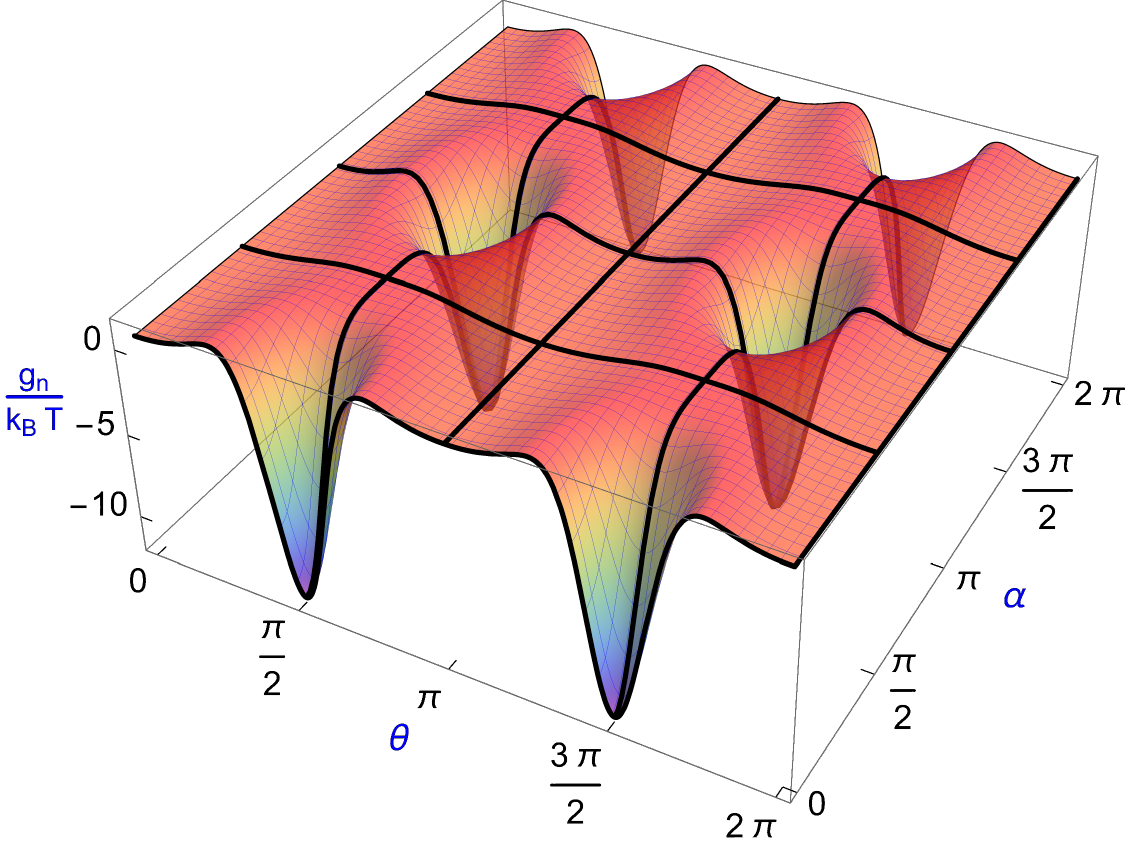}
		\end{minipage}
	}
	\subfigure[] 
	{
		\begin{minipage}{7.5cm}
			\centering       
			\includegraphics[width=73.00mm,height=58.00mm]{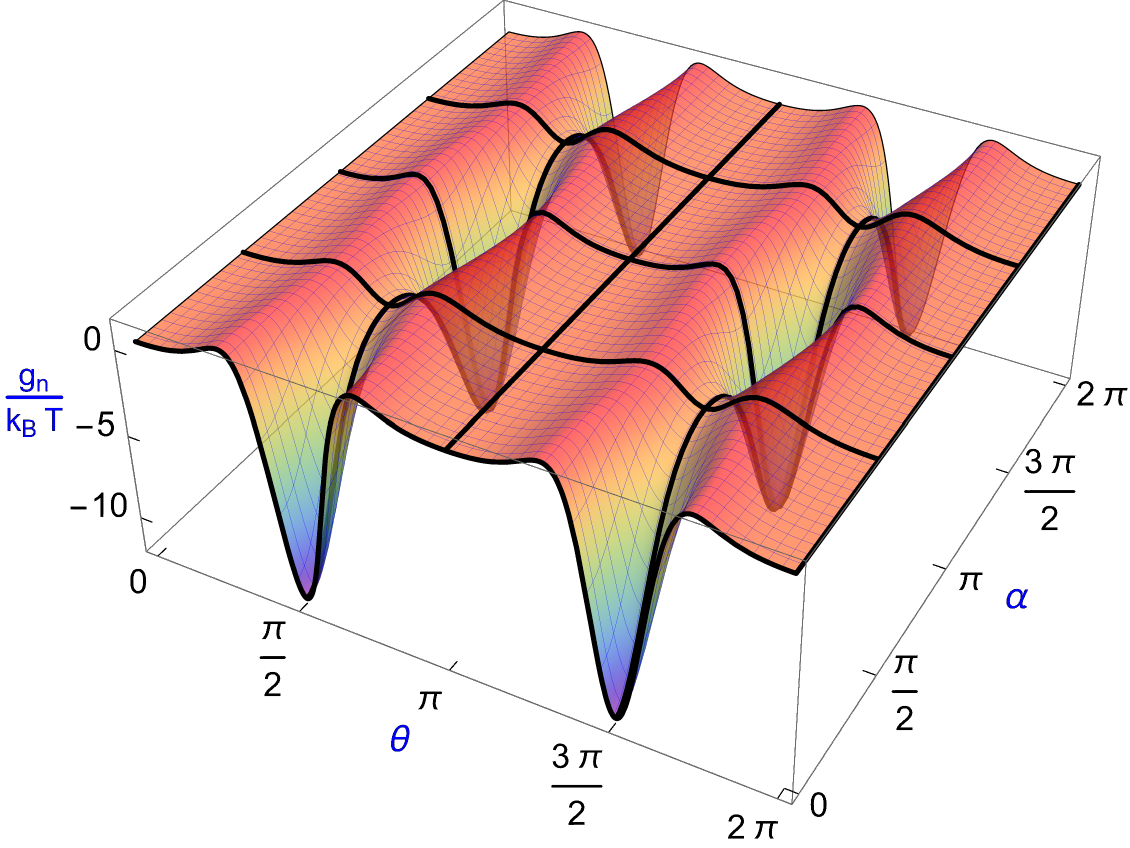}
		\end{minipage}
	}   
	\subfigure[] 
	{
		\begin{minipage}{7.5cm}
			\centering       
			\includegraphics[width=73.00mm,height=58.00mm]{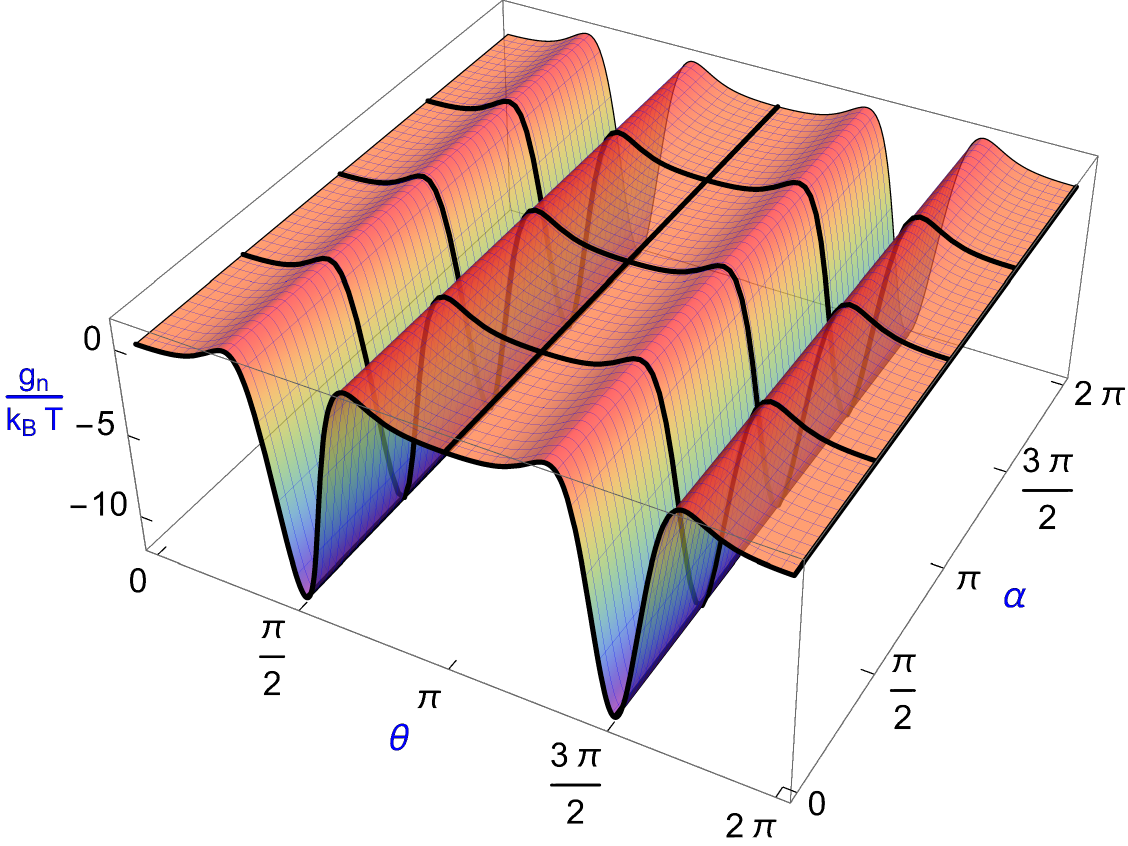}
		\end{minipage}
	} 
	\caption{
		Schematic representation of the orientation-dependent potential energy wells for chains under different loading conditions:  
		(a) uniaxial tension with $\gamma = -0.5$;  
		(b) pure shear with $\gamma = 0$;  
		(c) unequal biaxial tension with $\gamma = 0.5$; and  
		(d) equal biaxial tension with $\gamma = 1$.  
		All cases are computed with $N = 100$ and $h = 1.0$ for illustrative purpose.
	}
	\label{fig:direction-energy} 
\end{figure}

\begin{figure}[]
	\centering    
	\subfigure[] 
	{
		\begin{minipage}{7.0cm}
			\centering       
			\includegraphics[width=60.00mm,height=55.00mm]{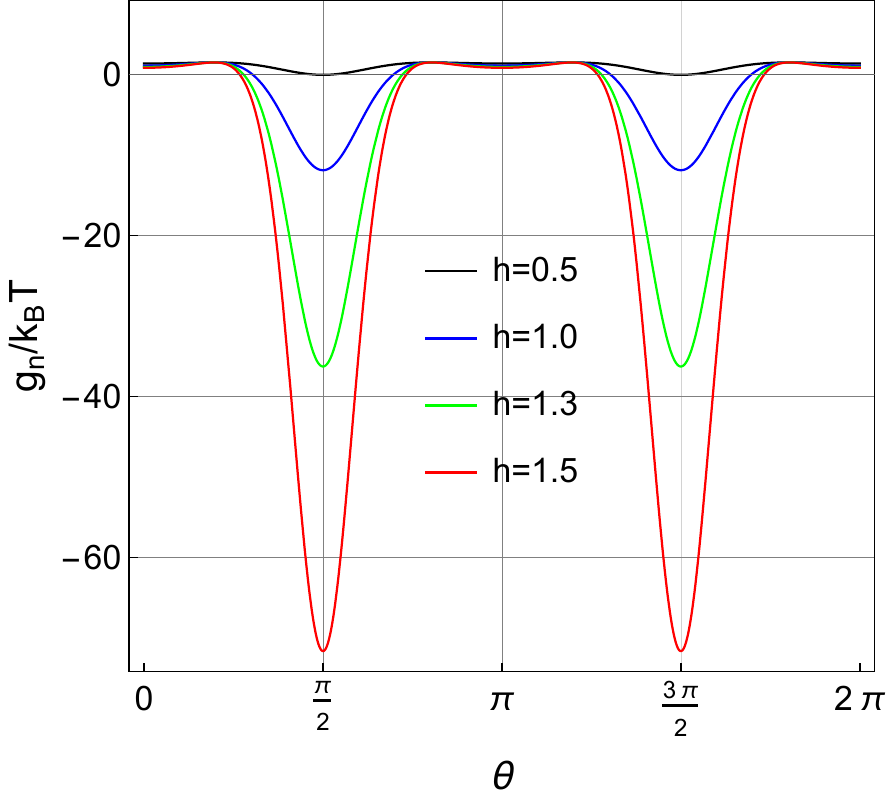}
		\end{minipage}
	}
	\subfigure[] 
	{
		\begin{minipage}{7.0cm}
			\centering       
			\includegraphics[width=60.00mm,height=55.00mm]{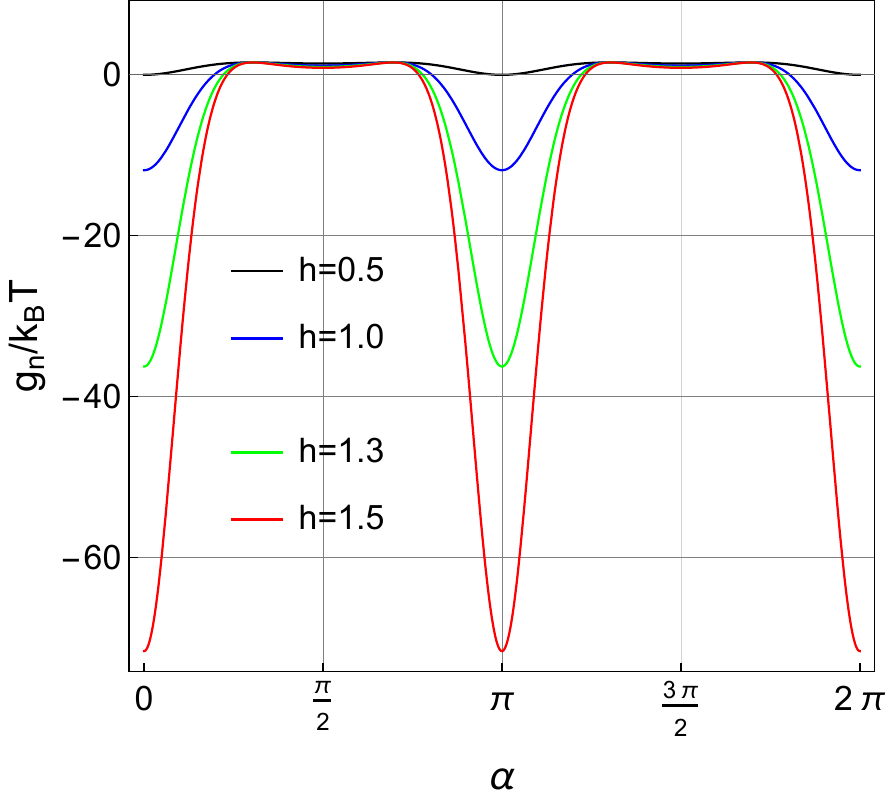}
		\end{minipage}
	}   
	\caption{
		Schematic plots of the chain potential energy under different strain magnitudes of uniaxial tension ($\gamma=-0.5$):  
		(a) in the $\bm e_1 - \bm e_3$ plane ($\alpha=0\, \&\, \pi$); and  
		(b) in the $\bm e_1 - \bm e_2$ plane ($\theta=\pi/2$).  
		All results are computed with $N = 100$ for illustrative purpose.
	}
	\label{fig:energy-plane} 
\end{figure}

For one individual chain, equilibrium would be achieved in the direction where the chain potential
\begin{equation}
	\label{Prtheta}
g_{\bm n}= w_{\bm n} - \dfrac{\partial w_{\bm n}}{\partial h_{\bm n}} h_{\bm n} =k_BTN\left( \dfrac{\lambda}{\sqrt N} \beta  + \ln \dfrac{\beta}{\sinh \beta}-\dfrac{\lambda}{\sqrt N} \beta h_{\bm n}\right).
\end{equation}
is the minimum.  Fig. \ref{fig:direction-energy} visualizes the $g_{\bm n}$ on the unit sphere surface, and the equilibrium orientations manifest clearly as distinct energy wells. Across all loading conditions, the primary stretched direction $\bm{e}_1 $ consistently corresponds to the global energy minimum, while other orientations emerge as local minima. In the special case of equal biaxial tension ($ \gamma = 1 $), all directions within the $ \bm{e}_1 - \bm{e}_2 $ plane become energetically equivalent and represent global minima. Fig.~\ref{fig:energy-plane} further depicts the potential energy under uniaxial tension for various strain magnitudes. As deformation increases, the energy well along the stretched direction $\bm{e}_1 $ deepens significantly, whereas those associated with contractive directions remain relatively shallow. This indicates an increasing likelihood for chains to align with the primary stretched direction under large deformation.

At the network level, it can be easily verified that the potential term diverges 
\begin{equation}
	\label{Prtheta_energy}
	\int P_{\bm{n}} g_{\bm n}  \mathrm{d}\bm{n} \rightarrow -\infty
\end{equation}
as the chains align perfectly the $\bm e_1$ direction and the chain stretch approaches the finite extensibility limit $\lambda/\sqrt N=1$. On the other hand, the orientation entropy is bounded:
\begin{equation}
	\label{Prtheta_entropy}
	-\dfrac{4\pi}{e}\leq \int P_{\bm{n}} \ln P_{\bm{n}}  \mathrm{d}\bm{n} \leq 0,
\end{equation}
as the function $x \ln x$ attains its minimum at $x = 1/e$ and vanishes at $x = 0$ or $x = 1$. Hence, at sufficiently large deformations, minimization of the Gibbs free energy in Eq. (\ref{gibbs mini}) is overwhelmingly dominated by the potential term $\tilde{W}=M\int P_{\bm{n}} g_{\bm n}  \mathrm{d}\bm{n}$, whose rapid decrease outweighs the relatively modest contribution from the orientation entropy. In this setting, chains tend to perfectly align the primary stretched direction $\bm e_1$ to attain the global minimum.

\subsection{Biaxial instability}
\begin{figure}[]
	\centering 
			\subfigure[] 
	{
		\begin{minipage}{8.5cm}
			\centering       
			\includegraphics[width=80.00mm,height=60.00mm]{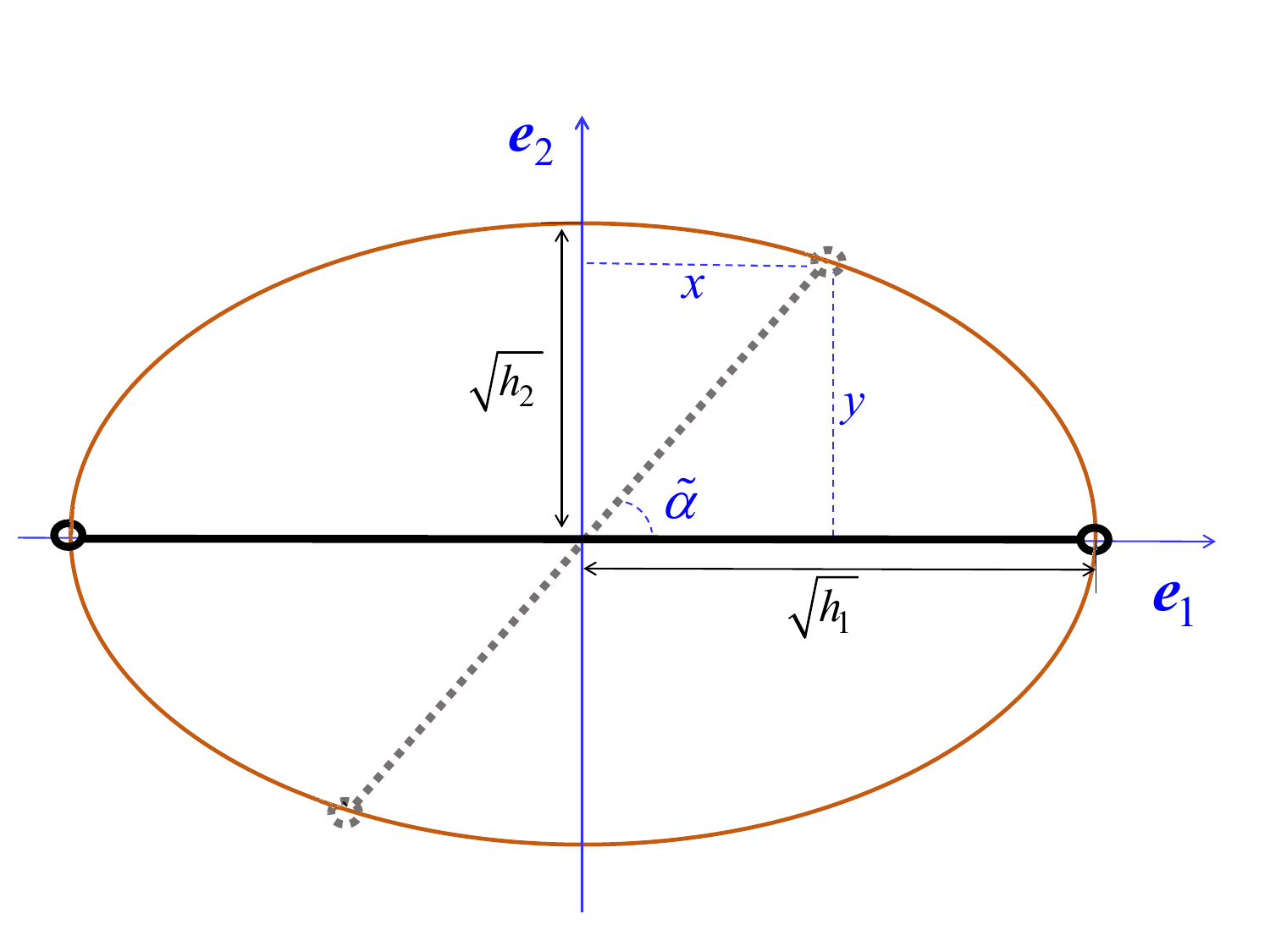}
		\end{minipage}
	}  
		\subfigure[] 
	{
		\begin{minipage}{6.0cm}
			\centering       
			\includegraphics[width=60.00mm,height=60.00mm]{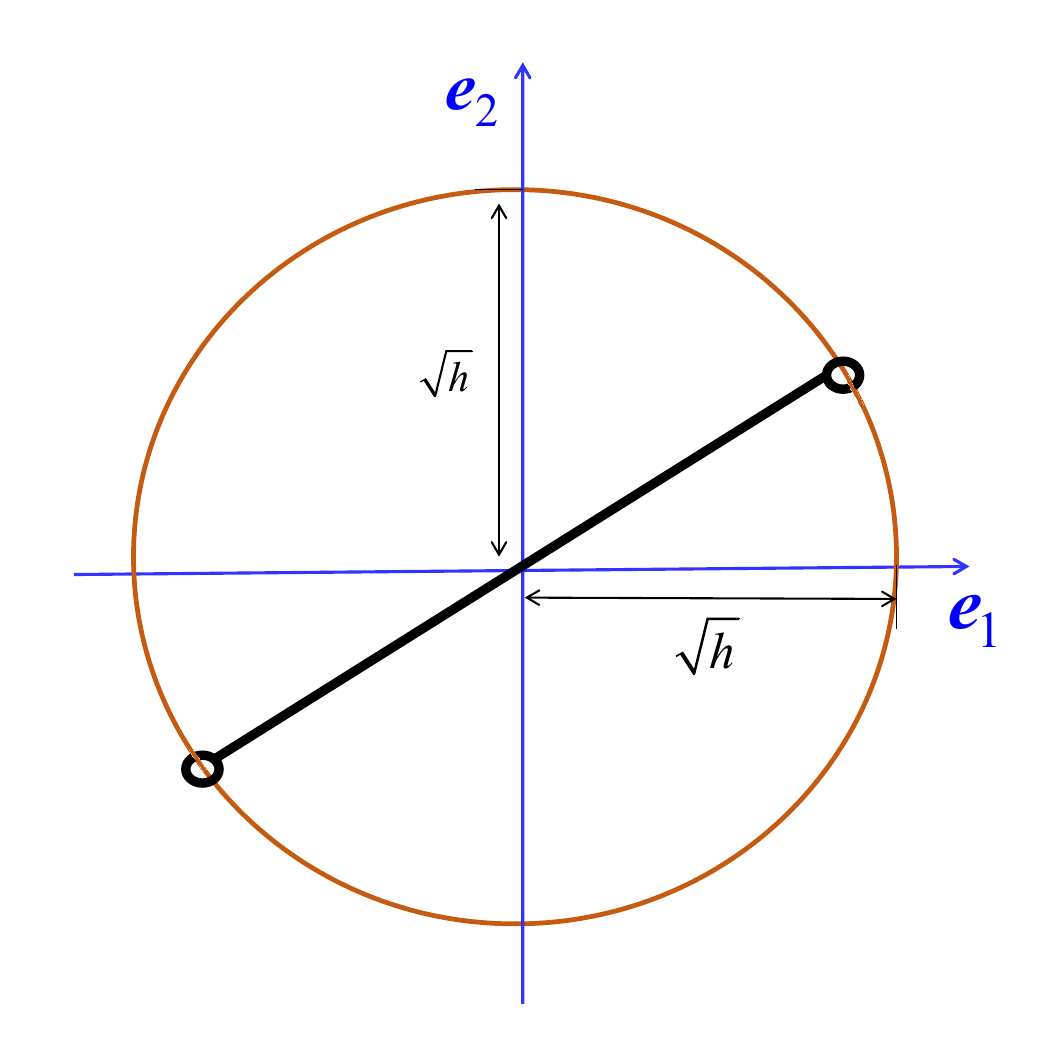}
		\end{minipage}
	}    
	\caption{A sketch for the orientation energy ellipse of biaxial tension.} 
	\label{fig:epli} 
\end{figure}

We now consider the chain orientation behaviors in the extension plane for biaxial tension, i.e., the $ \bm{e}_1 - \bm{e}_2 $ plane for $ \gamma > 0 $. In this case, $\theta=\pi/2$ and the logarithmic chain stretch in Eq.  \ref{chainthetaphi} can be recast to
\begin{equation}
	\label{chainn}
	h_{\bm n}= h_1 \cos^2 \alpha + h_2 \sin^2 \alpha = x^2 + y^2,
\end{equation}
where the coordinates \( x \) and \( y \) are given by
\begin{equation}
	\label{chainn}
	x = \sqrt{h_1} \cos \alpha, \quad y = \sqrt{h_2} \sin \alpha.
\end{equation}
It follows that $\sqrt {h_{\bm n}}$ represents the radial coordinate of an elliptical contour whose semi-long and short axes are $ \sqrt{h_1} $ and $ \sqrt{h_2} $, respectively. The associated elliptical equation reads
\begin{equation}
	\label{chainn}
	\dfrac{x^2}{h_1} + \dfrac{y^2}{h_2} = 1,
\end{equation}
as illustrated in Fig.~\ref{fig:epli}a.  It can be verified that the polar angle $\tilde\alpha$ associated with the ellipse is given by
\begin{equation}
	\label{polarangle}
	\tan\tilde\alpha = \sqrt{\gamma} \tan\alpha.
\end{equation}
This relation allows the chain orientation to be related directly to the polar angle of the ellipse. 

It is easy to verify that the potential energy $g_{\bm n}$  decreases monotonically with increasing $h_{\bm n}$ under the condition $h_{\bm n}>0$. Hence, one may intuitively analogize the chain as a nonlinear spring in compression state, with its effective length characterized by $h_n$. In this context, if the elliptical contour is assumed frictionless, the chain will naturally align along the long axis $\bm e_1 \otimes \bm e_1$, where the length is maximal and the potential is minimized, as is shown in Fig.~\ref{fig:epli}a.

An exception arises in the degenerate case of a circular contour, corresponding to the equal biaxial tension mode with $\gamma = h_2/h_1 = 1$, where all orientations in the $\bm e_1 - \bm e_2$ plane are energetically equivalent, and equilibrium may be achieved in any direction as shown in Fig. \ref{fig:epli}b. A particularly intriguing implication of this result is that a slight perturbation from this circular shape, i.e., a small deviation of the $\gamma$, can lead to a shift of the equilibrium orientation from Fig. \ref{fig:epli}b to \ref{fig:epli}a, potentially inducing significant changes in chain alignment and macroscopic material response.

Based on the discussions in Subsections 4.1 and 4.2 , we draw the following conclusions:
\begin{enumerate}
	\item For general deformation modes except for equal biaxial tension,  all chains tend to align along the primary stretch direction $\bm e_1$ as the deformation becomes large.
	
	\item In the case of equal biaxial tension under large deformation, instability may arise if the condition $\gamma = h_2/h_1 = 1$ is not strictly satisfied. Specifically, if a slight perturbation leads to $h_1 \gtrsim h_2$, the initially isotropic orientation within the $\bm e_1 - \bm e_2$ plane would abruptly shift to a concentrated alignment along the $\bm e_1 \otimes \bm e_1$ direction.
	
	\item The chain force, as the microscopic origin of macroscopic stress, aligns with the chain orientation in the equilibrium state. Consequently, if all chains align with $\bm e_1$, the stresses in the $\bm e_2$ and $\bm e_3$ directions would vanish, disregarding that the corresponding strains $h_2$ or $h_3$ may remain fairly large.
\end{enumerate}

\subsection{Numerical results}
The above conclusions can be verified using the analytical expression of orientation probability function $P_{\bm n}$ established in  Eqs. (\ref{Prn} - \ref{gn}). To quantitatively describe the anisotropy of chain orientation, we introduce the anisotropy tensor \citep{de1971short,zhan2025statistical} defined by
\begin{equation}
	\label{orientationtensor}
	\bm A\equiv\dfrac{3}{2}\left(\int P_{\bm n}\bm n\otimes\bm n\text d\bm n-\dfrac{1}{3}\bm I\right).
\end{equation}
Here, the subtraction of $\bm I/3$ ensures that $\bm A = \bm O$ for isotropic distributions. It is also straightforward to verify that $\bm A$ is traceless.

The degree of alignment along a given direction $\bm m$ is then characterized by a scalar order parameter $A_m$, defined by
\begin{equation}
	\label{orientationscalar}
	A_{ m}\equiv\bm m\bm A\bm m=\dfrac{3}{2}\int P_{\bm n}\left(\bm m\cdot\bm n\right)^2\text d\bm n-\dfrac{1}{2}.
\end{equation}
This scalar takes the value $A_m = 1$ when all chains are perfectly aligned along $\bm m$, $A_m = -0.5$ when all chains are perpendicular to $\bm m$, and $A_m = 0$ under isotropic orientation.

With Eqs. (\ref{lambdan} - \ref{gn}) and (\ref{orientationtensor}), the anisotropy tensor $\bm A$ can, in principle, be expanded as a polynomial series of the logarithmic strain tensor. As a result, it remains coaxial with the strain measure and can be represented by spectral decomposition:
\begin{equation}
	\label{anisotropy}
	\bm A=A_1\bm e_1\otimes\bm e_1+A_2\bm e_2\otimes\bm e_2+A_3\bm e_3\otimes\bm e_3,
\end{equation}
with the component along $\bm e_1$ given by
\begin{equation}
	\label{A1}
	A_1=\dfrac{3}{2}\int P_{\bm n}\left(\bm e_1\cdot\bm n\right)^2\text d\bm n-\dfrac{1}{2}=\dfrac{3}{2}\iint P_{\bm n}\sin^3\theta\cos^2\alpha\text d\theta\text d\alpha-\dfrac{1}{2},
\end{equation}
and the component along $\bm e_2$ by
\begin{equation}
	\label{A2}
	A_2=\dfrac{3}{2}\int P_{\bm n}\left(\bm e_2\cdot\bm n\right)^2\text d\bm n-\dfrac{1}{2}=\dfrac{3}{2}\iint P_{\bm n}\sin^3\theta\sin^2\alpha\text d\theta\text d\alpha-\dfrac{1}{2}.
\end{equation}
Since $\bm A$ is traceless, the third component satisfies
\begin{equation}
	\label{A3}
	A_3=-A_1-A_2.
\end{equation}

\begin{figure}[]
	\centering    
	\subfigure[] 
	{
		\begin{minipage}{7.0cm}
			\centering       
			\includegraphics[width=60.00mm,height=60.00mm]{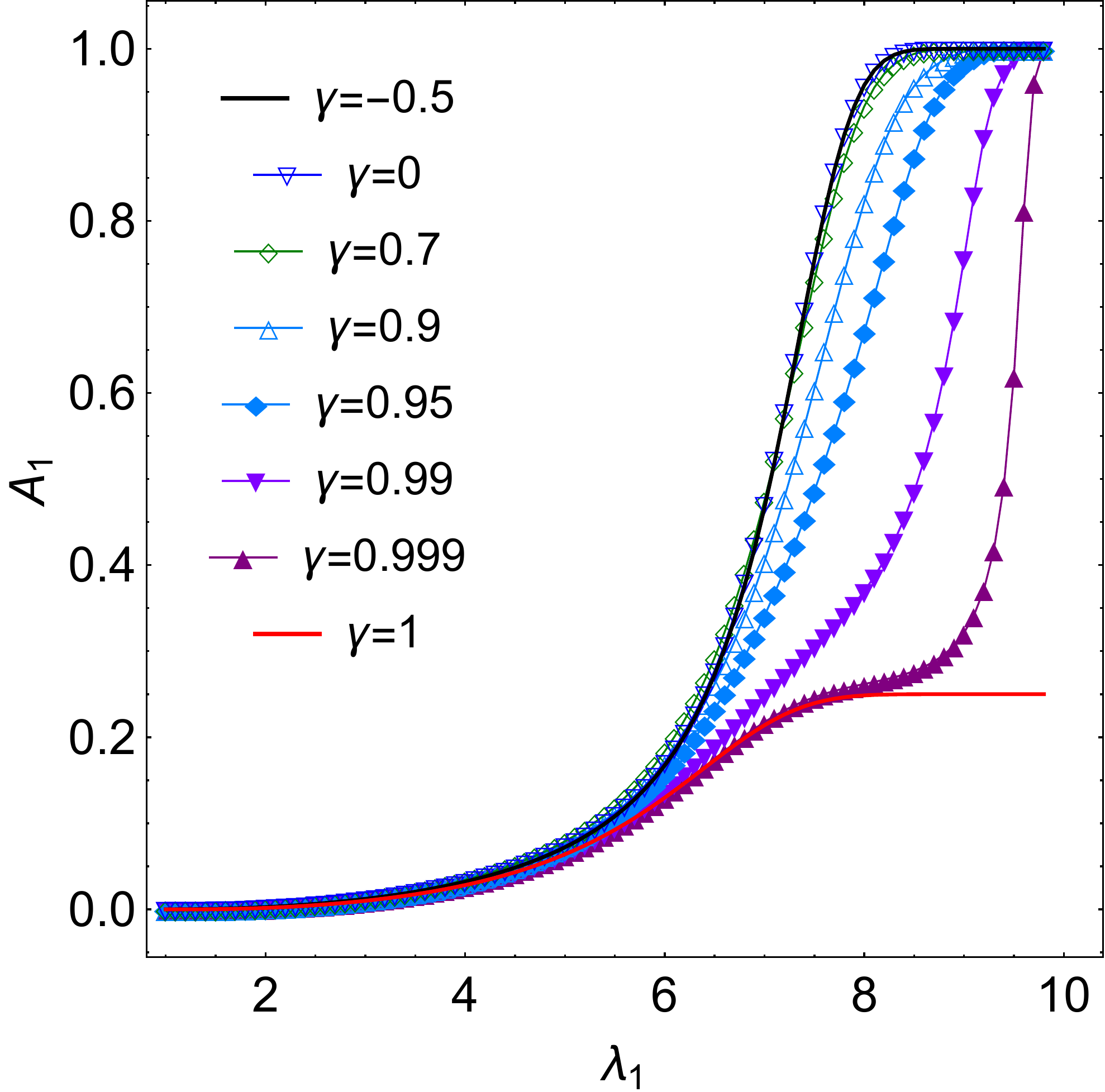}
		\end{minipage}
	}
	\subfigure[] 
	{
		\begin{minipage}{7.0cm}
			\centering       
			\includegraphics[width=60.00mm,height=60.00mm]{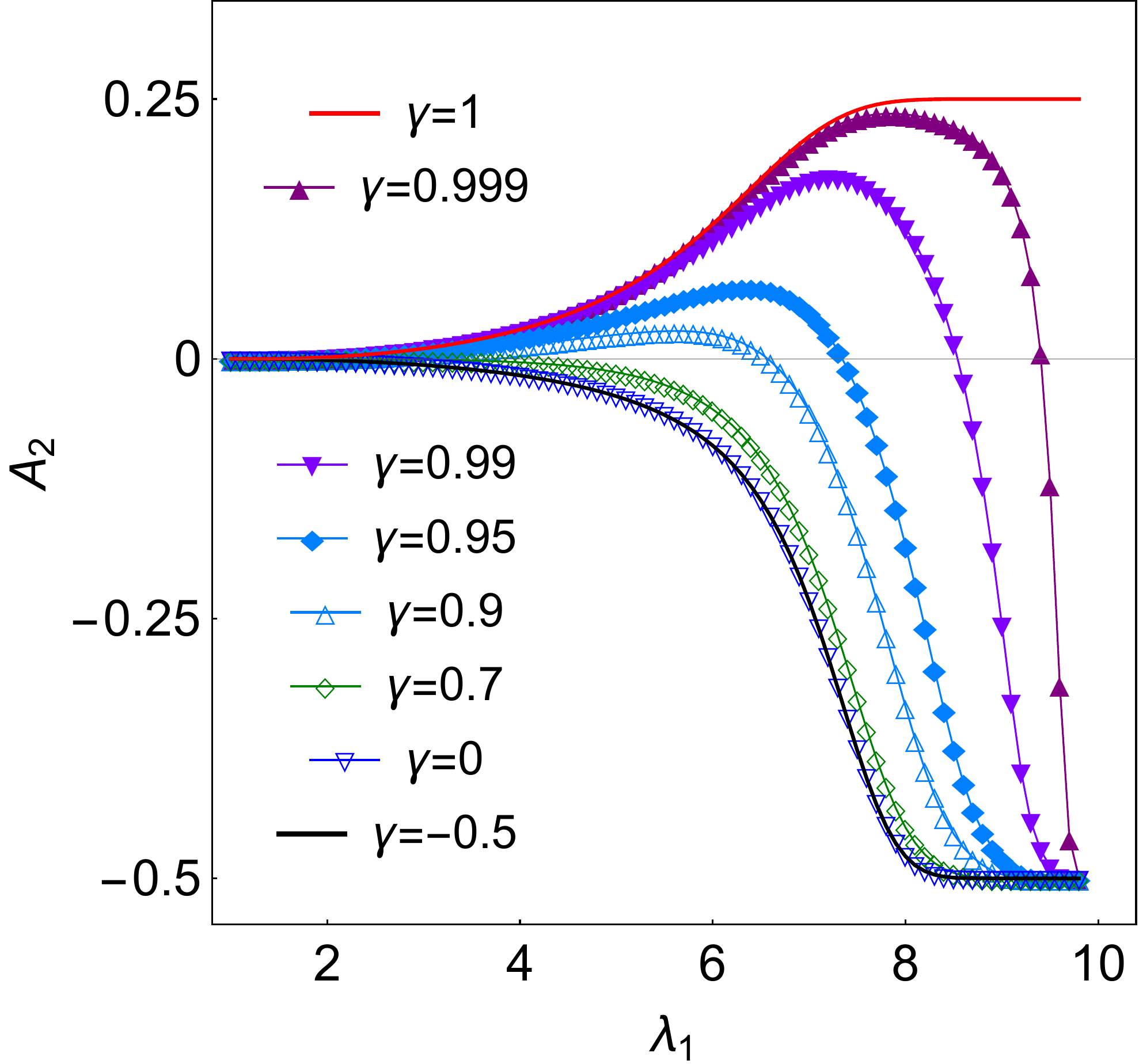}
		\end{minipage}
	}   
		\subfigure[] 
	{
		\begin{minipage}{7.0cm}
			\centering       
			\includegraphics[width=60.00mm,height=60.00mm]{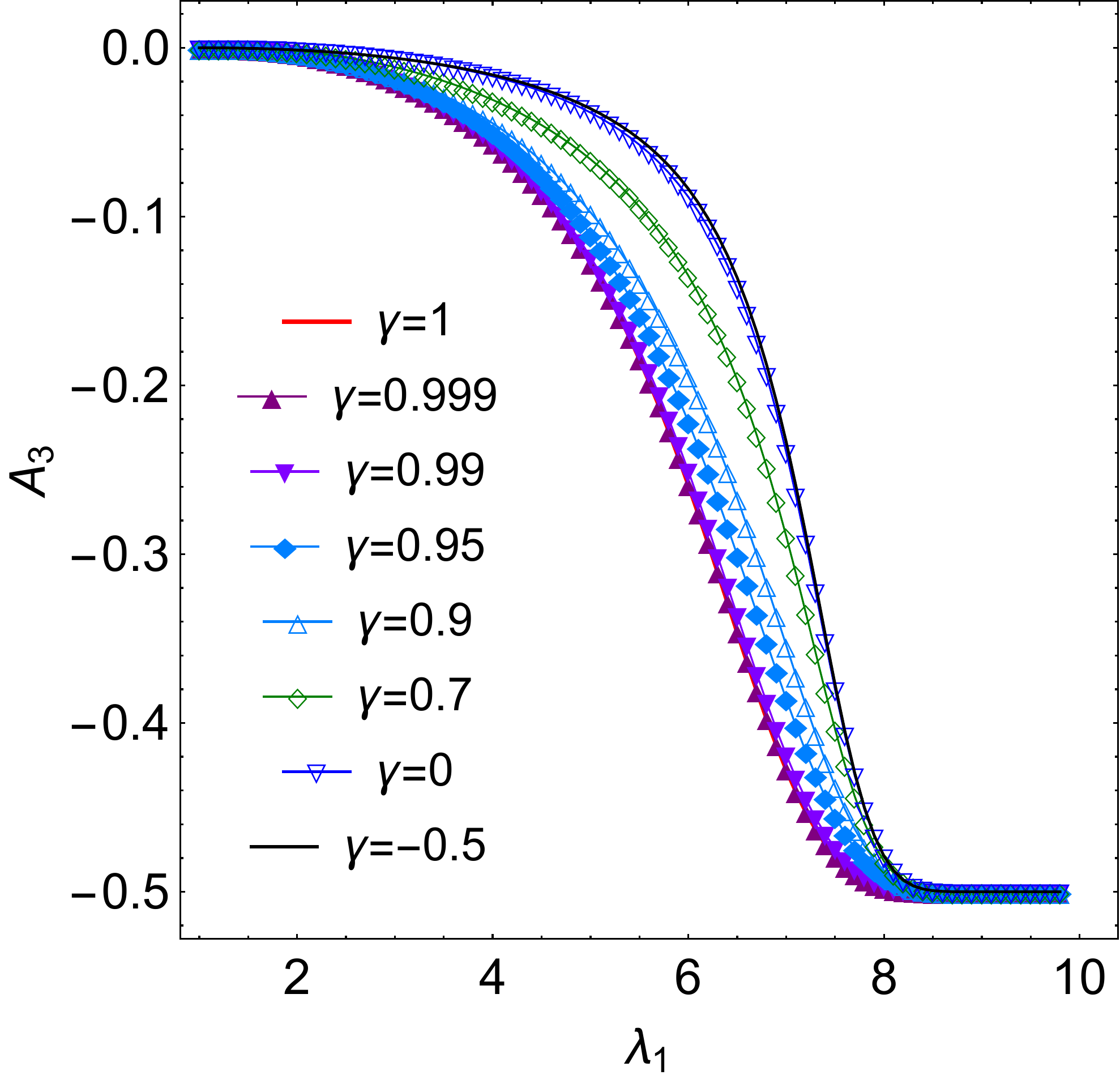}
		\end{minipage}
	}   
		\subfigure[] 
{
	\begin{minipage}{7.0cm}
		\centering       
		\includegraphics[width=60.00mm,height=60.00mm]{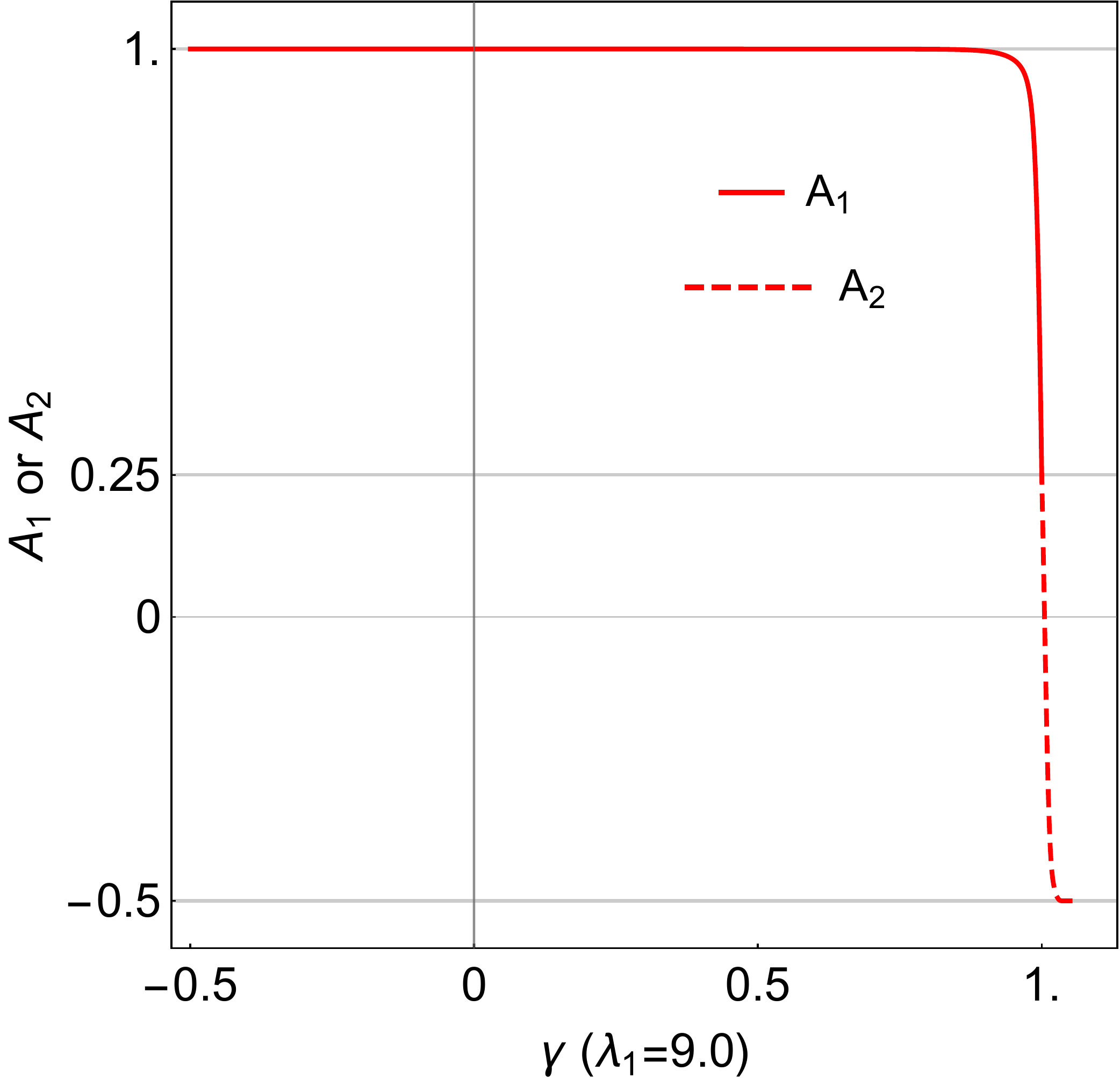}
	\end{minipage}
}  
	\caption{(a-c) Order parameters in three principle directions. (d) Order parameter as a function of the $\gamma$ with fixed primary stretch $\lambda_1 = 9.0$. All results are computed with $N = 100$ for illustrative purpose.} 
	\label{fig:order} 
\end{figure}

\begin{figure}[]
	\centering    
	\subfigure[] 
	{
		\begin{minipage}{7.0cm}
			\centering       
			\includegraphics[width=60.00mm,height=55.00mm]{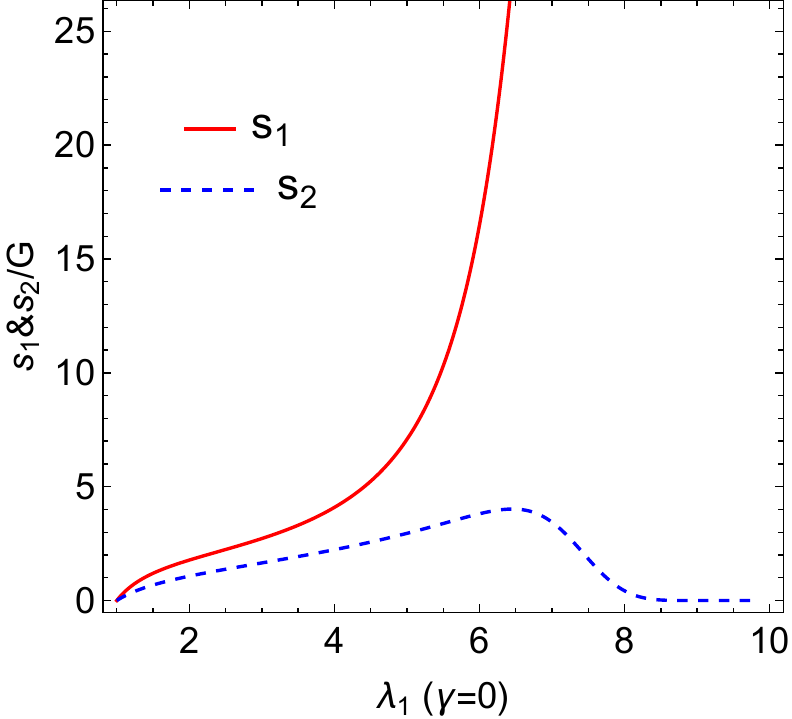}
		\end{minipage}
	}
	\subfigure[] 
	{
		\begin{minipage}{7.0cm}
			\centering       
			\includegraphics[width=60.00mm,height=55.00mm]{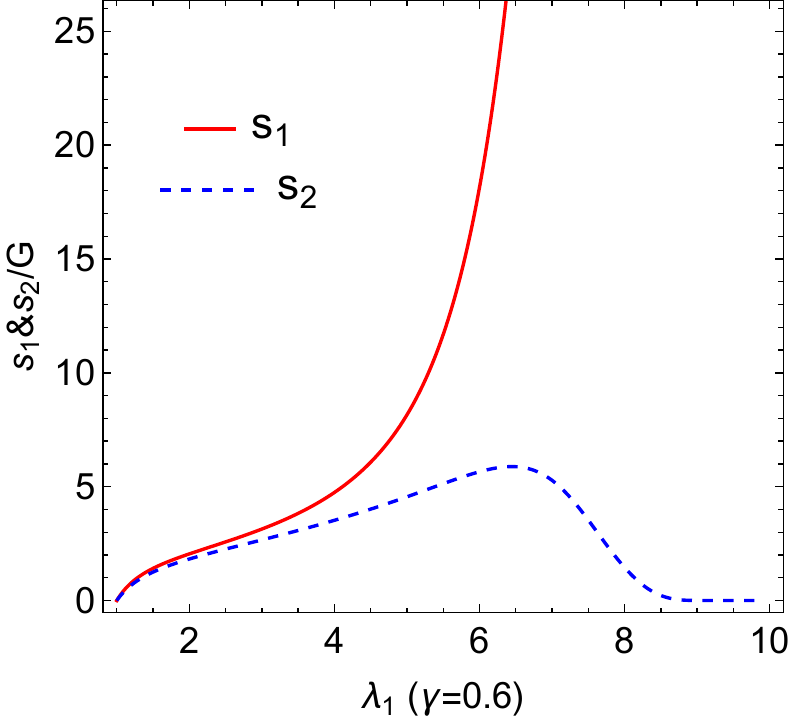}
		\end{minipage}
	}   
	\subfigure[] 
{
	\begin{minipage}{7.0cm}
		\centering       
		\includegraphics[width=60.00mm,height=55.00mm]{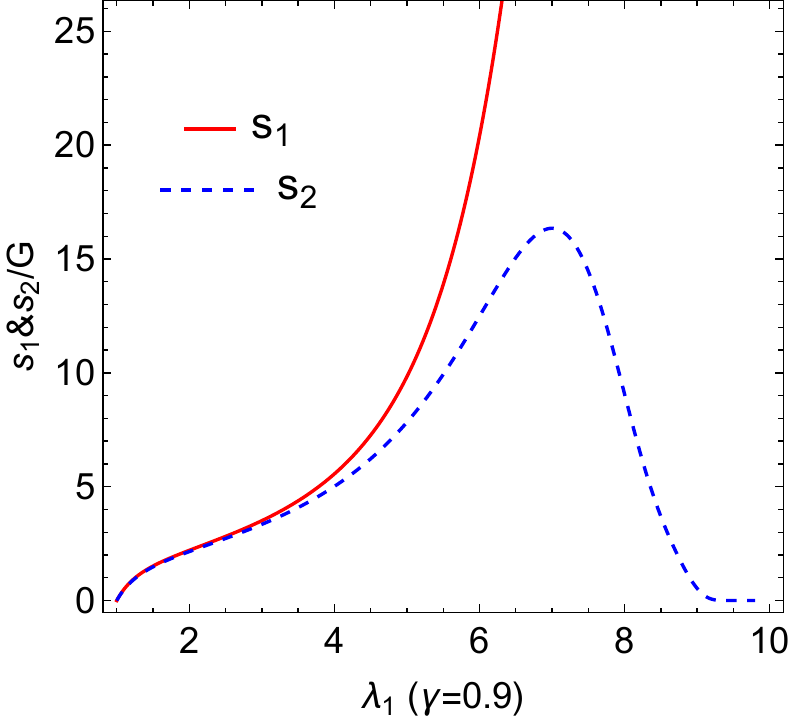}
	\end{minipage}
}     
	\subfigure[] 
{
	\begin{minipage}{7.0cm}
		\centering       
		\includegraphics[width=60.00mm,height=55.00mm]{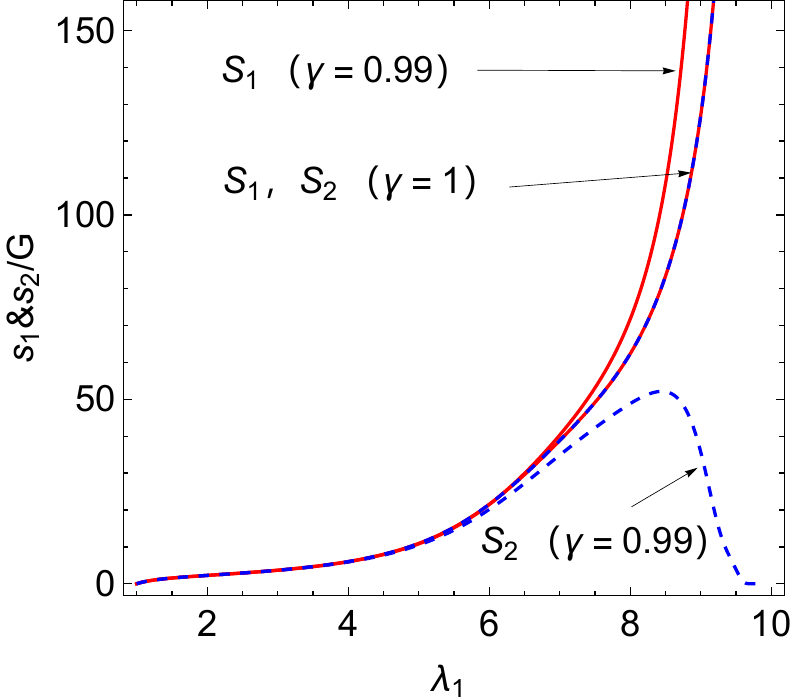}
	\end{minipage}
}   
	\caption{(a-c) Stress softening in unequal biaxial tensions. (d) Instability in equal biaxial tension. All results are computed with $N = 100$ for illustrative purpose. $s_1$ and $s_2$ represent the nominal stresses in the $\bm e_1$ and $\bm e_2$ directions, respectively.} 
	\label{fig:instability} 
\end{figure}

Fig.~\ref{fig:order} illustrates the numerical results for order parameters in three principal directions for biaxial tension  with different $\gamma$. It is observed from Figs. \ref{fig:order}a - b that the order parameters along $\bm e_1$ and $\bm e_2$ exhibit a sharp transition as the secondary stretch $\lambda_2$ slightly deviates from the primary stretch $\lambda_1$ in the large deformation region. Specifically, the order parameters shift from an isotropic state with $A_1 = A_2 = 1/4$ to a highly anisotropic configuration with $A_1 = 1$ and $A_2 = -0.5$. This behavior indicates that, in the regime of large deformation, the initially isotropic chain orientations in equal biaxial tension collapse into alignment with the primary stretch direction $\bm e_1$ as long as perfect biaxial symmetry is lost. The transition is clearly illustrated in Fig.\ref{fig:order}d, where the order parameter in the direction with fixed stretch $\lambda=9.0$ plotted as a function of $\gamma$ exhibits a sharp transition at $\gamma = 1$.  This confirms our predictions in the previous subsection.

These results suggest that equal biaxial tension at large strains corresponds to an unstable equilibrium state: even slight perturbations in the loading conditions can induce a rapid reorientation of polymer chains, thereby disrupting the initial isotropic configuration. This microscopic phase transition is accompanied by pronounced macroscopic consequences: the stress component in $\bm e_2$ direction vanishes when all chains orientate to the direction $\bm e_1$ with slightly higher stretch $\lambda_1$. As a result, equal biaxial loading becomes mechanically unstable under large deformation, since perfect equality between the $\lambda_1$ and $\lambda_2$ is nearly impossible to maintain experimentally due to inevitable external disturbances or imperfections in the loading apparatus. This phenomenon is clearly seen in the numerical results shown in Fig.\ref{fig:instability}d, where a small deviation from equal biaxial stretching leads to a significant shift in the stress response. Furthermore, Figs.~\ref{fig:instability}a – c confirm that, under large deformation, the transverse stress component $s_2$ progressively diminishes as the primary stretch $\lambda_1$ increases, marking a distinct departure from those observed in the moderate strain regime (see, for example, Figs. \ref{fig:pdms}, \ref{fig:jonestreloar} and \ref{fig:gel}).

The practical observation of this biaxial instability might be difficult due to material damage at large strains. However, in systems where chain orientation is additionally modulated by external stimuli, for example, the applied electric fields, the onset of biaxial instability may occur at considerably lower strain levels. This phenomenon may be related to the electro-mechanical instability in dielectric elastomers \citep{zhao2007method,suo2008nonlinear}. Our present theory has the potential to offer a physically grounded framework for linking these instabilities to microscopic orientation dynamics within a consistent statistical-mechanical formulation, which will be elaborated in a forthcoming study.

\section{Further discussions}

\subsection{Shifting from Lagrangian to Eulerian formulation}

In the present theory, segments/chains that share the same orientation are treated as statistically identical and indistinguishable. This indistinguishability implies that individual segments/chains do not need to be tracked, and only ensemble-average properties carry physical meaning. Such a statistical perspective naturally calls for a Eulerian formulation, wherein chain stretch and orientation are defined as spatial field variables over the current configuration, rather than being tied to specific chains. This shift parallels the transition in fluid mechanics from tracking individual particles to describing spatial fields within a Eulerian framework.
In the present study, the chain kinematics is governed by two factors: the deformation of the macroscopic network geometry and the random thermal motion. The stretch of a chain along the spatial direction $\bm n$ is described by the Eulerian kinematic relation $\ln\lambda=\bm n\bm h\bm n$. However, due to thermal fluctuations, a specific chain cannot be continuously tracked throughout the deformation, as it may randomly reorient to other directions with the probability $P_{\bm n}$ that derived from fundamental statistical principles.

This stands in contrast to most classical microstructural models, which depict the network structure through the trajectories of specific chains and are fundamentally rooted in a Lagrangian perspective. These models implicitly assume that the configuration of a chain evolves deterministically from its initial state, dismissing the inherent statistical freedom due to molecular-scale thermal motion.

\subsection{The privileged role of logarithmic strain in nonlinear elasticity theory}

An important outcome of the present theory is the natural emergence of the logarithmic strain as the governing measure of deformation. As shown in Section 2, the explicit relationship between molecular chain stretch and macroscopic deformation can be only obtained using the logarithmic strain. A more physical justification is provided in \cite{zhan2025statistical}, where the chain logarithmic stretch arises from a canonical transformation that symmetrically treats force and deformation as a joint variable, so as to eliminate the thermodynamic inconsistency in describing the parallel chain system. These insights suggest the possible statistical significance of logarithmic strain, particularly in systems composed of identical particles with indistinguishable microstates, where the logarithmic forms can naturally emerge from the requirement of thermodynamic consistency between intensive and extensive conjugate pairs.

In continuum mechanics, the logarithmic strain has long been appreciated for its mathematical, geometrical or numerical merits \citep{hoger1987stress,xiao1997logarithmic,criscione2000invariant,bruhns2001constitutive,xiao2002hencky,miehe2002anisotropic,neff2015exponentiated,montella2016exponentiated,zhan2022high}. Building on this foundation, the present study further contributes a physical rationale for the privileged role of logarithmic strain in nonlinear large deformation elasticity theory.

\subsection{The proper physical picture of polymer network.}
Due to the striking agreements between theoretical predictions and experimental observations, we are led to believe that the hierarchical statistical ensemble elaborated in Section 2 properly captures the essential features of molecular kinematics of the polymer network. As previously discussed, the network behaves statistically like a system of weakly interacting particles, analogous to the ideal gas. Each segment adopts its orientation to maximize the global entropy, akin to how gas molecules distribute their velocities and positions in thermodynamic equilibrium. The presence of chains, or equivalently the network junctions, imposes localized kinematic constraints on segments, much like nails anchoring a soft fluctuating mesh. The nails (junctions) themselves also undergo random thermal motion, while the continuum stress acts as a global constraint confines the collective behavior of these ``nails". Neither constraint alters the role of segments as the fundamental statistical units and their weak-coupling interaction.

A more interesting perspective may be drawn by analogy with the well-known blackbody radiation problem in statistical physics. Although light manifests as an electromagnetic wave, the thermal radiation from a blackbody — a system comprising a large collection of electromagnetic waves under thermodynamic equilibrium — can be accurately modeled by treating the system as a photon gas, where photons act as the basic units of energy exchange. It is only by applying statistical mechanics to individual photons that the correct Planck’s law can be obtained. Conversely, directly applying statistical analysis to electromagnetic waves yields the incorrect Rayleigh–Jeans law (see, for instance, \cite{blundell2010concepts}). 

In a similar vein, the thermodynamic equilibrium of a network composed of polymer chains can be viewed as analogous to that of a black body with electromagnetic waves, where segments play the role of photons, i.e., the fundamental carriers of thermal interaction. The energy of a photon is determined by the wave’s frequency, while the energy of a segment is determined by the chain length. Our results demonstrate that statistical treatment based on the potential energy of individual segments (expressed as $g_{\bm n}/N$ in Eq.  (\ref{Prn})) yields results that are quantitatively consistent with experimental data. In contrast, performing statistical analysis directly on the total chain potential $g_{\bm n}$ leads to predictions that deviate significantly from observations. This failure, underscores the need to redefine the statistical degrees of freedom in polymer networks, elevating the segment, not the chain, as the fundamental unit governing network behavior.

\section{Conclusions}
This work presents a new statistical-mechanical theory for the entropic elasticity of soft polymer networks, without invoking ad hoc assumptions or empirical parameters. Instead of directly extending well-established single-chain models to the continuum level, as in traditional approaches, the macroscopic behavior is derived from global thermodynamic equilibrium condition that maximizes the number of accessible microstates for all segments in the network.

The theory provides explicit expressions for chain stretch and orientation probability as functions of spatial direction and logarithmic strain. With only two physical parameters, the resulting hyperelastic model outperforms existing two-parameter models under a wide range of multiaxial loading conditions. For moderate deformations, it reduces to an analytical form equivalent to the Biot-chain model \citep{zhan2023new}; in the small-strain limit, it recovers Hencky’s classical quadratic strain energy.

The framework further predicts a biaxial mechanical instability as a collective phase transition in chain orientation. Under large deformations, chains spontaneously align with the primary stretch direction, diminishing their density in orthogonal directions. Consequently, stress in non-primary directions may decrease when density loss outweighs the force gain. This effect is especially pronounced in equal biaxial tension, where any infinitesimal asymmetry between principal stretches can trigger rapid chain reorientation and stress softening.

Taken together, these results constitute a conceptual and practical advance in the constitutive modeling of soft matter. We believe the present theory establishes a new perspective that bridges the gap between segment/chain statistical physics and the macroscopic network behavior, and provides a potential framework for further relevant studies.

\section*{Acknowledgements}
Lin Zhan would like to expresses sincere gratitude to Professor Xiaofeng Jin (Fudan University) for his inspiring open courses on Thermodynamics and Statistical Physics. This work is supported by the National Natural Science Foundation of China (Grant No. 12202378), and the Guangdong Basic and Applied Basic Research Foundation (Grant No. 2025A1515011469), and the Postdoctoral Fellowship Program of CPSF (Grant No. GZC20230963), and the China Postdoctoral Science Foundation (Grant Nos. 2024T170347, 2023M731318). 

\section*{Appendix: Further numerical examples for the hyperelastic model}
\begin{figure}[]
	\centering    
	\includegraphics[width=60.00mm,height=60.00mm]{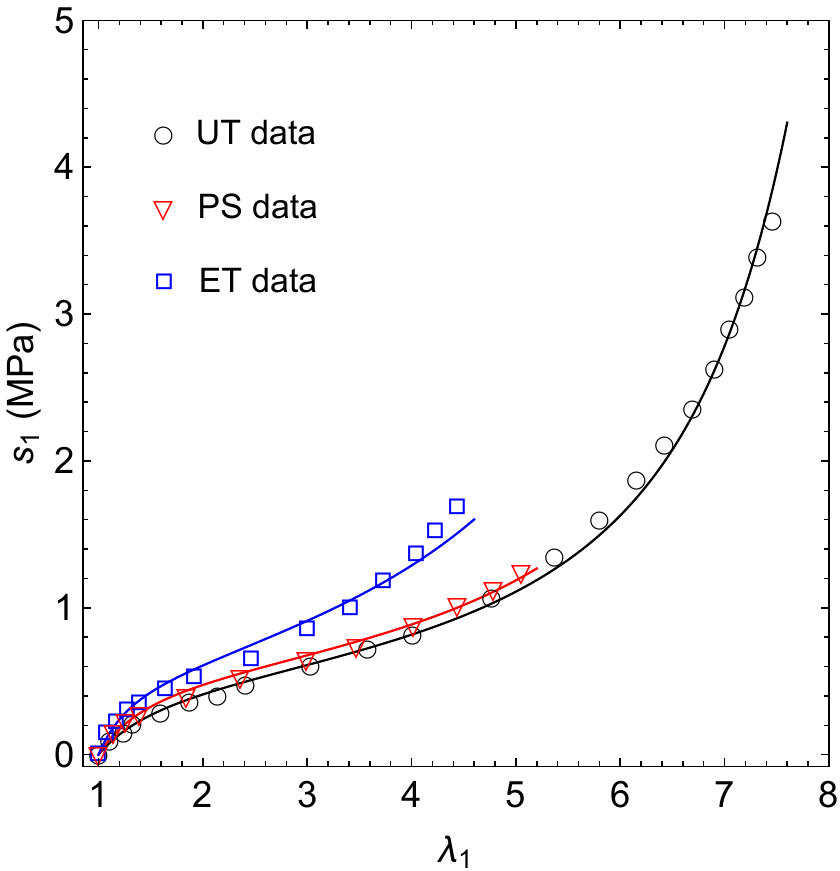}
	\caption{Comparing the predicted and measured nominal stresses for Entec Enflex S4035A TPE using the model in Eq.  (\ref{cauchy stress Langevin}) with $G=0.28$MPa, $N=166$. The experimental data is extracted from \cite{zhao2016continuum}.} 
	\label{fig:TPE} 
\end{figure}

\begin{figure}[]
	\centering    
	\subfigure[] 
	{
		\begin{minipage}{7.0cm}
			\centering       
			\includegraphics[width=60.00mm,height=55.00mm]{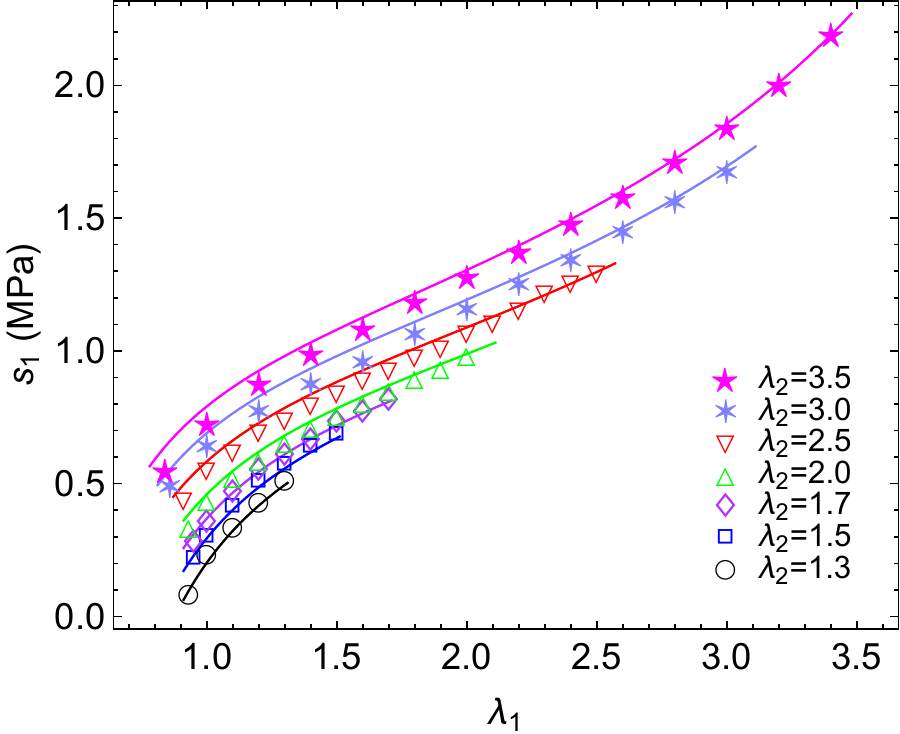}
		\end{minipage}
	}
	\subfigure[] 
	{
		\begin{minipage}{7.0cm}
			\centering       
			\includegraphics[width=60.00mm,height=55.00mm]{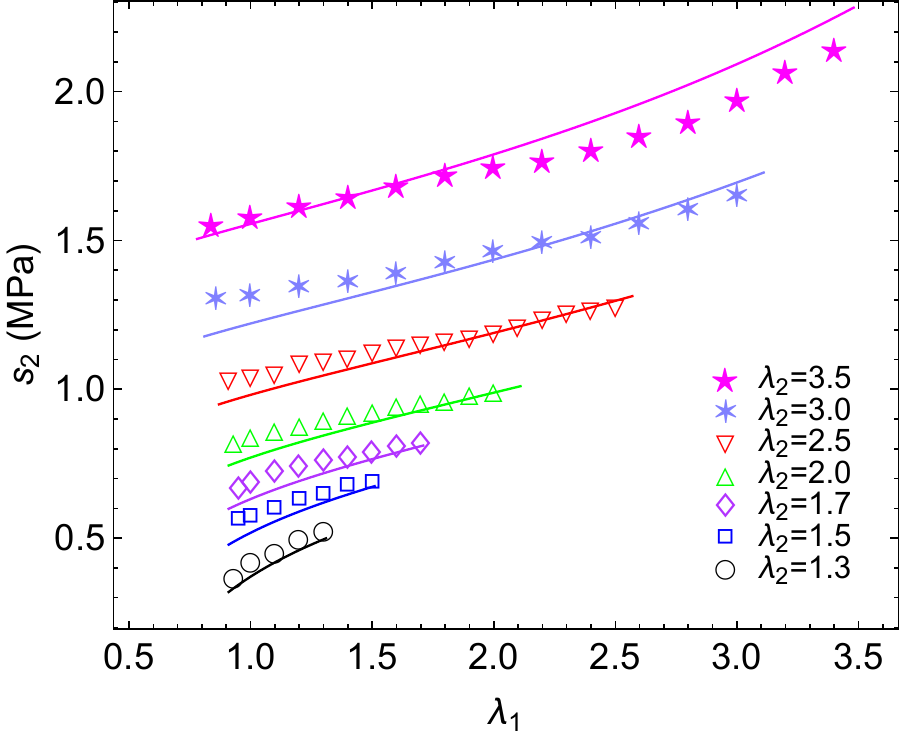}
		\end{minipage}
	}    
	\caption{Comparing the predicted and measured nominal stresses for natural rubber using the model of Eq.  (\ref{cauchy stress Langevin}) with $G=0.435$MPa, $N=56$. The experimental data is extracted from \cite{jones1975properties}.} 
	\label{fig:jonestreloar} 
\end{figure}

\begin{figure}[]
	\centering    
	\includegraphics[width=60.00mm,height=60.00mm]{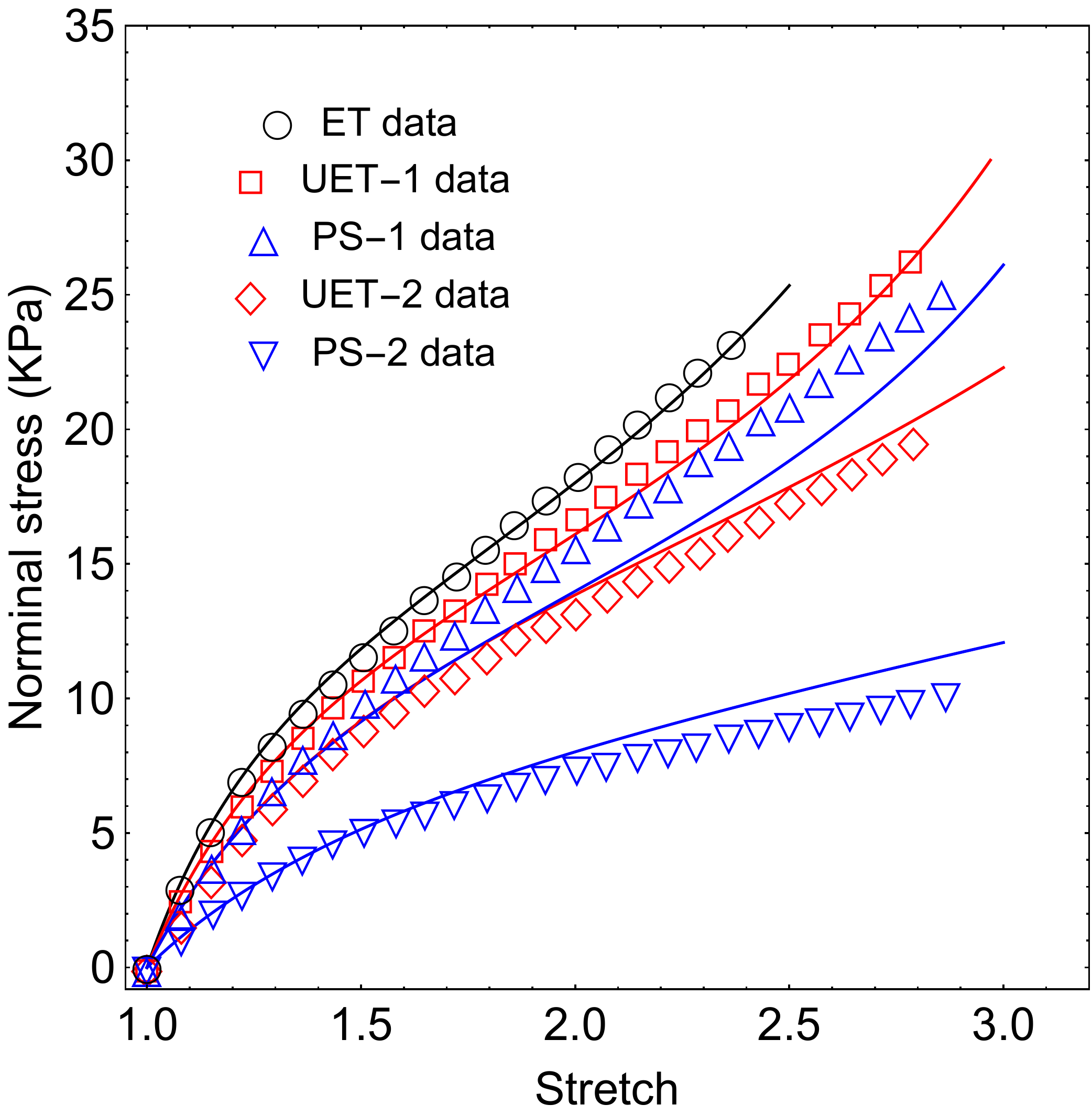}
	\caption{Comparing the predicted and measured nominal stresses for Tetra-PEG gel using the model of Eq.  (\ref{cauchy stress Langevin}) with $G=7.48$KPa, $N=32$. The experimental data is extracted from \cite{katashima2012strain}.} 
	\label{fig:gel} 
\end{figure}

\begin{figure}[]
	\centering    
	\includegraphics[width=60.00mm,height=60.00mm]{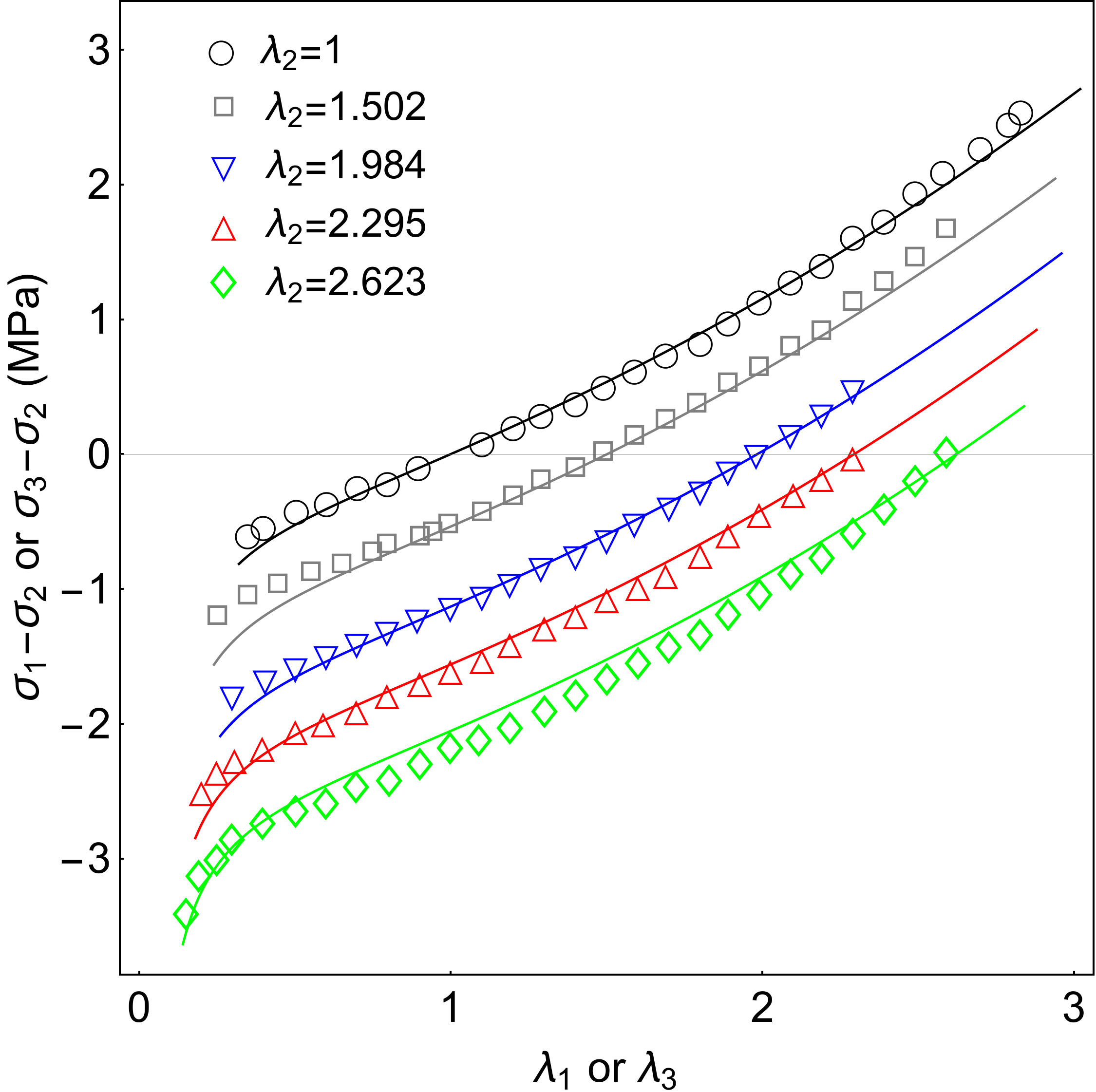}
	\caption{Comparing the predicted and measured true stresses for vulcanized rubber using the isotropic model of Eq.  (\ref{cauchy stress Gaussian}) with $G=0.504$MPa. The experimental data is extracted from \cite{james1975strain}.} 
	\label{fig:james} 
\end{figure}

In this Appendix, we present additional numerical examples to further validate the predictive capability of the proposed model. As shown in Figs.~\ref{fig:TPE} – \ref{fig:james}, the model exhibits remarkable agreements with experimental data using only two physical parameters.
A slight deviation is observed in the case of the Tetra-PEG gel (Fig.~\ref{fig:gel}), where the chain length is relatively short ($N = 32$). This discrepancy can be attributed to statistical fluctuations inherent to systems with small numbers of degrees of freedom. According to  statistical mechanics, macro fluctuations scale as $1/\sqrt{N}$ for a system with $N$ particles; see, for example, \cite{blundell2010concepts}. Therefore, the current model may lose quantitative accuracy when applied to polymer networks composed of very short chains.

\bibliographystyle{elsarticle-harv} 
\bibliography{ref}
\end{document}